\newcommand\submitms{y}		

\documentclass[apj]{emulateapj}
\usepackage{apjfonts}

\usepackage{ifthen}
\usepackage{natbib}
\usepackage{amssymb, amsmath}
\usepackage{appendix}
\usepackage{etoolbox}
\bibpunct[, ]{(}{)}{,}{a}{}{,}

\usepackage[
bookmarks=true,           
bookmarksnumbered=true,   
colorlinks=true,          
citecolor=blue,           
linkcolor=blue,           
menucolor=blue,           
urlcolor=blue,            
linkbordercolor={0 0 1},  
pdfborder={0 0 1},
frenchlinks=true]{hyperref}

\providecommand{\adsurl}[1]{\href{#1}{ADS}}

\makeatletter
\patchcmd{\NAT@citex}
  {\@citea\NAT@hyper@{%
     \NAT@nmfmt{\NAT@nm}%
     \hyper@natlinkbreak{\NAT@aysep\NAT@spacechar}{\@citeb\@extra@b@citeb}%
     \NAT@date}}
  {\@citea\NAT@nmfmt{\NAT@nm}%
   \NAT@aysep\NAT@spacechar\NAT@hyper@{\NAT@date}}{}{}

\patchcmd{\NAT@citex}
  {\@citea\NAT@hyper@{%
     \NAT@nmfmt{\NAT@nm}%
     \hyper@natlinkbreak{\NAT@spacechar\NAT@@open\if*#1*\else#1\NAT@spacechar\fi}%
       {\@citeb\@extra@b@citeb}%
     \NAT@date}}
  {\@citea\NAT@nmfmt{\NAT@nm}%
   \NAT@spacechar\NAT@@open\if*#1*\else#1\NAT@spacechar\fi\NAT@hyper@{\NAT@date}}
  {}{}
\makeatother

\makeatletter
\DeclareRobustCommand{\lowcase}[1]{\@lowcase#1\@nil}
\def\@lowcase#1\@nil{\if\relax#1\relax\else\MakeLowercase{#1}\fi}
\makeatother

\newcommand\chisq{\ifmmode{\chi\sp{2}}\else\math{\chi\sp{2}}\fi}
\newcommand\redchisq{\ifmmode{ \chi\sp{2}\sb{\rm red}}
                    \else\math{\chi\sp{2}\sb{\rm red}}\fi}

\DeclareSymbolFont{UPM}{U}{eur}{m}{n}
\DeclareMathSymbol{\umu}{0}{UPM}{"16}
\let\oldumu=\umu
\renewcommand\umu{\ifmmode\oldumu\else\math{\oldumu}\fi}
\newcommand\micro{\umu}
\renewcommand\micron{\micro m}
\newcommand\microns{\micron}

\let\oldsim=\sim
\renewcommand\sim{\ifmmode\oldsim\else\math{\oldsim}\fi}
\let\oldpm=\pm
\renewcommand\pm{\ifmmode\oldpm\else\math{\oldpm}\fi}
\newcommand\by{\ifmmode\times\else\math{\times}\fi}

\newbox{\wdbox}
\renewcommand\c{\setbox\wdbox=\hbox{,}\hspace{\wd\wdbox}}
\renewcommand\i{\setbox\wdbox=\hbox{i}\hspace{\wd\wdbox}}

\newcount\timect
\newcount\hourct
\newcount\minct
\newcommand\now{\timect=\time \divide\timect by 60
         \hourct=\timect \multiply\hourct by 60
         \minct=\time \advance\minct by -\hourct
         \number\timect:\ifnum \minct < 10 0\fi\number\minct}

\catcode`@=11

\newcommand\comment[1]{}

\newcommand\commenton{\catcode`\%=14}
\newcommand\commentoff{\catcode`\%=12}

\renewcommand\math[1]{$#1$}
\newcommand\mathshifton{\catcode`\$=3}
\newcommand\mathshiftoff{\catcode`\$=12}

\comment{alignment tab}
\let\atab=&
\newcommand\atabon{\catcode`\&=4}
\newcommand\ataboff{\catcode`\&=12}

\let\oldmsp=\sp
\let\oldmsb=\sb
\def\sp#1{\ifmmode
           \oldmsp{#1}%
         \else\strut\raise.85ex\hbox{\scriptsize #1}\fi}
\def\sb#1{\ifmmode
           \oldmsb{#1}%
         \else\strut\raise-.54ex\hbox{\scriptsize #1}\fi}
\newbox\@sp
\newbox\@sb
\def\sbp#1#2{\ifmmode%
           \oldmsb{#1}\oldmsp{#2}%
         \else
           \setbox\@sb=\hbox{\sb{#1}}%
           \setbox\@sp=\hbox{\sp{#2}}%
           \rlap{\copy\@sb}\copy\@sp
           \ifdim \wd\@sb >\wd\@sp
             \hskip -\wd\@sp \hskip \wd\@sb
           \fi
        \fi}
\def\msp#1{\ifmmode
           \oldmsp{#1}
         \else \math{\oldmsp{#1}}\fi}
\def\msb#1{\ifmmode
           \oldmsb{#1}
         \else \math{\oldmsb{#1}}\fi}
\def\supon{\catcode`\^=7}
\def\supoff{\catcode`\^=12}
\def\subon{\catcode`\_=8}
\def\suboff{\catcode`\_=12}
\def\supsubon{\supon \subon}
\def\supsuboff{\supoff \suboff}

\newcommand\actcharon{\catcode`\~=13}
\newcommand\actcharoff{\catcode`\~=12}

\newcommand\paramon{\catcode`\#=6}
\newcommand\paramoff{\catcode`\#=12}

\comment{And now to turn us totally on and off...}

\newcommand\reservedcharson{ \commenton  \mathshifton  \atabon  \supsubon 
                             \actcharon  \paramon}

\newcommand\reservedcharsoff{\commentoff \mathshiftoff \ataboff \supsuboff 
                             \actcharoff \paramoff}

\catcode`@=12
\reservedcharsoff

\reservedcharson

\comment{ Must have ONLY ONE of these... trust these macros, they work

}
\reservedcharsoff

\actcharon
\if\submitms y

\else

\comment{
\received{}
\revised{}
\accepted{}
}
\fi

\reservedcharson

\shorttitle{3D Retrieval}
\shortauthors{Blecic {\em et al.}}

\begin{document}

\slugcomment{Published in {\em ApJ}. 2017 October 23, 848:127 (24pp)}

\title {The Implications of 3D Thermal Structure on 1D Atmospheric Retrieval}

\author{Jasmina Blecic\altaffilmark{1},\altaffilmark{2}, Ian Dobbs-Dixon\altaffilmark{1},\altaffilmark{2}, Thomas Greene\altaffilmark{3}}

\affil{\sp1 NYU Abu Dhabi, Abu Dhabi, UAE}
\affil{\sp2 Max-Plank-Institut f{\"u}r Astronomie, K{\"o}nigstuhl 17, D-69117 Heidelberg, Germany}
\affil{\sp3 NASA Ames Research Center, Space Sciece and Astrobiology Division, M.S. 245-6; Moffett Field, CA 94035, USA} 

\email{jasmina@nyu.edu}

\begin{abstract}

Using the atmospheric structure from a 3D global radiation-hydrodynamic simulation of HD 189733b and the open-source
Bayesian Atmospheric Radiative Transfer (BART) code, we investigate
the difference between the secondary-eclipse temperature structure
produced with a 3D simulation and the best-fit 1D retrieved
model. Synthetic data are generated by integrating the 3D models over
the {\em Spitzer}, the {\em Hubble Space Telescope (HST)}, and {\em the James Web Space Telescope (JWST)} bandpasses, covering the
wavelength range between 1 and 11 {\micron} where most spectroscopically
active species have pronounced features. Using the data from different
observing instruments, we present detailed comparisons between the
temperature--pressure profiles recovered by BART and those from the 3D
simulations. We calculate several averages of the 3D thermal structure
and explore which particular thermal profile matches the retrieved
temperature structure. We implement two temperature parameterizations
that are commonly used in retrieval to investigate different thermal profile
shapes. To assess which part of the thermal structure is best
constrained by the data, we generate contribution functions for both our
theoretical model and each of our retrieved
models. Our conclusions are strongly affected by the spectral
resolution of the instruments included, their wavelength coverage, and
the number of data points combined. We also see some limitations in
each of the temperature parametrizations, as they are not able to
fully match the complex curvatures that are usually produced in hydrodynamic
simulations.  The results show that our 1D retrieval is
recovering a temperature and pressure profile that most closely
matches the arithmetic average of the 3D thermal structure.  When we use a
higher resolution, more
data points, and a parametrized temperature profile that allows more
flexibility in the middle part of the atmosphere, we find a
better match between the retrieved temperature and pressure profile
and the arithmetic average. The {\em Spitzer} and {\em HST} simulated
observations sample deep parts of the planetary atmosphere and provide
fewer constraints on the temperature and pressure profile, while the
{\em JWST} observations sample the middle part of the atmosphere,
providing a good match with the middle and most complex part of the
arithmetic average of the 3D temperature structure.

\end{abstract}


\section{INTRODUCTION}
\label{intro}

Well known from our own solar system planets and general circulation
models, planetary atmospheres are inherently 3D. The complex network
of atmospheric dynamics, chemistry, planetary rotation, circulation,
and stellar irradiation drives the planetary atmospheres to form
non-uniform temperature, chemical, and cloud structures not just in the
vertical direction, but also in the longitudinal and latitudinal
directions \citep[e.g.,][]{showman2009, Dobbs-DixonAgol2013-RHD}.

To compare 1D models with observations, one can use self-consistent
theory-driven forward models \citep[e.g.,][]{FortneyEtal2005apjlhjmodels, FortneyEtal2006apjAtmDynamics,
BurrowsSudarskyHubeny2006apjTheorySpectra,
KnutsonEtal2007natHD189733b} or observation-driven retrieval
models \citep{MadhusudhanSeager2009ApJ-AbundanceMethod, LineEtal2014-Retrieval-II, Waldmann2015-TAU, Blecic2016arXiv-dissertation}. In general, forward techniques try to
include all known physical and chemical processes to describe
atmospheric thermal structure and chemical composition. Retrieval
techniques, on the other hand, disfavor complex, time-consuming model
calculations because of their computationally demanding
iterative-statistical approach. Thus, one has to use simplified parametrized approaches
that can mimic a wide variety of possible physical and chemical
scenarios. In order to constrain model parameters with the
observations, the forward approach must choose a limited set of
tuning parameters and fix the remaining parameters. The retrieval approach,
on the other hand, performs robust exploration of the parameter
phase space using statistical algorithms \citep{MadhusudhanSeager2010, BennekeSeager2012-Retrieval, LeeEtal2012-CF, LineEtal2014-Retrieval-II, BennekeEtal2015ApjCOratios}.

The complexity of these models is largely determined by the nature of
the data at hand. For the low-resolution disk-integrated spectra usually
observed today, the initial assumption of a 1D temperature and
pressure profile and chemical composition seems to be
appropriate \citep{Swain2013-WASP12b, Burrows2014-review, Hansen2014features, LaughlinLissauer2015}. In addition, only a few data points are usually gathered during these observations,
which further limits the complexity of the retrieved model.

With the spectral resolution and coverage of current space telescopes,
one has to be careful not to overinterpret the broadband emission
spectra, claiming molecular bands rather than the astrophysical and
instrumental noise \citep{Burrows2014-review}. However, with the
advent of new spectral instruments, particularly the {\em James Web
Space Telescope (JWST)}, our prospects of performing detailed
atmospheric characterization are more promising, raising the question of
whether simplified assumptions like 1D thermal and chemical profiles
have become inadequate.

In this paper, we investigate how well these planet-averaged
assumptions correspond to the realistically complex atmospheric
dynamics and chemistry. To our knowledge, this is the first exploration of the limitations of 1D atmospheric retrieval that concentrates on assessing how well
the reverse approach can retrieve the inherently 3D temperature structure
during secondary eclipse. A recent paper
from \citet{FengLineEtal2016ApJ-2TPs} does investigate biases that
result from 1D assumptions within retrieval, but they use two thermal
profiles and study how non-isotropic temperature distributions can
impact retrieval results. Here, we use a complex 3D thermal structure
that comes from a hydrodynamic solution to generate a high-resolution
model. By simulating different observing instruments, we generate data
points with uncertainties and pass them to retrieval. We then
explore which particular temperature profile is revealed by the
retrieval using different data sets.  We discuss which results are
produced with the currently available and future instruments, and in
particular, how well the {\em JWST}
spectra will constrain the temperature and pressure profiles and the chemical
composition of transiting exoplanets.

The paper is organized as follows: in Section \ref{sec:Met} we
describe the method and tools used in this analysis; in
Section \ref{sec:model}, we present our high-resolution synthetic
model; in Section \ref{sec:retr} we describe the retrieval setup, two
temperature parametrizations, and the approach we use to average the
initial 3D thermal structure and contribution functions so we can
compare them with the retrieval output; in Section \ref{sec:res} we
show results of our retrieval analyses using {\em JWST} data set
alone and combined data sets of the {\em HST} and {\em Spitzer} and discuss
their implications. In Section \ref{sec:conc} we state our conclusions
and describe the reproducible-research license (RR) that accompanies the
software developed for this analysis. We also provide the webpage with the code, results, and
plots produced for this paper (\href{https://github.com/dzesmin/RRC-BlecicEtal-2017-ApJ-3Dretriev}{github.com/dzesmin/RRC-BlecicEtal-2017-ApJ-3Dretriev}). In
Appendix \ref{sec:Appendix-A} we explore possible thermal shapes of the 
two temperature parametrizations commonly used in retrieval, and in
Appendix \ref{sec:Appendix-B} we present our retrieval results using
all simulated data sets together and each of the data sets
separately.

\section{Method}
\label{sec:Met}

We use the output of our 3D radiative-hydrodynamic simulation of HD
189733b \citep{Dobbs-DixonAgol2013-RHD} to produce a
high-resolution secondary-eclipse emergent spectrum (see
Section \ref{sec:RHD}). We pass this spectrum through our
observational simulator (see Section \ref{sec:OBS}) to generate
the data points with associated uncertainties at a particular
resolution, assuming we have used {\em Spitzer}, {\em HST}, and {\em
JWST} observations \citep{Greene2016}. Using the Bayesian
Atmospheric Radiative Transfer code, BART (see
Section \ref{sec:BART}), we then run retrievals on different
combinations of data sets and explore the resulting posterior
distributions of model parameters to assess how well the observations
discriminate between different physical and chemical models
\citep[Chapters 5]{Blecic2016arXiv-dissertation,
Cubillos2016arXiv-dissertation}. Comparing our retrieved temperature
structure with our 3D inputs, we are particularly interested in which
temperature profile is revealed by the retrieval. Conclusions reached
in this analysis are independent of the validity of the detailed
solution provided by the hydrodynamic solution. Our goal was to
investigate how well a set of temperature--pressure profiles describing
a 3D structure of an object can be retrieved with the 1D retrieval
approach. In this paper, we exclude clouds and hazes from the
analysis.

We used the existing numerical tools tha have been well tested in the
literature \citep{Dobbs-DixonAgol2013-RHD,
Blecic2016arXiv-dissertation, BlecicEtal2016-TEA, Greene2016} and
developed several new ones to perform the 3D--1D comparison. All of the
tools are open-source software available to the community under the
RR license (see Section \ref{sec:conc}). Below, we
describe each of them and give webpage links where they can be
found. Additional information on the algorithms can be found in the
referenced papers.

\subsection{RHD}
\label{sec:RHD}

RHD is a 3D radiative-hydrodynamic atmospheric
simulator \citep{Dobbs-DixonAgol2013-RHD}. The code solves the
fully comprehensive Navier-Stokes equations coupled with the
wavelength-dependent radiative transfer to asses the hydrodynamics and
radiative capacity of the entire planet envelope. The equations are
solved in spherical coordinates with the resolution (\math{N\sb{r}},
\math{N\sb{\phi}}, \math{N\sb{\theta}}) = (100, 160, 64), where \math{r}
is the radial distance, \math{\phi} is the longitude,
and \math{\theta} is the latitude. Transfer of energy via radiation
employs a frequency-dependent two-stream
approximation \citep{Mihalas1978}. The full planetary spectrum is
divided into 30 bins using averaged frequency-dependent opacities
from \citet{sharp2007}. The code considers absorption due to the four
main and most spectroscopically active species in hot Jupites,
H\sb{2}O, CO, CO\sb{2}, and CH\sb{4}. To mimic the effect of clouds,
an additional opacity is added, consisting of both a gray and a Rayleigh
scattering component. There has been some criticism \citep[e.g.,][]{Amundsen2014, Amundsen2016} that this simplified averaged approach to radiative transfer will yield erroneous
thermal profiles when compared to more sophisticated techniques used
by other groups \citep{showman2009}. However, \citet{Dobbs-DixonAgol2013-RHD} undertook a detailed comparison between the simulations and available observations, showing that the results compared favorably for
transit, emission, and phase curve observations, suggesting that the
calculated thermal profiles are sufficiently adequate for our purposes here.
More on the
radiative-hydrodynamic solution of HD 189733b can be found
in \citet{Dobbs-DixonAgol2013-RHD}. The output of this code used in
this analysis can be found
on \href{https://github.com/dzesmin/RRC-BlecicEtal-2017-ApJ-3Dretriev}{github.com/dzesmin/RRC-BlecicEtal-2017-ApJ-3Dretriev}.

\subsection{BART}
\label{sec:BART}

BART \citep[][Chapters
5]{Blecic2016arXiv-dissertation, Cubillos2016arXiv-dissertation},
initializes a model for the atmospheric retrieval calculation,
generates thousands of theoretical model spectra using parametrized
pressure and temperature profiles and line-by-line radiative transfer
calculations, and employs a statistical package to compare the
models with the observations \citep[][in prep.]{BlecicEtal2017-BART,
CubillosEtal2017-BART, HarringtonEtal2017-BART}. Given transit or
eclipse observations at multiple wavelengths, BART retrieves the
thermal profile and chemical abundances of the selected atmospheric
species. It initializes a model atmosphere using the Thermochemical
Equilibrium Abundances (TEA) code \citep{BlecicEtal2016-TEA} or a
vertically uniform abundances-profile routine, calculates model
spectra using a radiative-transfer routine,
Transit \citep{Rojo2006PhDtransit}, and is driven through the
parameter space using the Multi-core Markov-chain Monte Carlo
statistical algorithm \citep{CubillosEtal2017apjRednoise}.

Transit's emission spectra models agree with models from Caroline Morley
within a few percent \citep[see Figures 5.5 and 5.6 in][Chapter
5]{Cubillos2016arXiv-dissertation}. Our opacity spectra are consistent
with those from \citet{sharp2007} \citep[see Figure 5.7 in][Chapter
5]{Cubillos2016arXiv-dissertation}. To perform a retrieval validation
test, we applied BART to synthetic observations of a
hot-Jupiter planet with the characteristics of the HD 209458
system \citep[see Figure 5.8 in][Chapter
5]{Cubillos2016arXiv-dissertation}. The best-fitting model and the
posterior distributions of the temperature profile and abundances
agree within the 1\math{\sigma} credible region of the input
values.

BART was applied to {\em Spitzer}, {\em HST} and ground-based
eclipse observations of the hot-Jupiter planet WASP-43b \citep[Section
5.4 in][Chapter 5]{Blecic2016arXiv-dissertation} and to the {\em
Spitzer} and {\em HST} transit observations of the Neptune-sized
planet HAT--P-11b \citep[Section 5.4 in][Chapter
5]{Cubillos2016arXiv-dissertation}. For the analysis of WASP-43b, our models
confirmed a decreasing temperature with pressure, a solar water
abundance and a mildly enhanced C/O ratio consistent with previous
analyses \citep{LineEtal2013-Retrieval-I, BlecicEtal2014apjWASP43b,  kataria2015atmospheric,
BennekeEtal2015ApjCOratios}. For HAT--P-11b,
we reproduced the conclusions of \citet{FraineEtal2014natHATP11bH2O}
by constraining the H\sb{2}O abundance and finding an atmosphere
enhanced in heavy elements.

In the following sections, we give more details about each of the
independently working routines of BART. BART is written in Python and
C, and is available to the community under an open-source
RR license
via \href{https://github.com/exosports/BART}{https://github.com/exosports/BART}.

\subsubsection{TEA}
\label{sec:TEA}

TEA \citep{BlecicEtal2016-TEA} calculates the mixing
fractions of gaseous molecular species following the method
by \citet{WhiteJohnsonDantzig1958JGibbs}
and \citet{Eriksson1971}. Given a \math{T-P} profile and elemental
abundances, TEA determines the species abundances by minimizing the
total Gibbs free energy of the system, using an iterative Lagrangian
steepest-descent method that minimizes a multivariate function under
constraint. To guarantee physically plausible positive mixing
fractions, TEA implements the lambda correction algorithm. TEA is
tested against the analytical models developed
by \citet{BurrowsSharp1999apjchemeq}, \citet{HengTsai2016ApJ-analuticalModels}, and the free
thermochemical equilibrium code Chemical Equilibrium with
Applications (CEA, \href{http://www.grc.nasa.gov/WWW/CEAWeb/}{http://www.grc.nasa.gov/WWW/CEAWeb/}, \citealp{GordonMcBride:1994}).
The code is open-source and available to the community under the
RR license
at \href{https://github.com/dzesmin/TEA}{https://github.com/dzesmin/TEA}.

\subsubsection{Transit}
\label{sec:Transit}

Transit is a 1D line-by-line radiative-transfer code originally
developed at Cornell University by Patricio Rojo and further modified
at the University of Central Florida \citep[][Chapters
5]{Blecic2016arXiv-dissertation, Cubillos2016arXiv-dissertation}. The
code can produce both transmission and hemisphere-integrated emission spectra
assuming hydrostatic balance, local thermodynamic equilibrium, and an 
ideal gas law. The opacities come from the HITRAN/HITEMP
database, \href{https://www.cfa.harvard.edu/hitran}
{https://www.cfa.harvard.edu/hitran}, where the line transitions are
due to electronic, rotational, and vibrational absorptions and
collision-induced absorption (CIA). The partition functions for the
HITRAN opacity sources were calculated based
on \citet{LaraiaEtal2011TIPS}. The CIA data come
from \citet{BorysowEtal2001-H2H2highT, Borysow2002-H2H2lowT} and \citet{RichardEtal2012-HITRAN-CIA}. Transit
takes an atmospheric model, line-list database, CIA, and molecular
information and calculates how the ray is traveling through the
planetary atmosphere for the desired geometry and wavelength range.

Transit performs a line-by-line opacity calculation by applying a
dynamical wavenumber sampling routine. The routine finds a minimum width
of the lines at every atmospheric layer to avoid undersampling of
narrow line profiles and oversampling of wide line profiles (which
would significantly slow down the line-by-line computation). The
resolution is then tuned down for the output spectrum to a
user-desired value (see the Transit User Manual at \href{https://github.com/exosports/transit/blob/master/doc/transit_user_manual.pdf}{https://github.com/exosports/transit/}). To speed up the spectrum calculation, Transit
provides an option to precalculate the opacity grid and interpolate
the opacities from the table.

The Transit code was used to detect water in the atmosphere of the
extrasolar planet HD 209458b using transit
spectroscopy \citep{Rojo2006PhDtransit, RojoEtal-HD209}. The code is
available to the community under the RR license
at \href{https://github.com/exosports/transit}{https://github.com/exosports/transit}.

\subsubsection{McCubed}
\label{sec:McCubed}

To explore the phase space of thermal profiles and species abundances
parameters, BART uses the Multi-core Markov-chain Monte Carlo
module \citep[McCubed,][]{Cubillos2016arXiv-dissertation}. McCubed is
an open-source fitting tool that uses Bayesian statistics to estimate
the best-fitting values and the credible regions for the model
parameters. It provides three routines to sample the parameter
posterior distributions: Differential-Evolution \citep[DEMC,
][]{terBraak2006markov}, Metropolis Random Walk (using multivariate
Gaussian proposals), or the Snooker-updater DEMC
algorithms \citep{terBraak2008differential}. The DEMC routine,
in particular, significantly improves the MCMC efficiency. By computing
the proposed jump for a given chain from the difference between the
parameter states of two other randomly selected chains, as the chains
approach convergence \citep{GelmanRubin1992}, DEMC adjusts the scale and
orients it along the desired distribution. McCubed is used in the
correlated-noise analyses applied to the exoplanet light curves
by \citet{CubillosEtal2017apjRednoise}. The code is written in Python
using several C-routines and is documented and available to the
community
via \href{https://github.com/pcubillos/}{https://github.com/pcubillos/MCcubed}.

\subsection{Observation Simulation Tool}
\label{sec:OBS}

\begin{figure}[t!]
\centering
\includegraphics[width=.55\textwidth, clip=True]{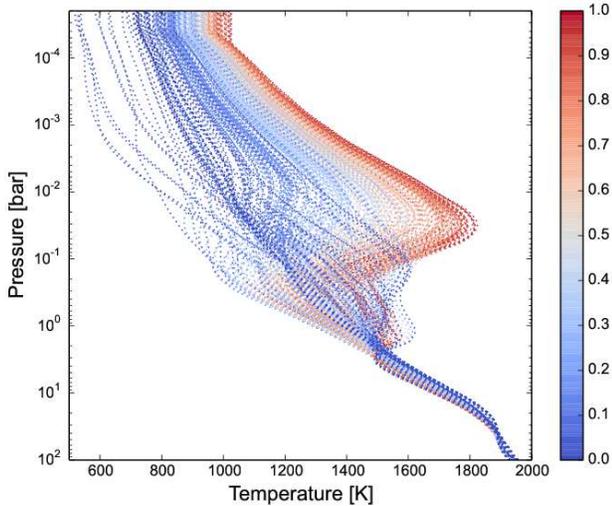}
\caption{\math{T-P} profiles of the HD 189733b dayside sampled on every 10 degrees latitude and longitude, generated using RHD. The colo bar shows \math{\mu} = cos(long)\,*\,cos(lat), with \math{\mu} = 1.0 (red) at substellar point, and \math{\mu} = 0.0 (blue) at terminator.}
\label{fig:3D T-P profiles}
\end{figure}

We developed a code to simulate observations of theoretical emission
spectra following \citet{Greene2016}. In addition to the {\em JWST}
(NIRISS, NIRCam/NIRSpec, and MIRI LRS) simulated observations
described in \citet{Greene2016}, we extended this code to simulate
{\em HST} (WFC3 G141) and {\em Spitzer} (IRAC Channel 1 and 2)
observations of modeled planets and their host stars. This code
computes the signal in electrons expected to be collected over a given
period, accounting for all observatory wavelength-dependent
throughputs as well as the duty cycle set by the readout options
without saturation in each instrument mode. System throughputs were
obtained from the JWST instrument teams or the STScI website (when
available), the \href{www.ipac.caltech.edu}{www.ipac.caltech.edu} website ({\em Spitzer} IRAC
channels 1 and 2), and the HST WFC3 documentation on the STScI
website. Noise from photo-electron Poisson statistics, detector
readouts, detector dark currents, and observatory backgrounds was also
included. A noise floor was also added to approximate the best results
achieved in the transit community literature or expected ({\em JWST})
for each mode (20 ppm for {\em HST} G141 and {\em JWST} NIRISS, 30 ppm
for {\em Spitzer} IRAC channels 1, 2 and {\em JWST} NIRCam, and 50 ppm
for {\em JWST} MIRI LRS).  These noise terms were combined in
quadrature to estimate the 1\math{\sigma} uncertainties in each
spectral bin.

\section{Theoretical Model}
\label{sec:model}

To start our analysis, we take one snapshot from the
radiative-hydrodynamic simulation of HD 189733b (when the upper atmosphere, \math{p \lesssim} 10 bars, of the simulation has reached a steady state) and process the 3D temperature and pressure
structure. The thermal structure of HD 189733b is strongly affected by
the presence of supersonic winds that efficiently advect the energy
from the day- to the nightside of the planet. The super-rotational
equatorial jet is present between 10\sp{-5} to 10 bars, and
counter-rotational jets are present at the higher latitudes.

The 3D temperature and pressure structure obtained from the
radiative-hydrodynamic solution is then interpolated on a constant
pressure grid (constrained between 2x10\sp{-5} and 10\sp{2} bars and
sampled 100 times uniformly in log space), and the temperature
profiles are extracted on every 10 degrees longitude and latitude
along the dayside of the planet. Figure \ref{fig:3D T-P profiles}
shows the ensemble of temperature and pressure profiles from the
dayside atmosphere of HD 189733b that was used in the analysis.

Utilizing TEA, we then used these temperature--pressure profiles to
calculate chemical species abundances. Assuming the solar elemental
composition from \citet[][]{AsplundEtal2009-SunAbundances}, TEA
included H, He, C, N, and O elemental species and the following molecular
species: H\sb{2}, CO, CO\sb{2}, CH\sb{4}, H\sb{2}O, N\sb{2}, HCN,
NH\sb{3}, C\sb{2}H\sb{2}, and C\sb{2}H\sb{4}. The mixing ratios of all
included species are calculated at each 3D location on the dayside hemisphere of HD189733.

\begin{figure}[h!]
\centering
\includegraphics[width=.5\textwidth, clip=True]{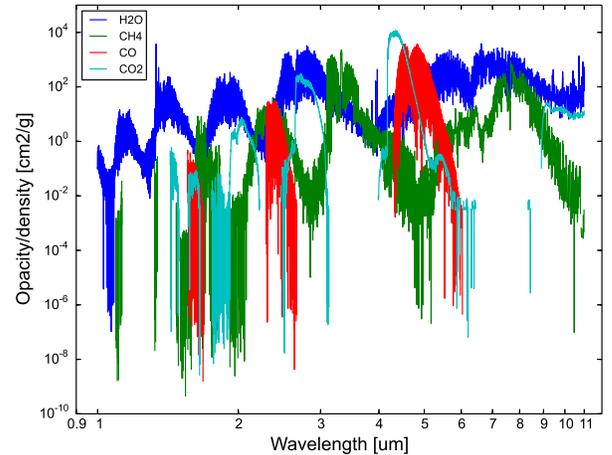}
\vspace{10pt}
\caption{Opacities of the four main molecular species (H\sb{2}O, CO, CO\sb{2}, and CH\sb{4}) used in our analysis, calculated at a temperature of 1500 K and a pressure of 1 bar. The gaps in the opacity curves come from the limited y-axis range (the values not shown are well below 10\sp{-12} cm\sp{2}\,g\sp{-1}). The line lists come from the HITRAN or HITEMP databases.}
\label{fig:4MajorOpacity}
\end{figure}

To produce planetary intensities at every location on the planet
surface, we used the Transit radiative-transfer code together with the
results from TEA and the line-by-line opacity data. We employed the HITRAN
and HITEMP databases to include the influence of the line-list
data. We choose four main molecular species, H\sb{2}O, CO, CO\sb{2}, and
CH\sb{4}, as they have the most significant spectral features on the
wavelength range of our interest (Figure \ref{fig:4MajorOpacity}). For
H\sb{2}O, CO, and CO\sb{2}, we used the HITEMP
database, \citet{rothman2010-hitemp} and for
CH\sb{4}, we used \citet{rothman2013-hitran2012}. The CIA opacities for H\sb{2}--H\sb{2}
and H\sb{2}--He came from \citet{RichardEtal2012-HITRAN-CIA}.

\begin{figure}[hb!]
\centering
\includegraphics[width=.52\textwidth, clip=True]{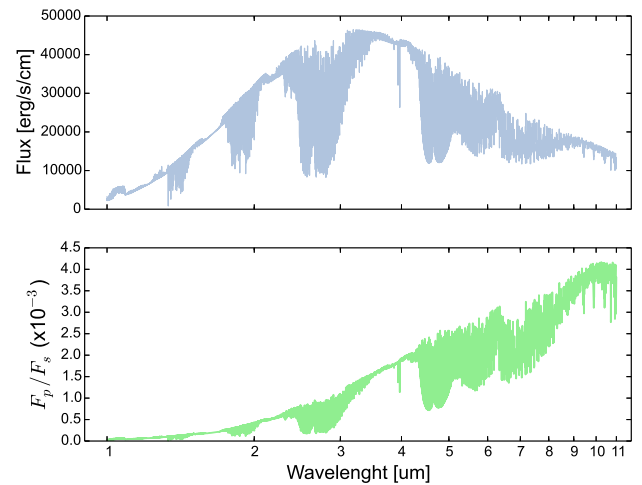}
\vspace{10pt}
\caption{Top: the HD 189733b planetary model spectra (flux) produced using the 3D \math{T-P} profiles from Figure \ref{fig:3D T-P profiles}. TEA was used to calculate species abundances, and Transit with HITRAN opacities was used to produce the spectra. Bottom: the high-resolution model spectrum (flux ratio) produced using the planetary model spectrum from the top panel and the Kurucz stellar model chosen based on the system parameters from Table \ref{tab:system}. These spectra are henceforth treated as a real high-resolution observation}.
\label{fig:fluxes}
\end{figure}

\begin{table}[ht!]
\centering
\caption{System Parameter Values}
\centering
\label{tab:system}
\begin{tabular}{lll}
\hline
Parameter & Value & Source \\
\hline
\hline
\math{R\sb{p} (R\sb{J}})    & 1.178    &  \citet{Triaud2009-HD189733b}  \\
\math{M\sb{p} (M\sb{J}})    & 1.138    &  \citet{Triaud2009-HD189733b}  \\
\math{log\,g\sb{p}} (cgs)        & 2.03     &  Calculated  \\
\math{R\sb{*} (R\sb{Sun}})  & 0.766    &  \citet{Triaud2009-HD189733b}  \\
\math{M\sb{*} (M\sb{Sun}})  & 0.840    &  \citet{Southworth2010}  \\
\math{T\sb{*}} (K)          & 5050     &  \citet{Bouchy2005-HD189733b}  \\
\math{log\,g\sb{*}} (cgs)        & 4.610    &  \citet{Southworth2010}  \\
Fe/H                 & -0.03    &  \citet{Triaud2009-HD189733b}  \\
\math{a}       (au)         & 0.0312    & \citet{Triaud2009-HD189733b}  \\
\hline
\end{tabular}
\end{table}

The planetary flux at the observer's location is calculated by
integrating the intensities at each location accounting for the
observer's angle. The top panel in Figure \ref{fig:fluxes} shows the
calculated dayside flux from the planet. The high-resolution emergent
model, flux ratio, shown in the bottom panel of Figure \ref{fig:fluxes} is calculated using the stellar grid models
from \citet{CastelliKurucz-2004new}. We adopted the system parameters
(planetary mass and radius, stellar metallicity, effective temperature,
mass, radius, and gravity, and the semimajor axis) listed in
Table \ref{tab:system}. These synthetic data are then used as our
high-resolution model spectrum for the subsequent observational and
retrieval analyses. From this point on, we treat this spectra as a
{\em real high-resolution observation}.

The software developed to couple the output from the RHD, TEA, and
Transit, and generate the high-resolution input model for the
observational tools and retrieval analysis is available at \href{
https://github.com/dzesmin/RRC-BlecicEtal-2017-ApJ-3Dretriev}{github.com/dzesmin/RRC-BlecicEtal-2017-ApJ-3Dretriev}
under the open-source license.

\subsection{1D Validation}
\label{sec:Valid}

To assess how well a 1D representation of HD 189733b matches the spectrum of BART, we compared the output of the radiative-transfer routine
from \citet{Dobbs-DixonAgol2013-RHD} to the Transit output. We used the 1D substellar
temperature and pressure profile from the HD 189733b hydrodynamic
simulation and passed it through the RHD radiative-transfer routine to
compare it to the substellar output from Transit using the same setup as
described above. To calculate the emission spectra from a RHD model,
\citet{Dobbs-DixonAgol2013-RHD} integrated through the modeled atmosphere,
calculating the net emergent flux as a function of wavelength at each
location on the planet. As these spectral calculations use a
snapshot from the RHD simulations, we are able to use a much
higher resolution in wavelength (as opposed to the necessarily
simplified radiative routine used while running concurrently with
the hydrodynamics). For the current calculations, we use 5000
wavelength points logarithmically spaced between 1.0 and 11.0 \math{\microns}. Further details and associated equations can be found in
\citet{Dobbs-DixonAgol2013-RHD}.

\begin{figure}[ht!]
\centering
\includegraphics[width=0.99\linewidth, clip=True]{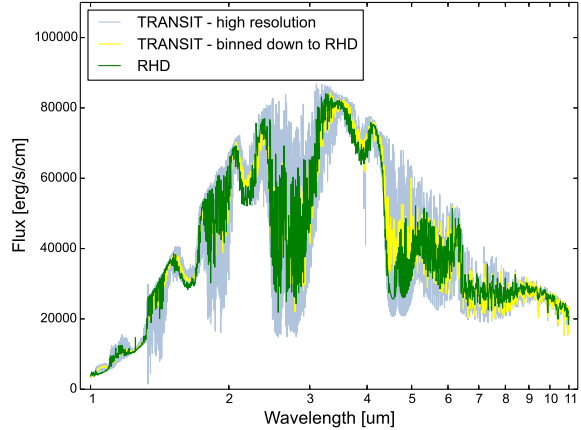}
\vspace{10pt}
\caption{Comparison between the substellar model spectra produced using the radiative-transfer rutine, Transit, and RHD. The high-resolution spectrum of Transit is shown in light blue, the Transit spectrum binned down to the RHD resolution is shown in light yellow, while the RHD spectrum is shown in green.}
\label{fig:Validation}
\end{figure}

Transit generates a spectrum with a high resolution, using a
line-by-line dynamical sampling (see Section \ref{sec:BART}), and tunes down the
resolution for the output results to 1 cm\sp{-1} in the wavenumber
space. Between 1 and 11 {\microns} this generates 9091 data points. To compare
outputs from Transit and RHD, we binned down the Transit resolution to the
same resolution as RHD, Figure \ref{fig:Validation}, showing a nice match between the two
radiative-transfer routines.

\subsection{Emission Spectrum Simulations}
\label{sec:synth}

We produced simulated secondary-eclipse observations from the
high-resolution 3D model given in Figure \ref{fig:fluxes} (bottom panel)
using the code described in Section~\ref{sec:OBS}. This was done for
the following instruments: the {\em JWST} instruments NIRISS, NIRCam (or NIRSpec), and MIRI LRS; for the {\em HST} (WFC3 G141) covering the wavelength
range from \math{\lambda} = 1.1 - 1.7 {\micron}, and for {\em
Spitzer} channels 1 (3.6 {\micron}) and 2 (4.5 {\micron}). We
focus our investigation on the wavelength range between 1 and
11 \math{\mu}m, as the most abundant molecular species show dominant
absorption features in this region.

Several recent studies have provided an assessment of how well {\em
JWST} will preform \citep{Doyon2012, Beichman2014, Barstow2015, Batalha2015,Kendrew2015P, Greene2016,HoweEtal2016,
 BatalhaLine2017AJ-JWST, Molliere2017-JWST} and speculated on which modes should be used
to answer certain questions. We decided to use the modes listed in
Table \ref{tab:jwst} because they provide large simultaneous
wavelength coverage, adequate spectral resolution, slitless operation,
and bright limits sufficient to observe the HD189733 system with high
throughput. Slitless spectra with good spatial sampling and stable
detectors have provided the best spectrophotometric
precision \citep{Kreidberg2014}. They also have the best spatial
sampling available on {\em JWST} over their wavelength ranges; good
sampling should minimize systematic errors that are due to intrapixel response
variations in the presence of pointing jitter \citep[e.g.,][]{Deming2009}.

\begin{table}[h!]
\strut\hfill\atabon
\caption{Selected Instrument Modes}
\vspace{5pt}
\label{tab:jwst}
\begin{tabular}{l@{ }ccc@{ }c@{ }c@{ }}
\hline
Instrument & Mode & Optics & \math{{\lambda} ({\micron})} & Native &
           Sampling \\ & & & & Resolution & (Pixels) \\
\hline
\hline
NIRISS & bright SOSS & GR700XD     & 1.0 - 2.5  & ~ 700  & \sim 25 \\
NIRCam & LW grism    & F322W2      & 2.5 - 3.9  & ~1400  & \sim 2 \\
NIRCam & LW grism    & F444W       & 3.9 - 5.0  & ~1550  & >2 \\
MIRI   & SLITLESS    & LRS prism   & 5.0 - 11   & ~100   & >2 \\
{\em HST} WFC3 & SLITLESS  & G141        & 1.1 - 1.7  & ~140   & 125 \\ 
{\em Spitzer} & IRAC       & Channel 1   & 3.2 - 3.9  & 0.65   & \sim 2 \\
{\em Spitzer} & IRAC       & Channel 2   & 4.0 - 5.0  & 0.74   & \sim 2 \\
\hline
\end{tabular}
\ataboff\hfill\strut
\end{table}

\begin{figure*}[ht!]
\centering
\includegraphics[width=0.9\linewidth, clip=True]{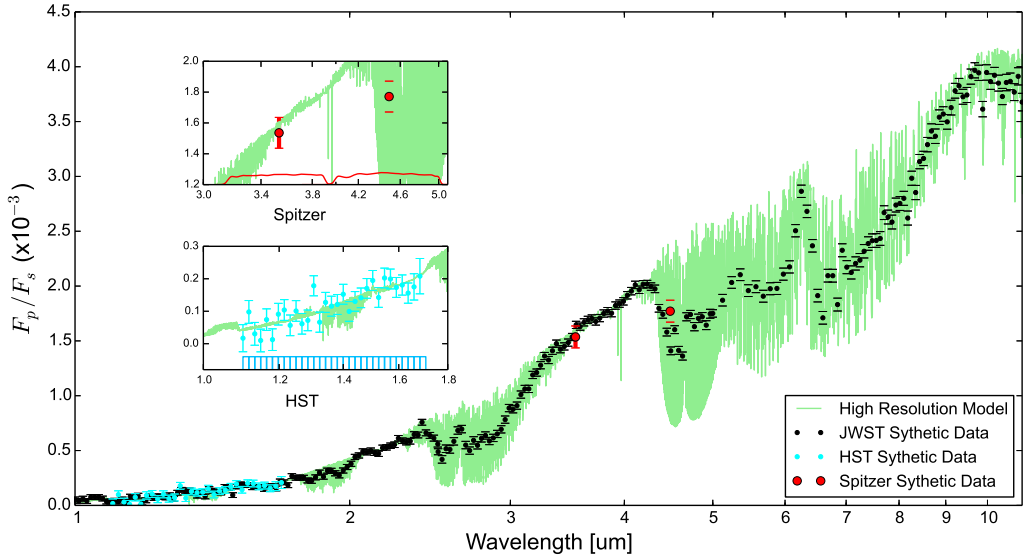}
\vspace{10pt}
\caption{High-resolution model spectrum from Figure \ref{fig:fluxes}, bottom panel, with synthetic data points and uncertainties for the {\em JWST} (black points), {\em HST} (blue points), and {\em Spitzer} (red points), produced using the observational tools from Section \ref{sec:OBS}. In the inset plots we zoom in on the {\em HST} and {\em Spitzer} data points and show their uncertainties along with the bandwidths of their channels/bins (2 for {\em Spitzer} in red, and 31 for {\em HST} in blue). We do not plot the bandwidths for the {\em JWST} in the main plot for clarity, as there are 229 tightly packed bins.}
\label{fig:SytheticModel}
\end{figure*}

All spectra were simulated for a single 1.8 hr secondary eclipse of
HD 189733b with equal time spent on the star and planet.  Signal
(\math{F\sb{p}/F\sb{*}}) and 1\math{\sigma} noise values are computed
for each spectral resolution element of each instrument mode.  The
simulated {\em JWST} data were binned to spectral resolving
power \math{R \leq 100} (only MIRI LRS has \math{R < 100}
at \math{\lambda \leq 7.5 {\micron}}).  Simulated {\em HST} WFC3 G141
data were binned to \math{R = 70}. These simulated signal and noise
values were then used as inputs for the retrieval
process. Figure \ref{fig:SytheticModel} shows the 3D model spectra
from all instruments together with the corresponding data points and
uncertainties.

\section{Retrieval Setup}
\label{sec:retr}

We performed retrieval runs with the goal to provide the observational
constraints derived from the {\em Spitzer}, {\em HST} and {\em JWST} instruments
separately and then again for different combinations of instruments.  The
data points and uncertainties for each of our runs come from the
convolution between our high-resolution model and the observation
simulation code described in Section \ref{sec:synth}.

The initial temperature and pressure model is chosen by running
several short trial runs or by taking the \math{T-P} parameter values
from the literature. The equilibrium mixing fractions of the desired
molecular species are calculated using TEA, providing a realistic
initial model atmosphere for our retrieval runs.

\begin{figure}[b!]
\vspace{-5pt}
\centering
\includegraphics[width=.55\textwidth, clip=True]{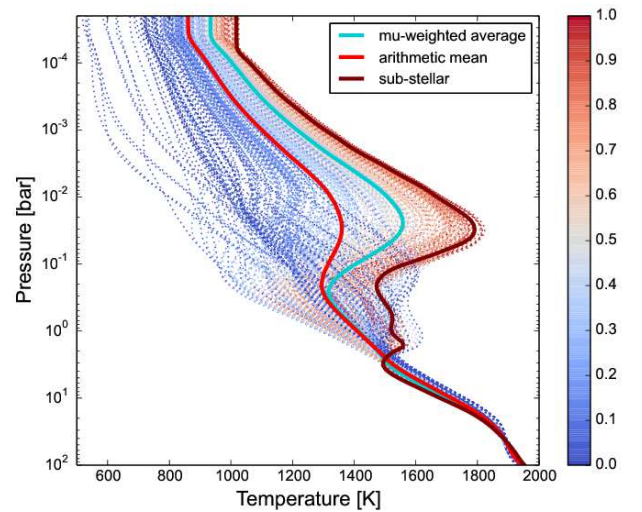}
\caption{\math{T-P} profiles of the HD 189733b dayside sampled every 10 degrees in latitude and longitude, generated using RHD. The colorbar shows \math{\mu} = cos(long)\,*\,cos(lat), with \math{\mu} = 1.0 (red) at the substellar point, and \math{\mu} = 0.0 (blue) at the terminator. In turquoise color, we show the \math{\mu}-weighted \math{T-P} profile, in red the arithmetic mean, and in magenta the substellar \math{T-P} profile.}
\label{fig:averages}
\vspace{-10pt}
\end{figure}

We apply two parametrization schemes to explore the shapes of the
best-fit \math{T-P} profiles and compare them to our initial 3D
thermal structure (see Section \ref{sec:TPs}). Each parametrization
carries a particular set of free parameters. The remaining four free
parameters come from the species scaling factors that we decided to
vary in our analysis: H\sb{2}O, CO, CO\sb{2}, and CH\sb{4}. We used
flat priors on all parameters, with boundary limits set to account for
all possible physically plausible solutions without constraints, to
allow McCubed to explore the parameter phase space
thoroughly. The temperature range is constrained between 300 and 3000
K, which corresponds to the range allowed by the HITRAN/HITEMP databases'
partition functions.

\begin{figure*}[th!]
\centering
\hspace{-20pt}\includegraphics[width=.49\textwidth]{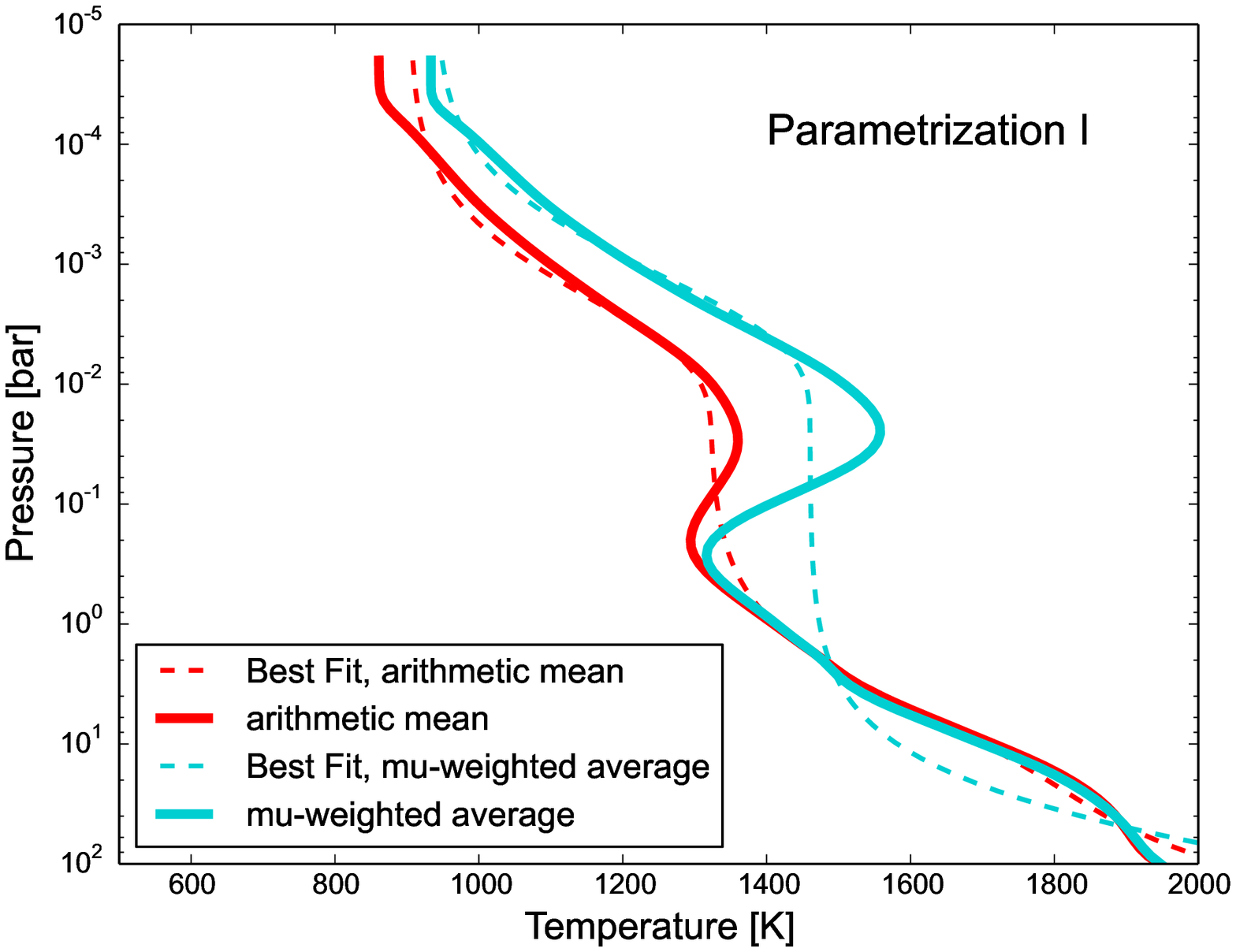}
\includegraphics[width=.49\textwidth]{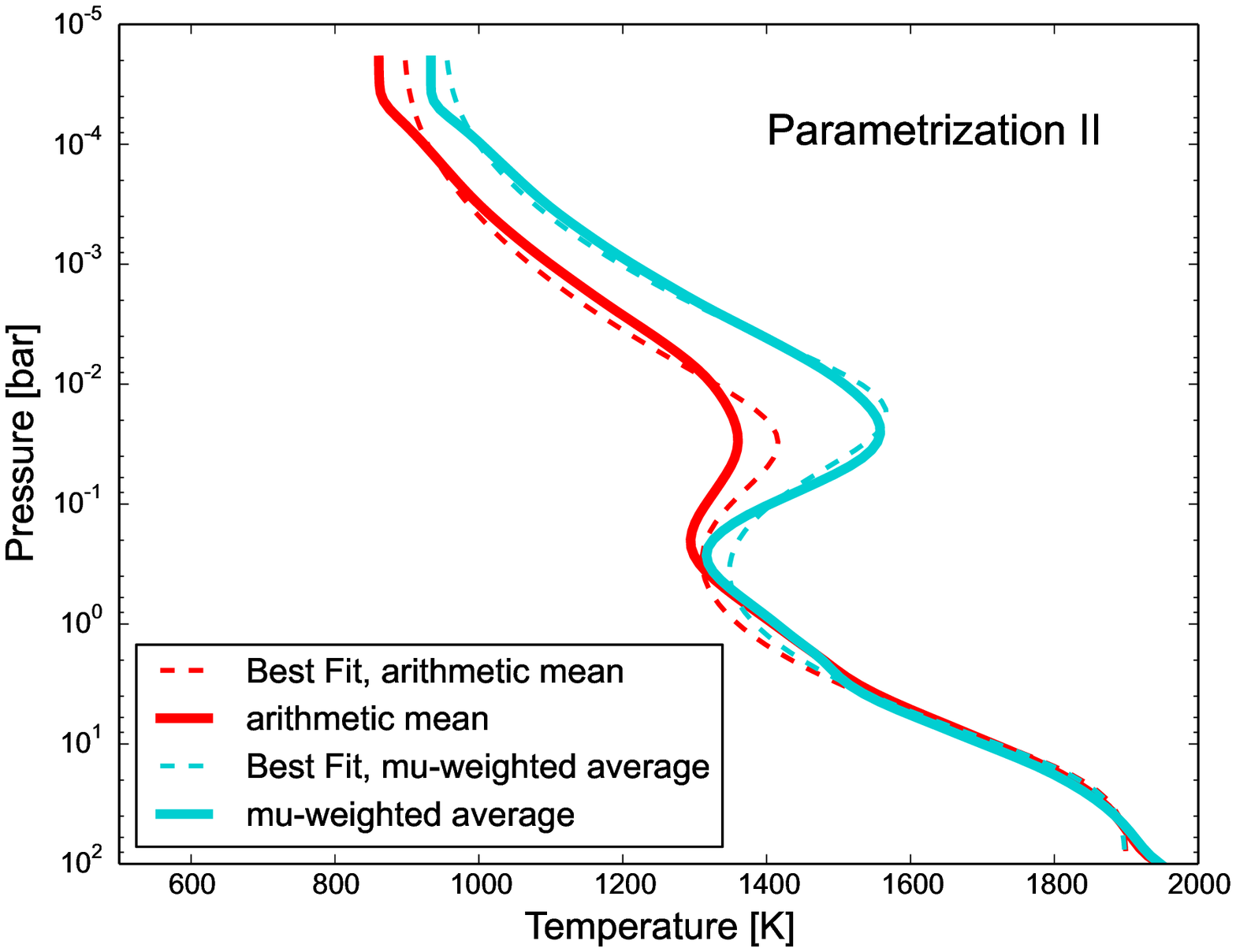}
\caption{Direct fit of the \math{\mu}-weighted average and arithmetic mean \math{T-P} profiles using each of the parametrizations. Parametrization I, left panel, is unable to generate the mid-atmosphere peak seen in both the arithmetic average and the \math{\mu}-weighted average, while Parametrization II, right panel, can potentially generate this curvature.}
\label{fig:MC3-PT-fits}
\end{figure*}

For each of our runs we use ten independent chains and enough
iterations (in the range of tens of thousands) until the Gelman and
Rubin convergence test for all free parameters dropped below
1\% \citep{GelmanRubin1992}. The transmission response functions for the  {\em JWST} and {\em HST} are top-hat functions, while the {\em
Spitzer} response functions for the channel 1 and 2 subarray
observations were taken from the {\em Spitzer} website,  \href{
http://irsa.ipac.caltech.edu}{http://irsa.ipac.caltech.edu}.  The system parameters used in the
analysis for the parametrized temperature model and to generate the
stellar spectrum from \citet{CastelliKurucz-2004new} are listed in
Table \ref{tab:system}.

Before presenting the results of the retrieval and comparing them to
the inputed 3D thermal structure, we apply several averages on our
theoretical model to understand which average is best represented by
the retrieval results. We compare these averages with the two temperature
parametrizations and calculate theoretical (derived from our 3D structure) and retrieved
contribution functions (derived from the best-fit retrieved models)
for each of our observational instruments.

In the following sections, we explore several methods of averaging our
3D \math{T-P} profiles (Section \ref{sec:aver}), present the two
temperature parametrization (Appendices \ref{sec:Line}
and \ref{sec:Madhu}), and discuss the way in which we calculate and average
theoretical and retrieved contribution functions
Sections \ref{sec:theo-cf} and \ref{sec:ret-cf}.

\subsection{Averaging}
\label{sec:aver}

To investigate how the retrieved \math{T-P} profile compares to the radiative-hydrodynamic model, we average the 3D dayside thermal structure of HD
189733b (Figure \ref{fig:3D T-P profiles}) in several
ways. Considering that the emergent flux is influenced by the
observer's angle, we first calculate \math{\mu}-weighted average,
where \math{\mu} is

\begin{equation}
\mu = \cos\,(lat)\,*\,\cos\,(long) \,,
\label{mu}
\end{equation}

\noindent and the weighted average is calculated as

\begin{equation}
X_{w} = \frac{\sum_{i}{X_{i}\,*\,w_{i}}}{\sum_{i}{w_{i}}} \,,
\label{weighted-aver}
\end{equation}

\noindent where \math{w} is the weight. For example, if we want to
average the 3D thermal structure provided by RHD, the weight
is \math{\mu} and since pressure is the independent
variable, \math{X\sb{i}}'s are temperatures at each location on the
dayside of HD 189733b.

We additionally consider the pure arithmetic average given by

\begin{equation}
X_{aver} = \frac{1}{n}\,\sum_{i=1}^{n}{X_{i}} \,.
\label{arithm-aver}
\end{equation}

Figure \ref{fig:averages} shows all \math{T-P} profiles generated with
RHD with the \math{\mu}-weighted \math{T-P} profile in turquoise,
arithmetic average in red, and substellar point \math{T-P} profile in
magenta.

\subsection{Temperature parametrization}
\label{sec:TPs}

Commonly, two temperature parametrizations are used in
retrievals \citep{MadhusudhanSeager2009ApJ-AbundanceMethod, LineEtal2013-Retrieval-III}. One approach is based on the parametrization described
in \citet{MadhusudhanSeager2009ApJ-AbundanceMethod} and the other has originally been developed by \citet{Guillot2010A-LinePTprofile}. We have
implemented both approaches to test which one produces a better match
to our simulated data.

In Appendix \ref{sec:Appendix-A}, we describe each
parametrization in detail and explore their possible shapes, revealing
the advantages and limitation of each approach. Here, we perform an
optimization test, where we directly fit our averaged \math{T-P}
profiles (the arithmetic average and \math{\mu}-weighted average) with
both parameterizations, looking for the best fit.

To fit these profiles, we used the McCubed general fitting tool and
allowed the MCMC to explore the phase space of both parametrizations
until it found the best fit. We used 5x10\sp{6} iterations in total
for each run within ten chains, and discarded the first 1000 burn-in
iterations. We allowed a wide range of our parameters and started our
chains in different positions of the phase space to avoid local
minima.

\begin{figure*}[t!]
\centering
\includegraphics[width=.5\textwidth, clip=True]{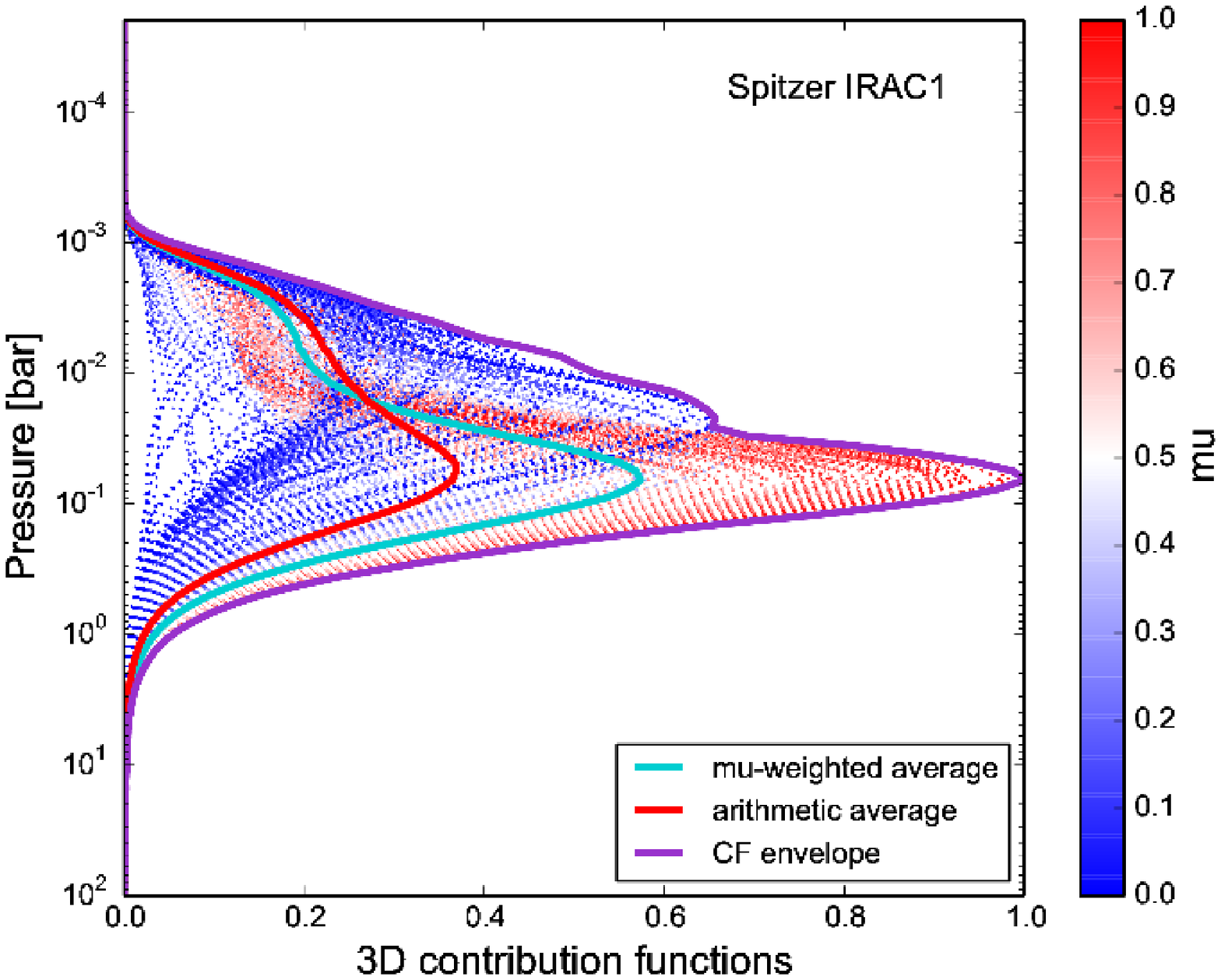}\hspace{-30pt}
\includegraphics[width=.5\textwidth, clip=True]{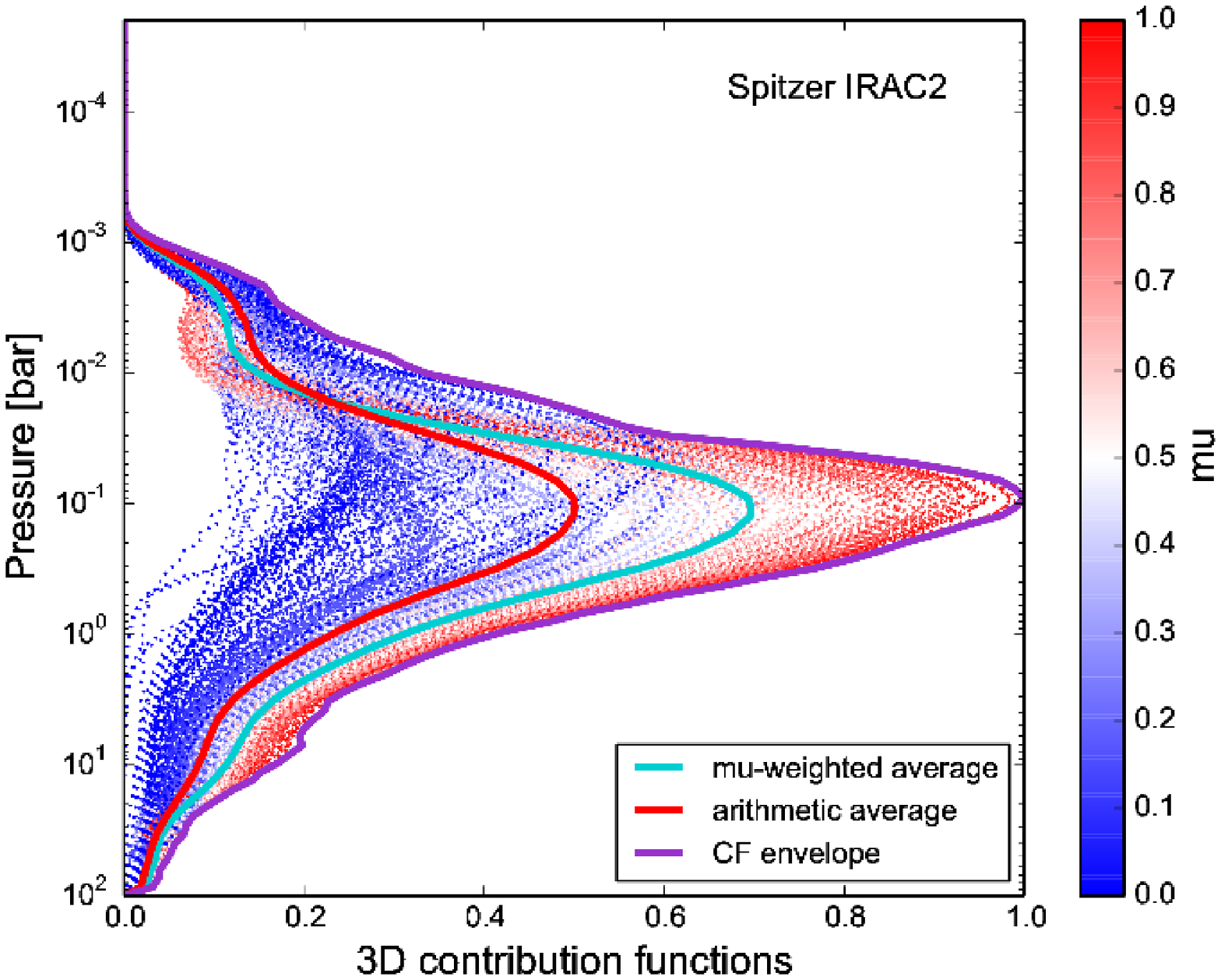}\hspace{-65pt}
\includegraphics[width=.5\textwidth, clip=True]{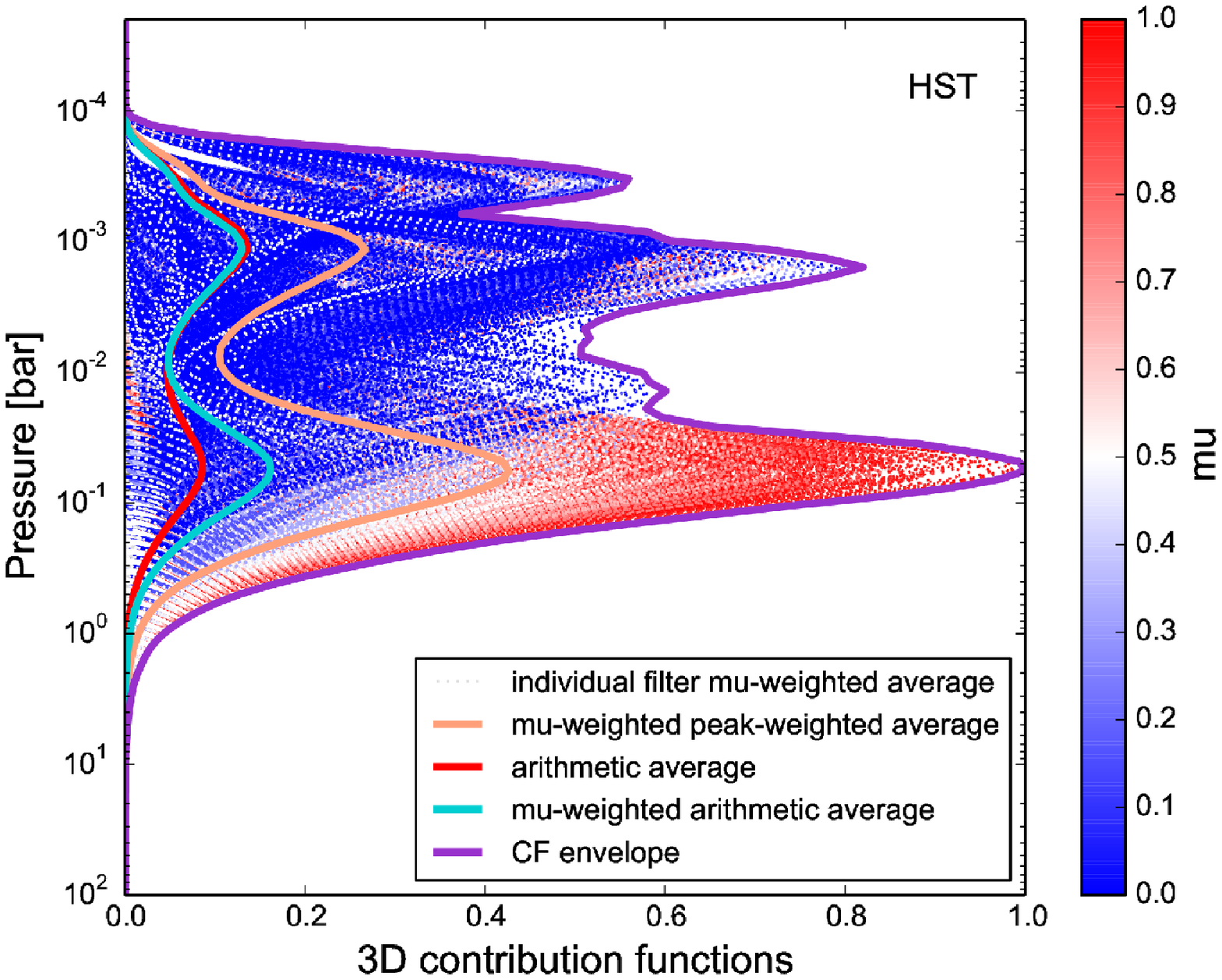}\hspace{-30pt}
\includegraphics[width=.5\textwidth, clip=True]{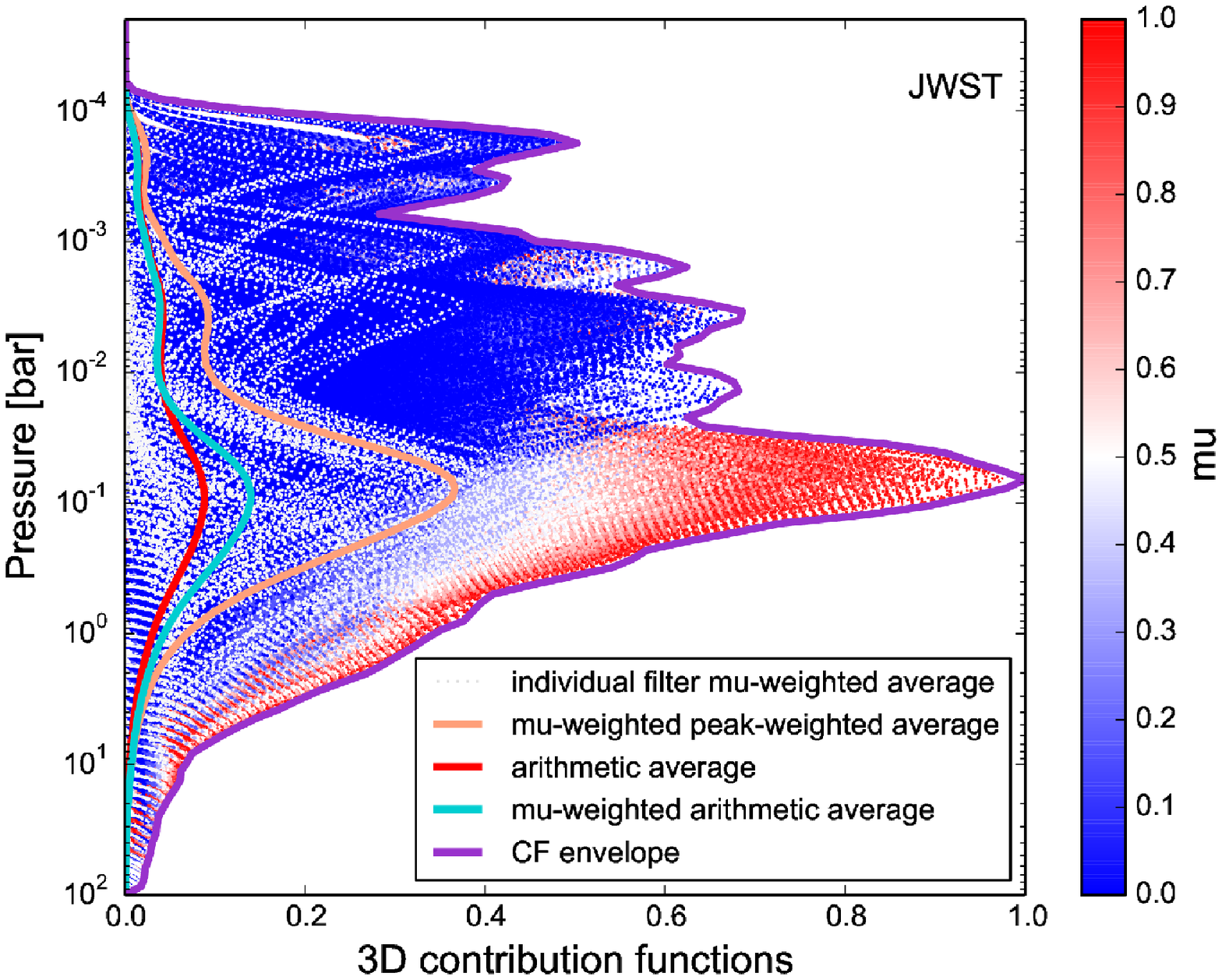}
\caption{3D theoretical contribution functions for the {\em JWST}, {\em HST}, and {\em Spitzer} overplotted with several averages (see Section \ref{sec:theo-cf}). All contribution functions are first scaled to their maximum peak value, so their maximum value is 1. The color code is based on the observer's angle \math{\mu}. For the {\em Spitzer} bandpasses, the turquoise curve is the \math{\mu}-weighted average, while the red curve is the plain arithmetic average. For the {\em HST} and {\em JWST}, we apply a pure arithmetic average for each filter and then again we use the  arithmetic average between the filters (red curve). For a more realistic average, we first apply the \math{\mu}-weighted average (the white dotted curves) for each bandpass (bin), and then the overall average is calculated either by applying the plain arithmetic average between bandpasses (turquoise curve), or by applying the peak-weighed average (orange curve). To encompass the full contribution from all 3D contribution functions, we also evaluate the contribution functions envelope by calculating the maximum contribution at each pressure level.}
\label{fig:Theor-CF-aver}
\end{figure*}

Figure \ref{fig:MC3-PT-fits} summarizes the advantages and limitations
of both parameterizations by showing the best-fit models to the
arithmetic average and the \math{\mu}-weighted average. The left panel
displays the limited capability of Parametrization I to reproduce
the complex curvatures seen in these profiles. Although an inversion is possible within both of these schemes, for certain sets of parameters (see Appendix \ref{sec:Line}, the types of inversions produced by the RHD simulations seen in Figure \ref{fig:averages}
cannot be generated.  Note also that as Appendix \ref{sec:Line}, Equation (\ref{LineTP})
gives a radiative solution, the profile must necessarily approach the
isothermal solution at high pressures.  The gradual steepening of
the \math{T-P} profile when approaching the radiative-convective boundary
cannot be captured in this model \citep[see
also][]{Guillot2010A-LinePTprofile}. However, as the
radiative-convective boundary is well below the photosphere at all
wavelengths, this fact does not affect the results of our retrieval.

The right panel of Figure \ref{fig:MC3-PT-fits} shows how well
Parametrization II is able to reproduce the complex curvatures seen in
both the \math{\mu}-weighted and the arithmetic averages. Using the inversion
set of equations (see Appendix \ref{sec:Madhu}), this approach is capable of generating the mid-atmosphere inversion peaks, as seen in
Figure \ref{fig:averages}. The approach allows the
exploration of a wide range of pressures where an inversion can occur,
and can differentiate mild and strong inversions (see Appendix \ref{sec:Madhu}). However, the main disadvantage of
this approach is that the inversion and non-inversion cases cannot be
covered with one set of equations, and each solution must be explored
separately in the retrieval (Appendix \ref{sec:Madhu}). This
limitation forbids a statistical assessment of models produced with
different numbers of free parameters.

\begin{figure*}[t!]
\centering
\hspace{-20pt}\includegraphics[width=.42\textwidth, height=6cm]{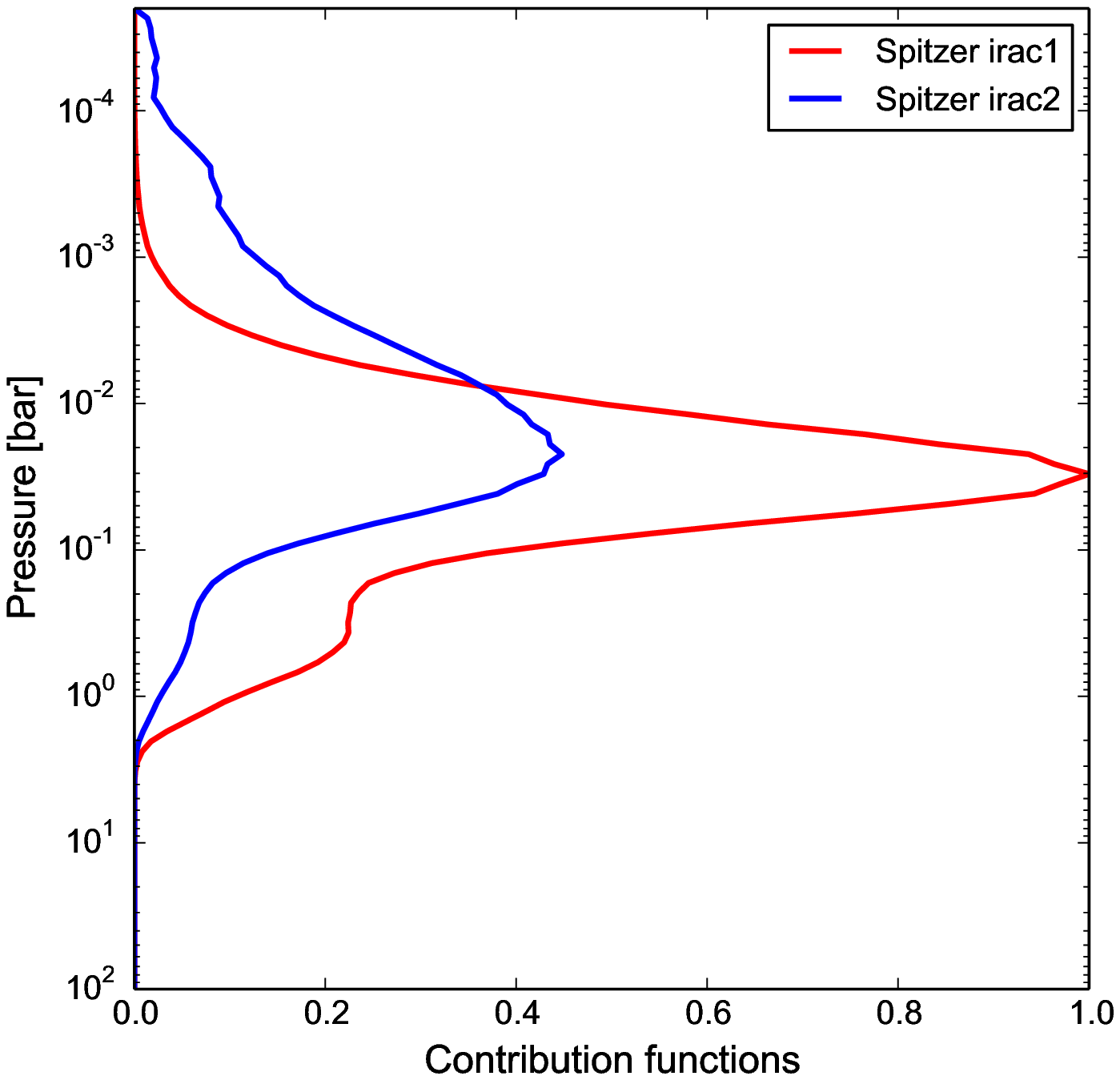}\hspace{-60pt}
\includegraphics[width=.38\textwidth, height=6cm]{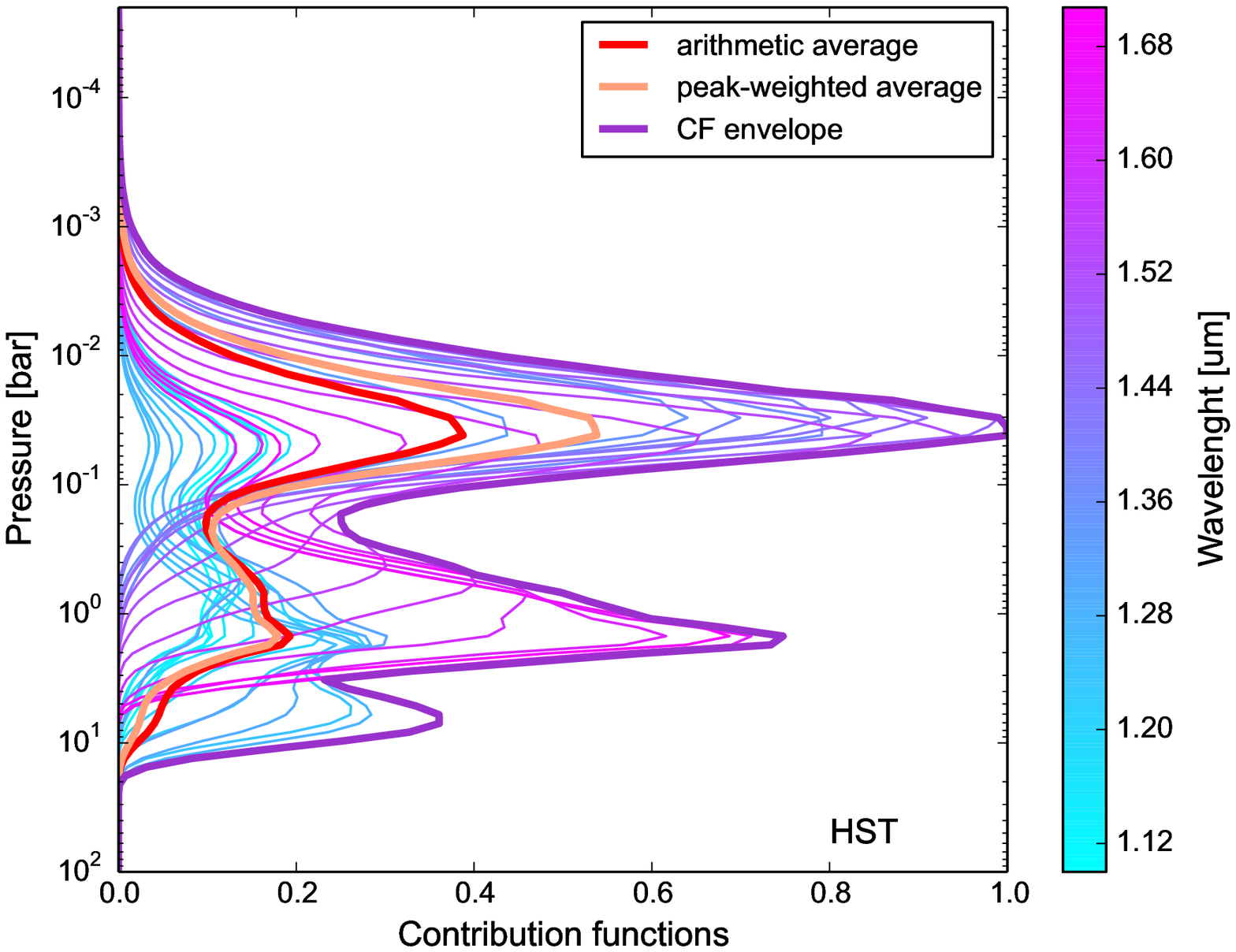}\hspace{-20pt}
\includegraphics[width=.38\textwidth, height=6cm]{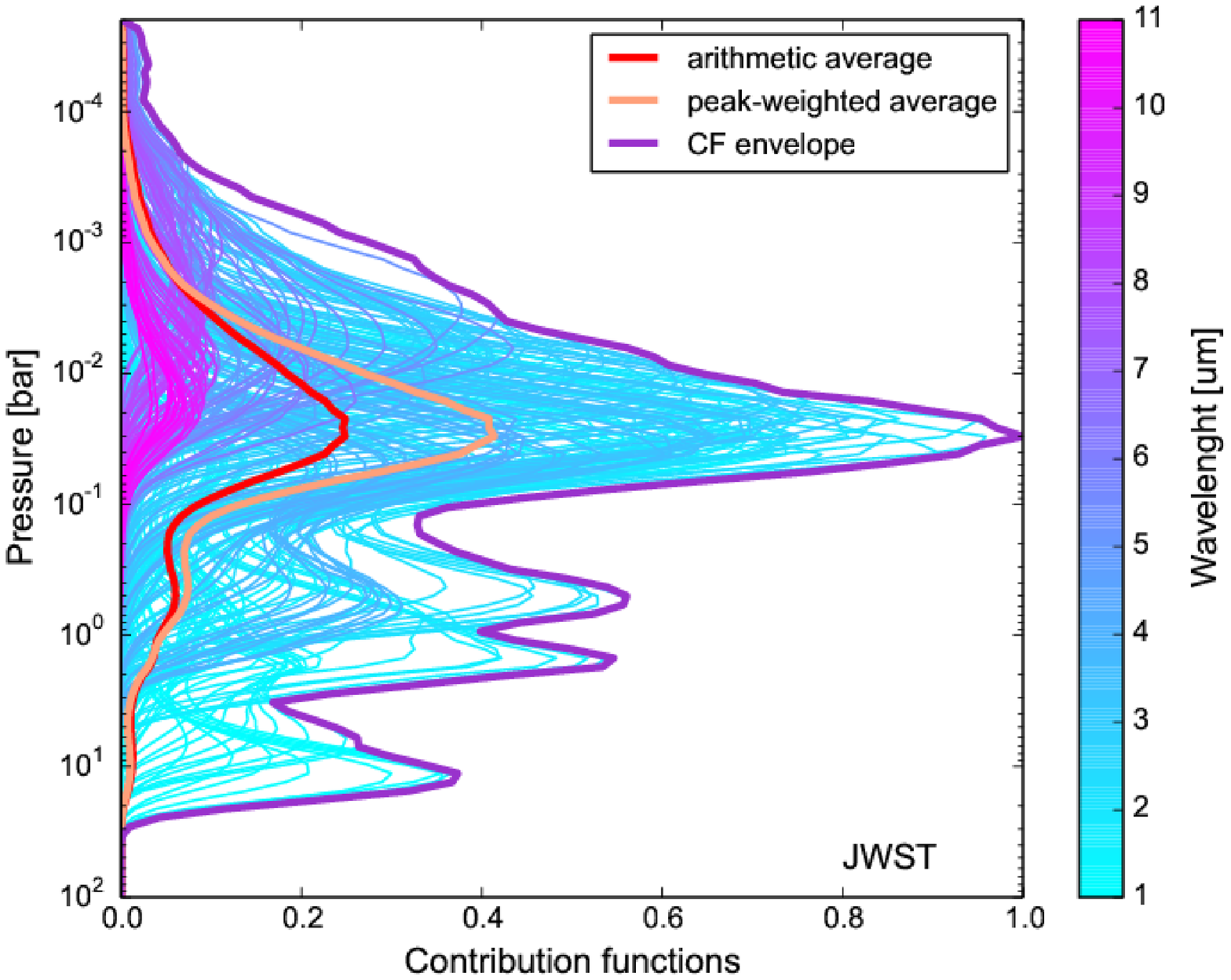}
\caption{1D retrieved contribution functions for {\em Spitzer}, left panel, the {\em HST}, middle panel, and the {\em JWST}, right panel, for the substellar model atmosphere. The contribution functions are overplotted with the peak-weighted contribution function (orange curve), calculated by scaling all the contribution functions to the maximum peak value, and by the plain arithmetic average (red curve). We plot the contribution function envelope in purple. The color code is wavelength based, with turquoise equal to the lower wavelength boundary and purple equal to the higher wavelength boundary for the instrument in question ({\em HST} in turquoise \math{\lambda}\sb{low} = 1.1 {\micron}, and in purple \math{\lambda}\sb{high} = 1.7 {\micron}, {\em JWST} in turquoise \math{\lambda}\sb{low} = 1.0 {\micron} and in purple \math{\lambda}\sb{high} = 11 {\micron}). For {\em Spitzer} we scale IRAC channel 2 (blue) to the peak value of the IRAC channel 1 (red).}
\label{fig:Retriev-CF}
\end{figure*}

\subsection{Theoretical Contribution Functions}
\label{sec:theo-cf}

We calculate theoretical contribution functions using our 3D
thermal structure to determine which parts of the atmosphere are
probed by the observations (our synthetic data points), i.e., from which pressure layers most of the flux is coming from. Our goal is
to compare these theoretical contribution functions to the contribution functions from the retrieval and assess the difference, i.e. how well retrieval
probes the same pressure layers as the 3D structure. To do so, we use
Equation (2) from \citet{KnutsonEtal2009ApJ-redistribution} and
calculate the contribution functions for each of the instruments
separately ({\em JWST}, {\em HST}, and {\em Spitzer}), and investigate
which part of the atmosphere would theoretically be possible to probe
with each of the instruments. We calculate the contribution functions
for each of the 361 \math{T-P} profiles provided by our RHD
simulation. In addition, to estimate where on average each
instrument is probing the atmosphere of HD 189733b, we average the
instrumental contribution functions in several
ways (see below). Figure \ref{fig:Theor-CF-aver} shows the contribution functions
for {\em Spitzer} channels 1 and 2, for the {\em HST}, and the {\em JWST}. The
contribution functions are normalized to the maximum bandpass value of
all contribution functions for that instrument.

The planetary emergent intensity during secondary eclipse calculated
at each location (latitude and longitude) on the planet surface is influenced by
the observer's angle. The emergent flux calculated as a hemispheric
average is thus strongly influenced
by \math{\mu=\cos\,(lat)\,*\,\cos\,(long)}. When averaging theoretical
contribution functions coming from the suite of our 3D thermal profiles
(361), for each bandpass we calculate the \math{\mu}-weighted
average. We also calculate the arithmetic average for the sake of
comparison.

{\em Spitzer} has two broad bandpasses, while the {\em JWST} and {\em
HST} spectra both have been binned into multiple narrow bands, 229 and 31,
respectively. For each filter (bandpass/bin), we calculate
the \math{\mu}-weighted average and the arithmetic average as
described in Section \ref{sec:aver}. The top two panels in
Figure \ref{fig:Theor-CF-aver} show both the \math{\mu}-weighted and
arithmetic averages for each {\em Spitzer} bandpass.

When we have many bandpasses (bins), we calculate
the \math{\mu}-weighted averages in two different ways. Our first
approach ({\em \math{\mu}-weighted arithmetic average, turquoise
lines}) is to calculate the contribution functions for each \math{T-P}
profile and each wavelength bandpass for each instrument. We then
normalize the contribution functions to the maximum peak value of all
contribution functions of that instrument.  This simply scales the
contribution functions to the value of 1.  This approach accounts for
the fact that if more flux is coming to the instrument,
the signal-to-noise ratio is higher, the error bars are smaller, and this will
contribute more to the result. To calculate the \math{\mu}-weighted
average for each bandpass, we apply Equations (\ref{mu}) and
(\ref{weighted-aver}). We then calculate the arithmetic average of the
normalized contribution functions.

Our second approach to averaging ({\em \math{\mu}-weighted
peak-weighted average, orange lines}) is calculated by accounting for
the different contributions of each bandpass. After performing
the \math{\mu}-weighted average (thin white lines), we use a
peak-weighted average depending on the peak bandpass value of each
contribution function.

Our third average ({\em arithmetic average, red lines}) comes from
simply arithmetically averaging all 361 contribution functions of each
bandpass of each instrument, and then arithmetically averaging them
again based on the number of bandpasses in each instrument.

Finally, we plot the instrument contribution function envelope ({\em
CF envelope, purple lines}) that accounts for the maximum contribution
at each pressure level.

For our analysis, we use the \math{\mu}-weighted peak-weighted average and
the contribution function envelope, as they seem to best represent the suite
of the contribution functions.

\newpage
\subsection{Retrieved Contribution Functions}
\label{sec:ret-cf}

After we retrieve the best-fit temperature profile, we calculate
contribution functions for each instrument based on the best-fit model
(Figure \ref{fig:Retriev-CF}).  Since we have only 2 bandpasses for
{\em Spitzer}, but 31 and 229 bins for the {\em HST} and {\em JWST},
respectively, we also average the {\em HST} and {\em JWST} contribution
functions to estimate where most of the flux is coming from.

We again perform two types of averaging. First, we simply scale all
contribution functions for each instrument to its peak value, so that all
contribution function are scaled to 1. Then, we calculate the plain {\em
arithmetic average} (red curves) and the {\em peak-weighted average}
(orange lines). We also plot the contribution function envelope to
account for all contributions at each atmospheric layer ({\em CF
envelope, purple lines}).

Figure \ref{fig:Retriev-CF} (middle and right panels) shows an example
of how we average the contribution functions for the {\em HST} and {\em
JWST}. We used the substellar model atmosphere for this example. We
calculated contribution functions for each of the bandpasses/bins of
each instrument. The {\em Spitzer} contribution functions are not
averaged; instead, each bandpass contribution function was calculated
separately, and two of them were scaled to their mutual maximum value
({\em Spitzer IRAC 1} has a higher value).

For our analysis, we use the peak-weighted average and the
contribution function envelope, as they seem to best represent the
suit of contribution functions for each instrument.

\begin{figure*}[t!]
\centering
\hspace{-30pt}\includegraphics[width=.75\textwidth, height=6.5cm]{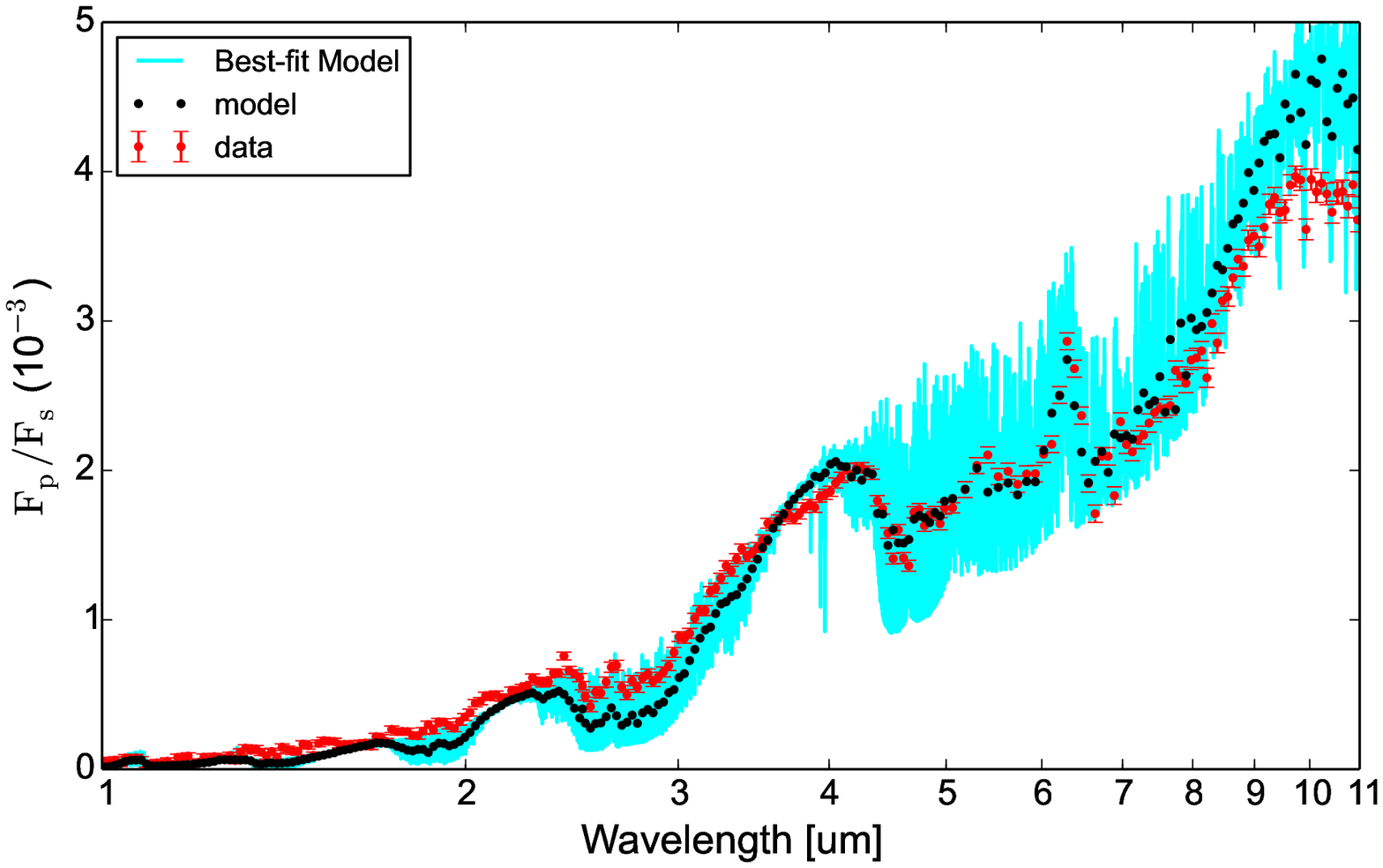}\hspace{-30pt}
\includegraphics[width=.28\textwidth, height=6.5cm]{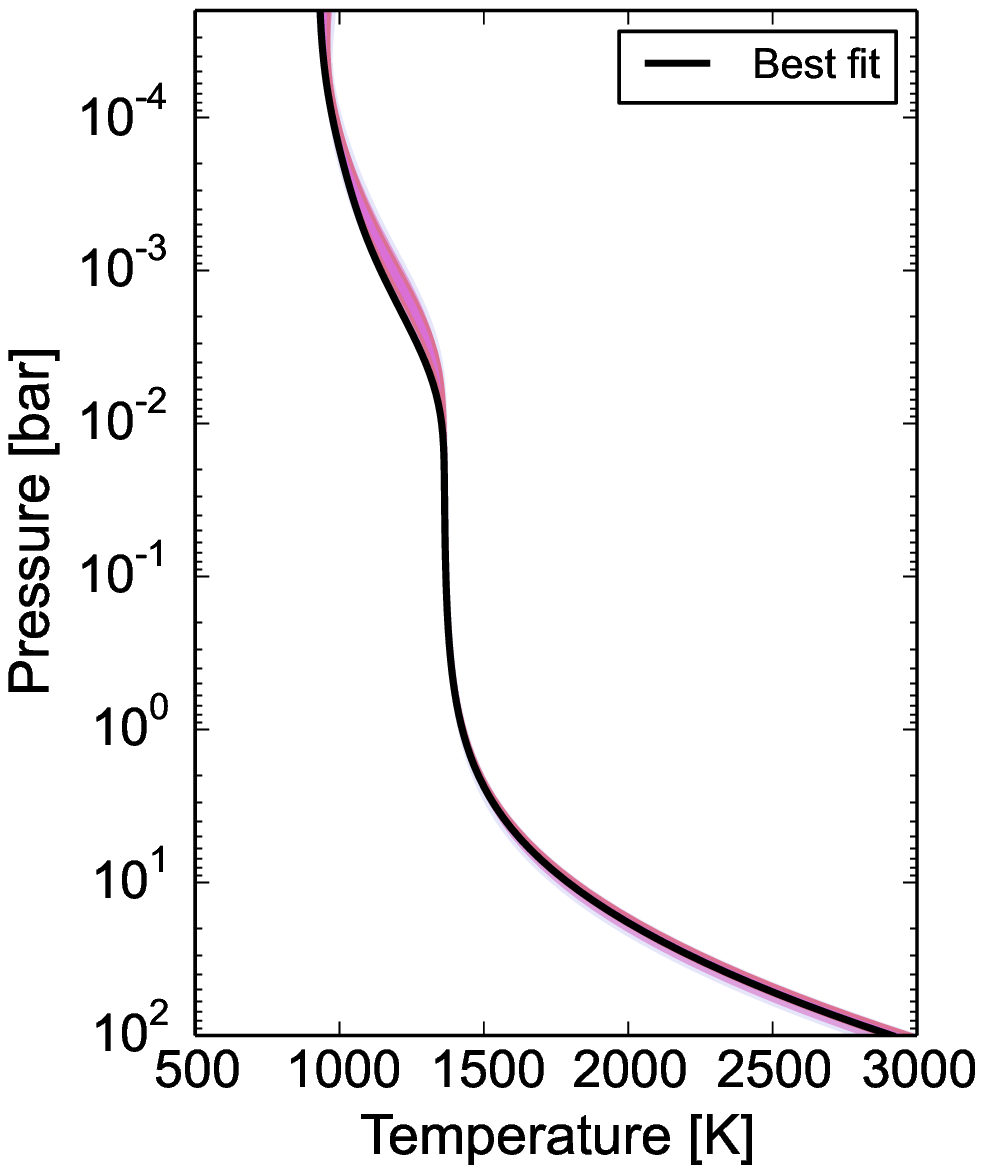}
\caption{Left: The retrieved best-fit spectra (blue) for the case when {\em only the {\em JWST}} synthetic data are included and the temperature profile is generated using the temperature {\em Parametrization I}, Appendix \ref{sec:Line}. In red are plotted the data points (eclipse depths) with error bars. In black we show the model points integrated over the bandpasses of our synthetic model. Right: the best-fit \math{T-P} profile with 1\math{\sigma} and 2\math{\sigma} confidence regions.}
\label{fig:jwst}
\end{figure*}

\begin{figure*}[t!]
\vspace{-5pt}
\centering
\includegraphics[height=3cm, clip=True]{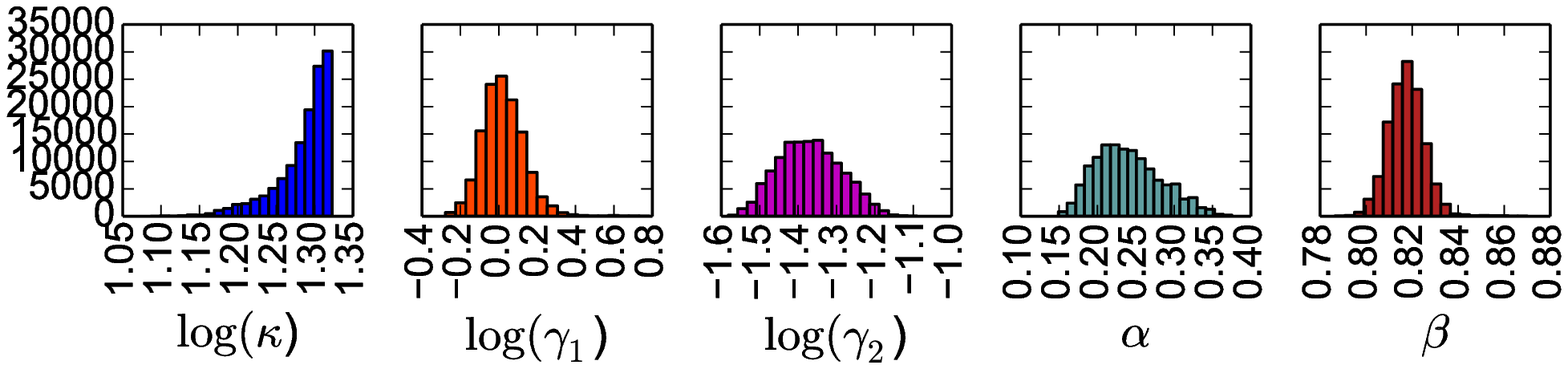}
\vspace{-5pt}
\caption{Histograms of the temperature profile parameters for the case when {\em only the {\em JWST}} synthetic data points are included and the temperature profile is generated using the temperature {\em Parametrization I}, Appendix \ref{sec:Line}. Figures show the \math{T-P} profile parameters, where some of them are expressed as \math{log\sb{10}(X)}, with \math{X} being the free parameter of the model.}
\vspace{-5pt}
\label{fig:jwst-hist}
\end{figure*}


\begin{figure*}[t!]
\centering
\hspace{-15pt}\includegraphics[height=7.0cm, clip=True]{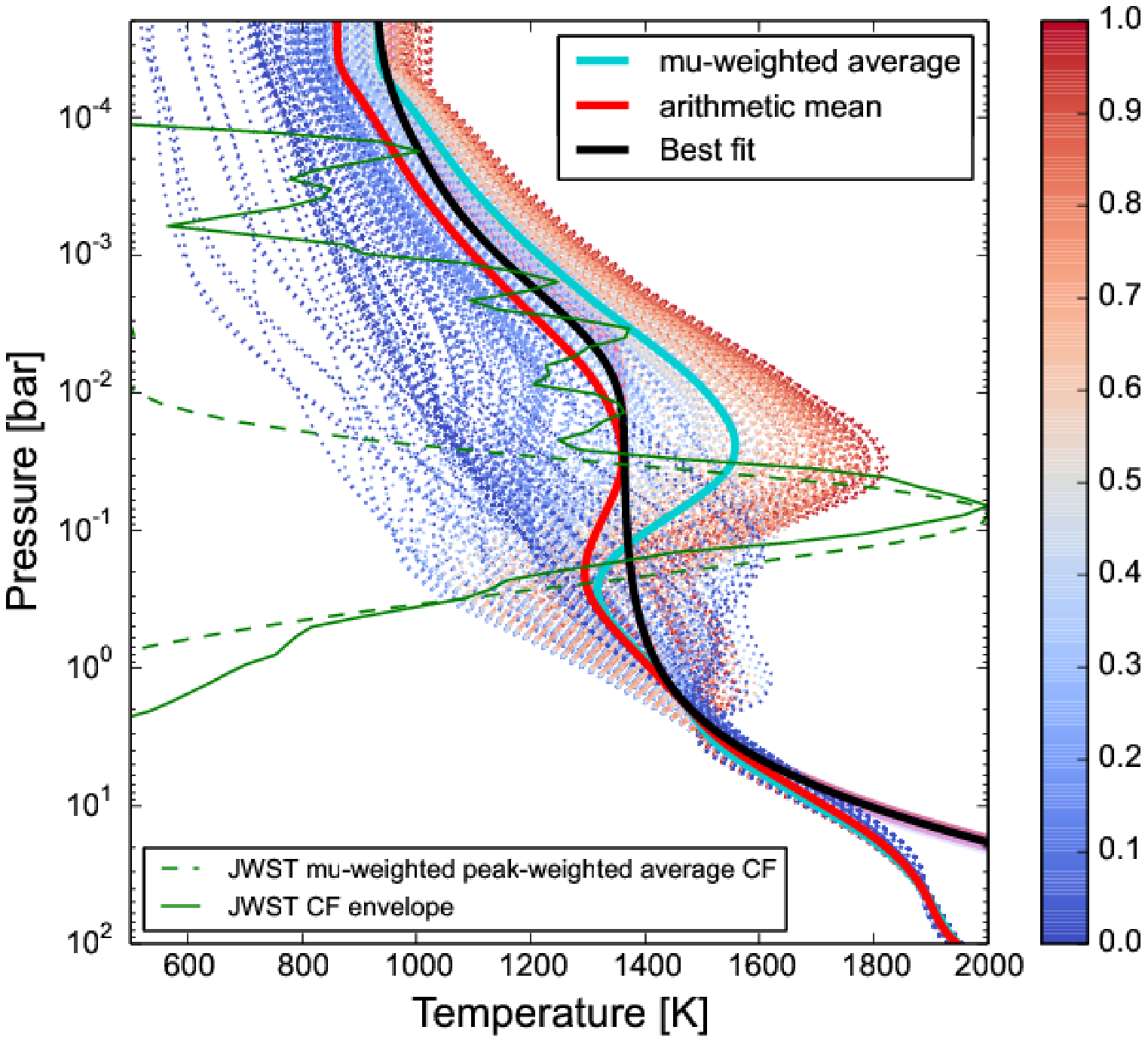}\hspace{-35pt}
\includegraphics[height=7.0cm, clip=True]{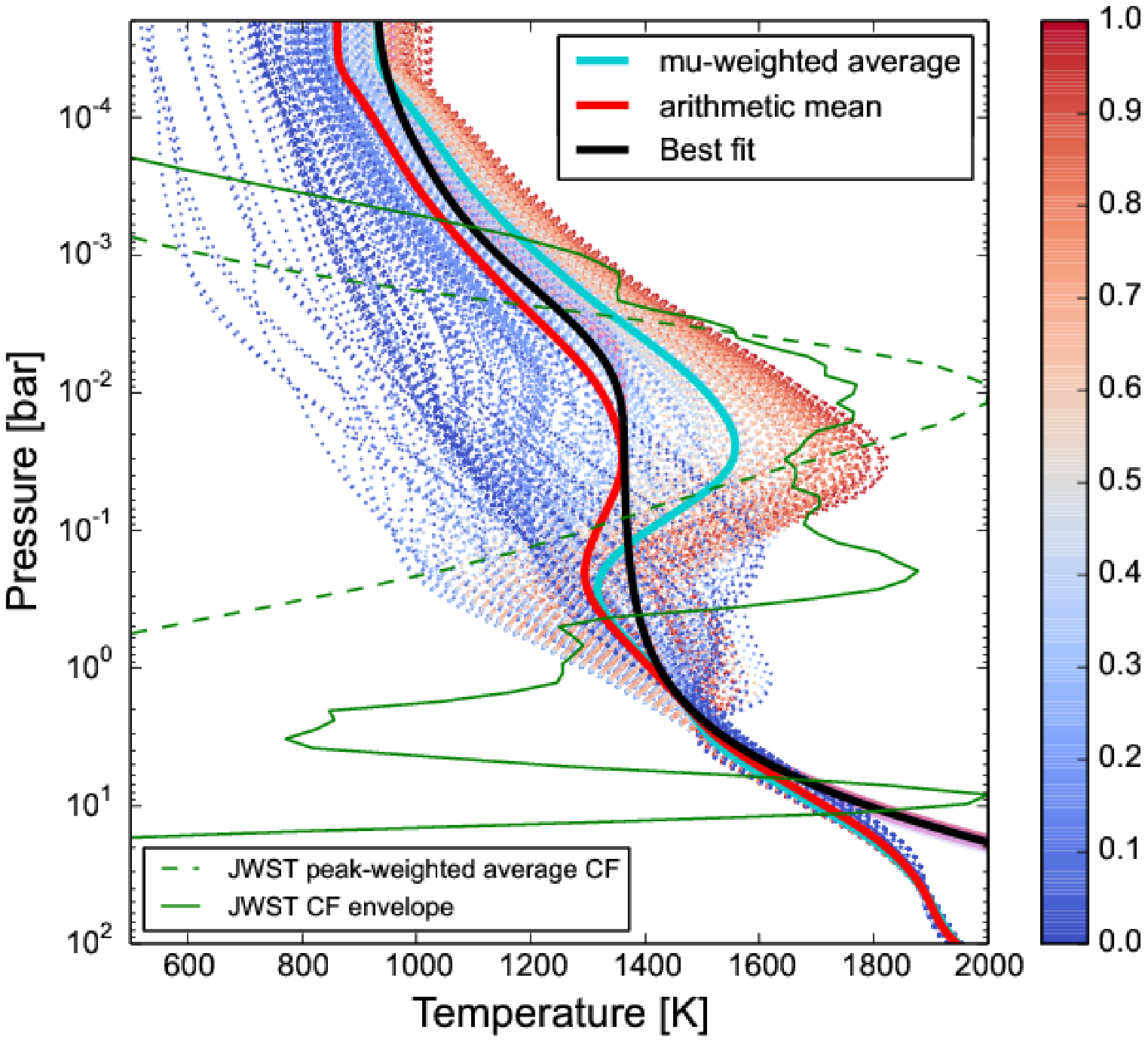}
\caption{Left: the 3D \math{T-P} profile dayside structure of HD 189733b, with the retrieved best-fit temperature profile (black curve) from Figure \ref{fig:jwst}, right panel, and the 3D thermal structure averages (red and turquoise curves), overplotted with {\em only the {\em JWST} theoretical} contribution functions, normalized to 2000, and generated using {\em Parametrization I}, Appendix \ref{sec:Line}. The dotted green curve is the {\em JWST} \math{\mu}-weighted peak-weighted average, while the solid green curve is the {\em JWST} contribution function envelope (see Figure \ref{fig:Theor-CF-aver}).  Right: the 3D \math{T-P} profile dayside structure of HD 189733b, with the retrieved best-fit temperature profile (black curve) and 3D thermal structure averages (red and turquoise curves), overplotted with {\em only the {\em JWST} retrieved} contribution functions, normalized to 2000, and generated using {\em Parametrization I}, Appendix \ref{sec:Line}. The dotted green curve is the {\em JWST} peak-weighted average, while the solid green curve is the {\em JWST} contribution function envelope.}
\label{fig:jwst-cf}
\end{figure*}

\begin{figure*}[t!]
\centering
\hspace{-30pt}\includegraphics[width=.75\textwidth, height=6.5cm]{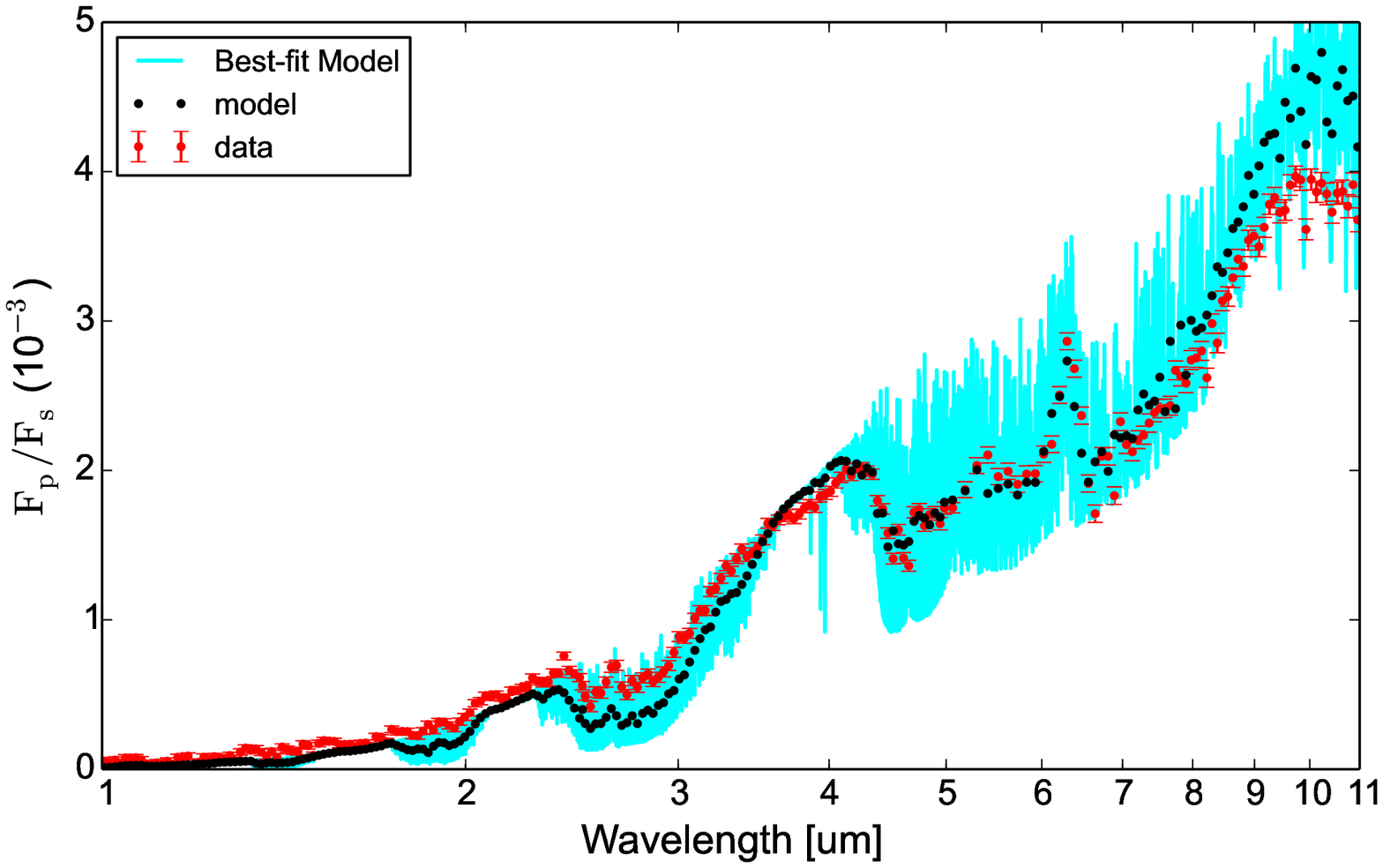}\hspace{-30pt}
\includegraphics[width=.28\textwidth, height=6.5cm]{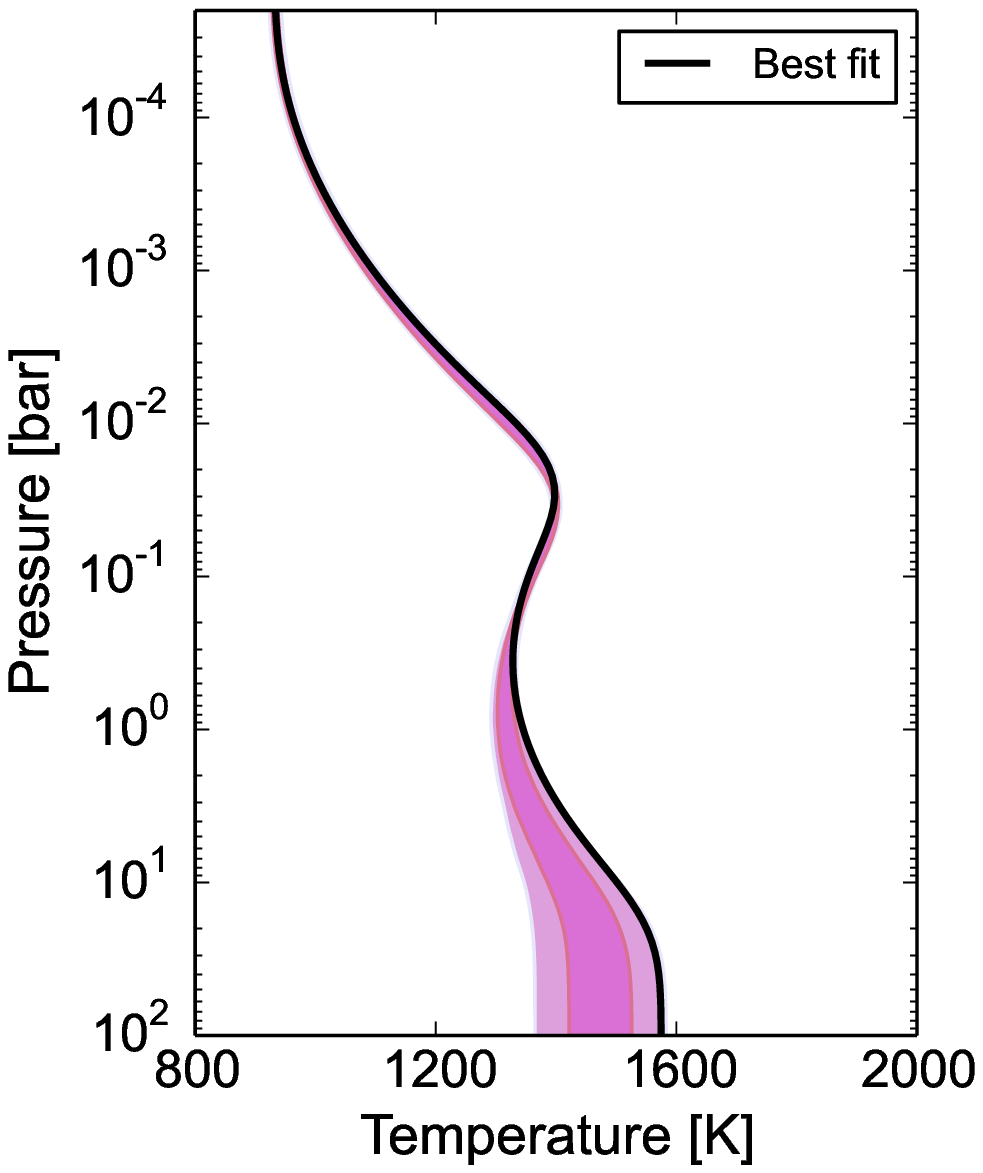}
\caption{Left: the retrieved best-fit spectra (blue) for the case when {\em only the {\em JWST}} synthetic data are included and the temperature profile is generated using the temperature {\em Parametrization II}, Appendix \ref{sec:Madhu}. In red are plotted the data points (eclipse depths) with error bars. In black we show are the model points integrated over the bandpasses of our synthetic model. Right: the best-fit \math{T-P} profile with 1\math{\sigma} and 2\math{\sigma} confidence regions.}
\label{fig:jwst-madhu}
\end{figure*}

\begin{figure*}[t!]
\vspace{-5pt}
\centering
\includegraphics[height=3cm, clip=True]{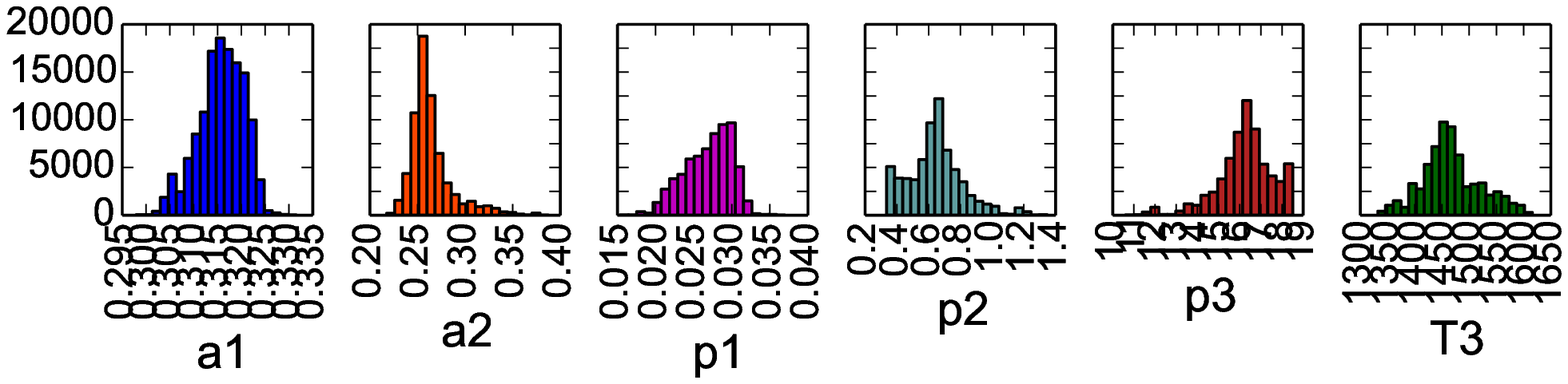}
\vspace{-5pt}
\caption{Histograms of the temperature profile parameters for the case when {\em only the {\em JWST}} synthetic data points are included and the temperature profile is generated using the temperature {\em Parametrization II}, Appendix \ref{sec:Madhu}. The panels show the \math{T-P} profile parameters.}
\vspace{-10pt}
\label{fig:jwst-hist-madhu}
\end{figure*}


\begin{figure*}[t!]
\centering
\hspace{-15pt}\includegraphics[height=7.0cm, clip=True]{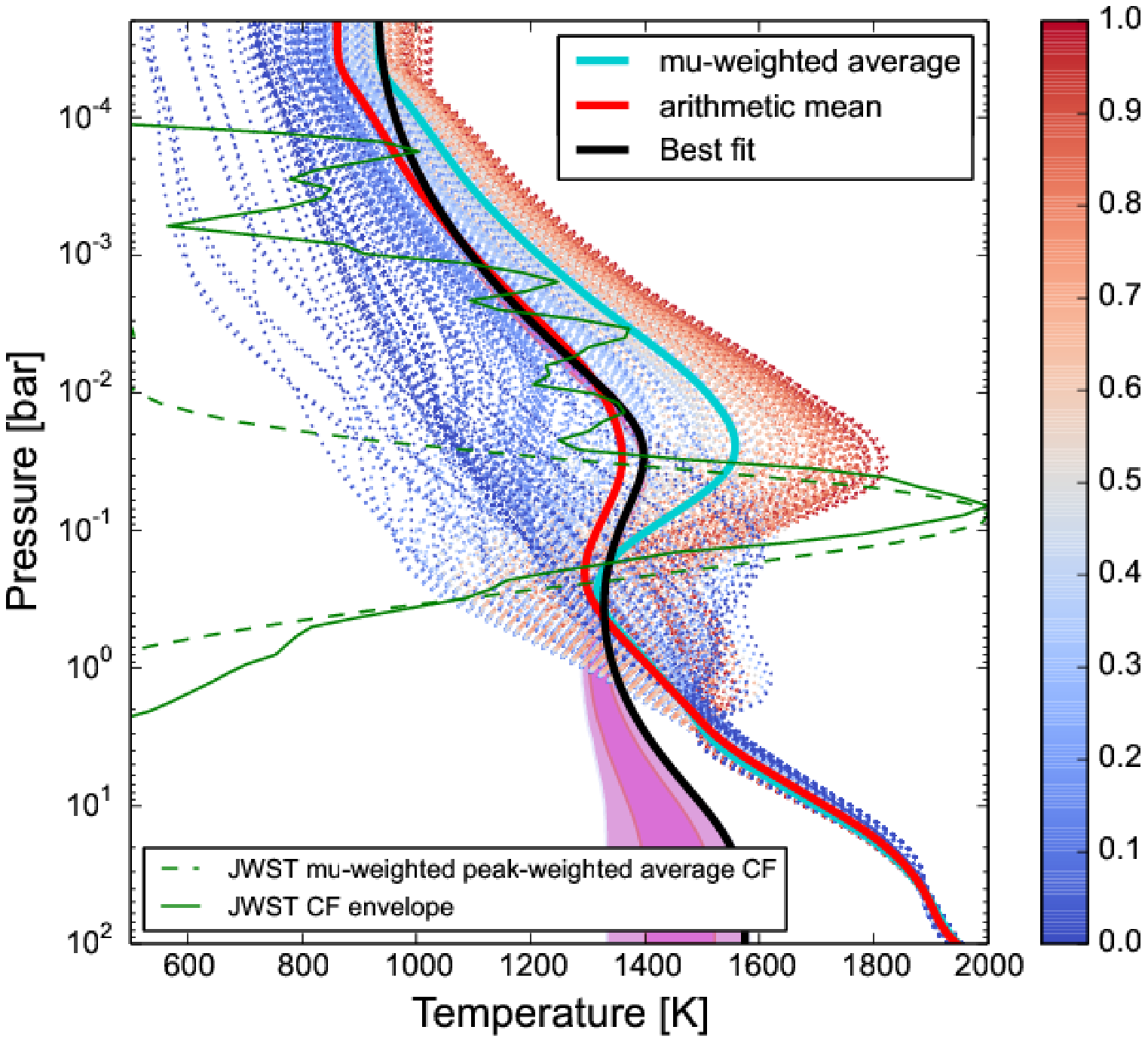}\hspace{-35pt}
\includegraphics[height=7.0cm, clip=True]{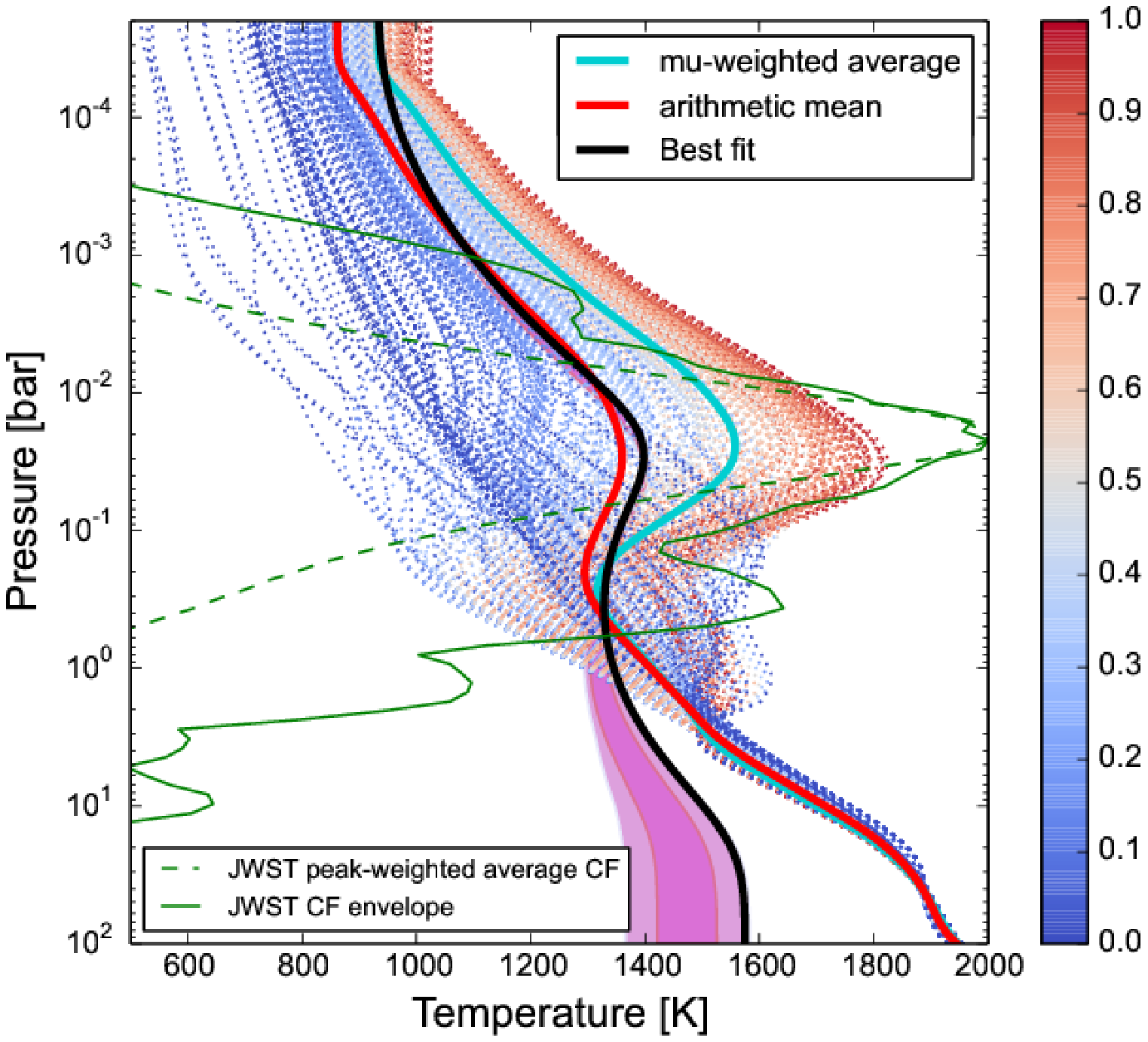}
\caption{Left: the 3D \math{T-P} profile dayside structure of HD 189733b, with the retrieved best-fit temperature profile (black curve) from Figure \ref{fig:jwst-madhu}, right panel, and the 3D thermal structure averages (red and turquoise curves), overplotted with {\em only the {\em JWST} theoretical} contribution functions, normalized to 2000, and generated using {\em Parametrization II}, Appendix \ref{sec:Madhu}. The dotted green curve is the {\em JWST} \math{\mu}-weighted peak-weighted average, while the solid green curve is the {\em JWST} contribution function envelope (see Figure \ref{fig:Theor-CF-aver}). Right: the 3D \math{T-P} profile dayside structure of HD 189733b, with the retrieved best-fit temperature profile (black curve) and 3D thermal structure averages (red and turquoise curves), overplotted with {\em only the {\em JWST} retrieved} contribution functions, normalized to 2000, and generated using {\em Parametrization II}, Appendix \ref{sec:Madhu}. The dotted green curve is the {\em JWST} peak-weighted average, while the solid green curve is the {\em JWST} contribution function envelope.}
\label{fig:jwst-cf-madhu}
\end{figure*}

\section{Results}
\label{sec:res}

To compare the 3D temperature and pressure structure with the results
from retrieval, we perform several analyses. First, we discuss
the results when we only include the {\em JWST} synthetic points and then
we discuss the results for the {\em HST} and {\em Spitzer} together, as this
combination of data is often found in the exoplanetary literature. We
also investigate two cases for the {\em JWST} analysis by using
different temperature parametrization, allowing the retrieval to
explore more complex thermal shapes. We first use the temperature
Parametrization I, Appendix \ref{sec:Line}, and then the temperature Parametrization II,
Appendix \ref{sec:Madhu}. In
Appendix \ref{sec:Appendix-B}, we discuss the retrieval results when
we include all the synthetic data points and uncertainties for the {\em
JWST}, {\em HST} and {\em Spitzer} together and for the {\em HST} and {\em
Spitzer}, separately.

\subsection{JWST}
\label{sec:jwst}

In this section, we present the results when only {\em JWST} synthetic
data are used. We have 229 data points available, sampling the whole
wavelength range between 1 and 11 {\micron} with a
resolution of \textasciitilde100.  First, we discuss the results when
we use temperature Parametrization I (Appendix \ref{sec:Line}), and then
we present the results using temperature Parametrization II
(Appendix \ref{sec:Madhu}).

\begin{figure*}[t!]
\centering
\hspace{-30pt}\includegraphics[width=.75\textwidth, height=6.5cm]{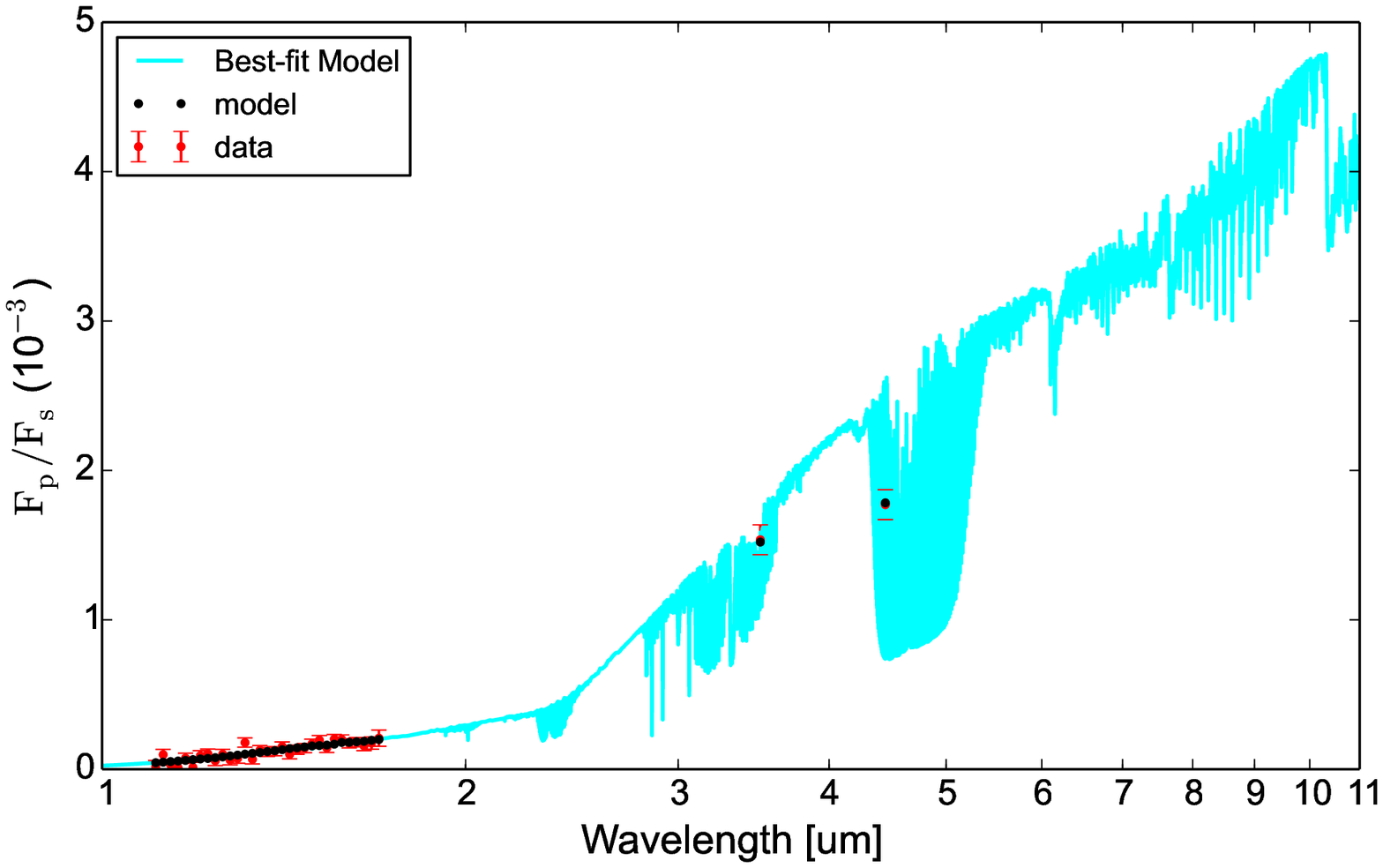}\hspace{-30pt}
\includegraphics[width=.28\textwidth, height=6.5cm]{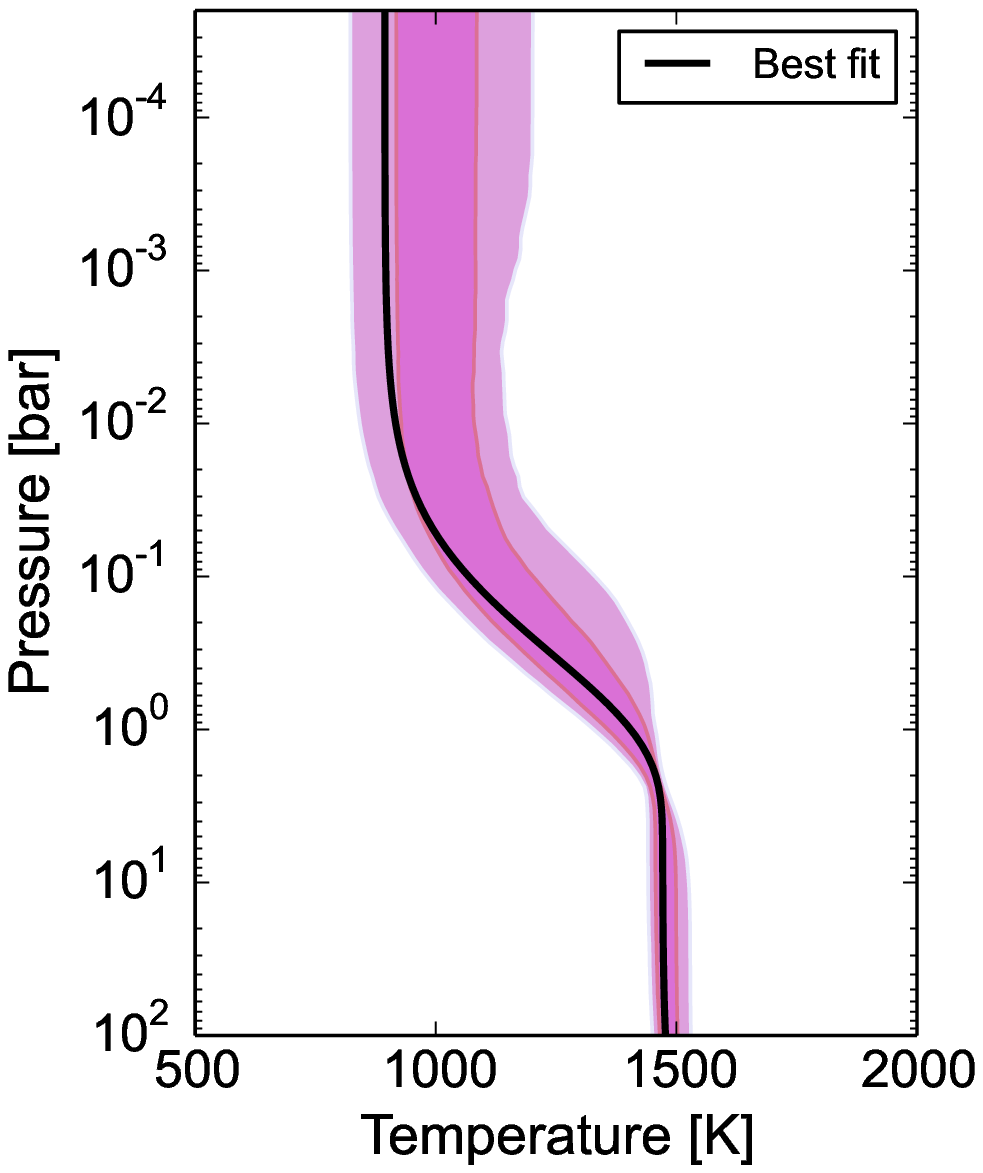}
\caption{Left: the retrieved best-fit spectra (blue) for the case when {\em only the {\em HST} and {\em Spitzer}} synthetic data are included and the temperature profile is generated using the temperature {\em Parametrization I}, Appendix \ref{sec:Line}. In red are plotted the data points (eclipse depths) with error bars. In black we show the model points integrated over the bandpasses of our synthetic model. Right: the best-fit \math{T-P} profile with 1\math{\sigma} and 2\math{\sigma} confidence regions.}
\label{fig:HST-Sp}
\end{figure*}

\begin{figure*}[t!]
\vspace{-5pt}
\centering
\includegraphics[height=3cm, clip=True]{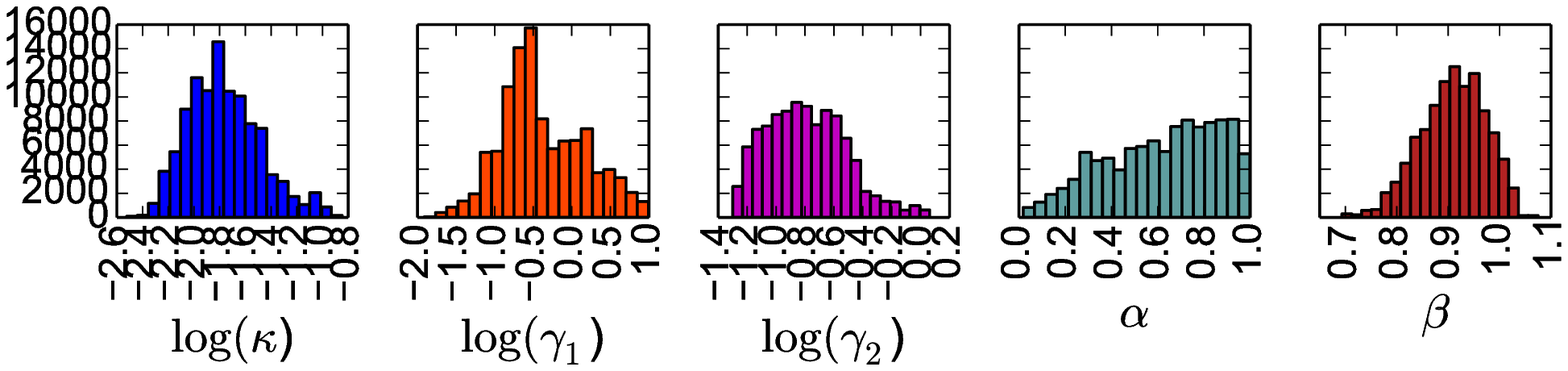}
\vspace{-5pt}
\caption{Histograms of the temperature profile parameters for the case when {\em only the {\em HST} and {\em Spitzer}} synthetic data points are included and the temperature profile is generated using the temperature {\em Parametrization I}, Appendix \ref{sec:Line}. Figures show the \math{T-P} profile parameters, where some of them are expressed as \math{log\sb{10}(X)}, with \math{X} being the free parameter of the model.}
\vspace{-10pt}
\label{fig:HST-Sp-hist}
\end{figure*}


\begin{figure*}[t!]
\centering
\hspace{-15pt}\includegraphics[height=7.0cm, clip=True]{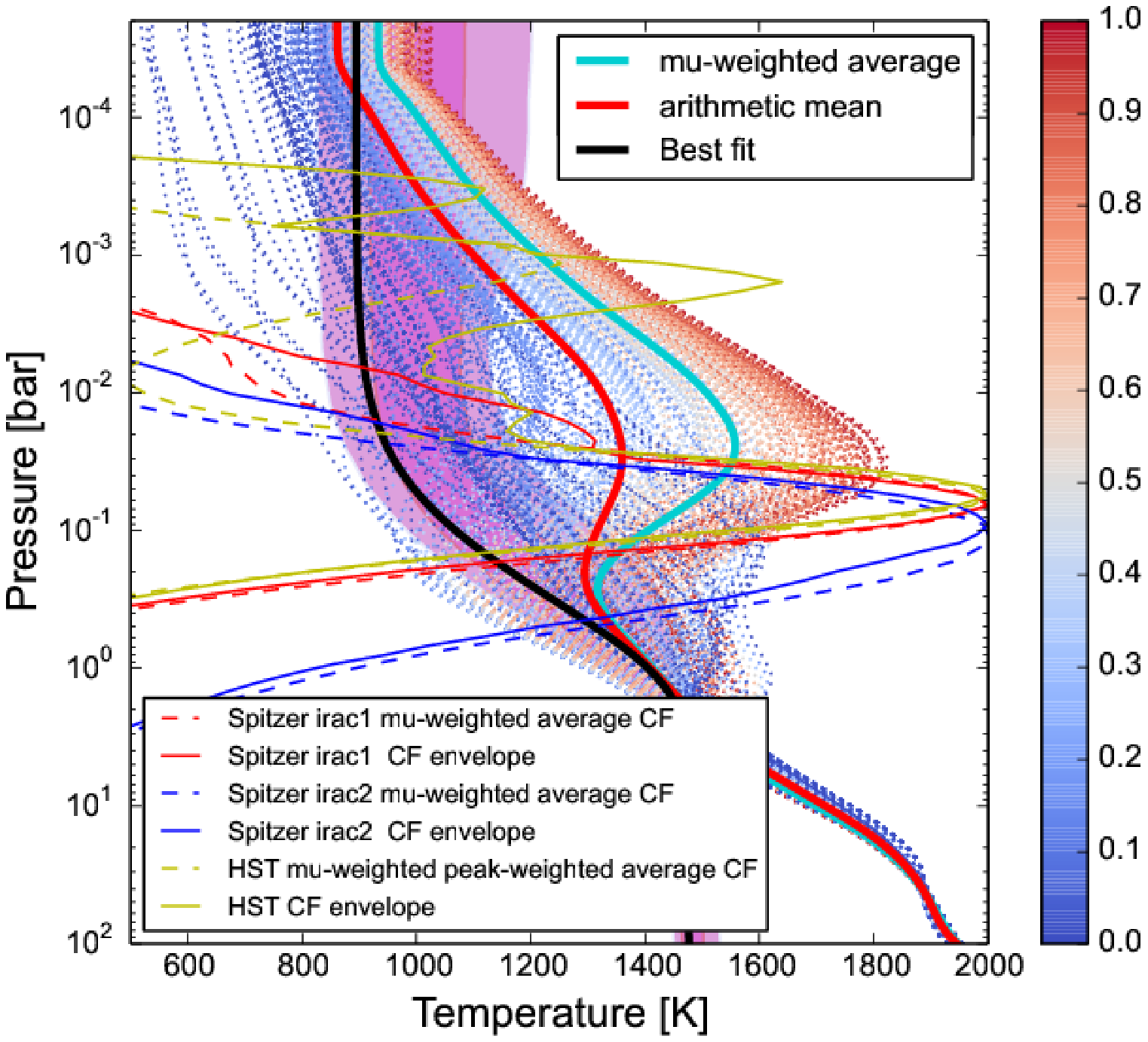}\hspace{-35pt}
\includegraphics[height=7.0cm, clip=True]{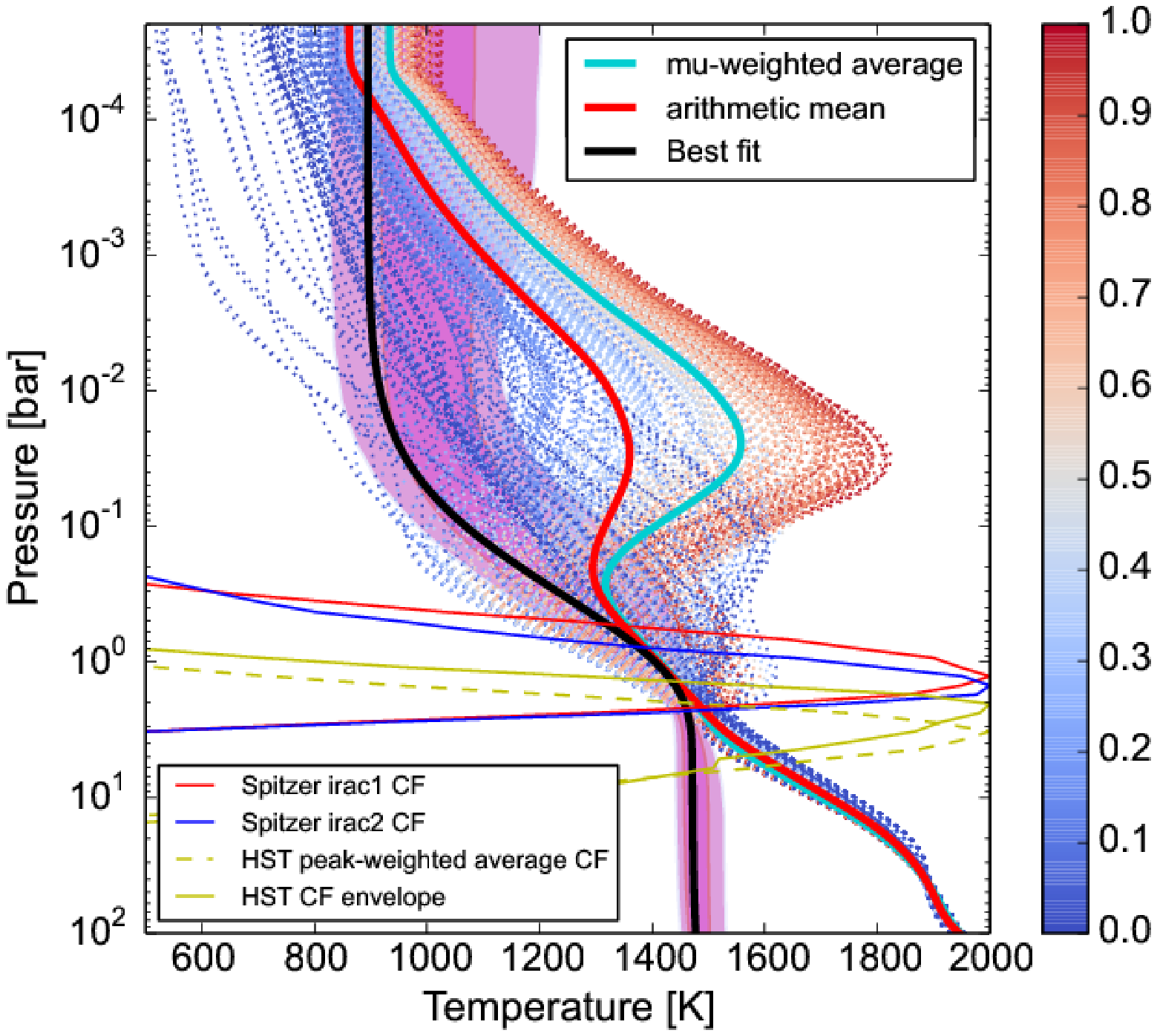}
\caption{Left: the 3D \math{T-P} profile dayside structure of HD 189733b, with the retrieved best-fit temperature profile (black curve) from Figure \ref{fig:HST-Sp}, right panel, and the 3D thermal structure averages (red and turquoise curves), overplotted with {\em only {\em HST} and {\em Spitzer} theoretical} contribution functions, normalized to 2000, and generated using {\em Parametrization I}, Appendix \ref{sec:Line}. Red and turquoise curves are {\em Spitzer} theoretical contribution functions, while the yellow dotted curve is the {\em HST} theoretical \math{\mu}-weighted peak-weighted average, and the solid yellow curve is the {\em HST} contribution function envelope (see Figure \ref{fig:Theor-CF-aver}). Right: the 3D \math{T-P} profile dayside structure of HD 189733b, with the retrieved best-fit temperature profile (black curve) and 3D thermal structure averages (red and turquoise curves), overplotted with {\em only {\em HST} and {\em Spitzer} retrieved} contribution functions, normalized to 2000, and generated using {\em Parametrization I}, Appendix \ref{sec:Line}. Red and turquoise curves are {\em Spitzer} contribution functions, while the yellow dotted curve is the {\em HST} peak-weighted average, and the solid yellow curve is the {\em HST} contribution function envelope.}
\label{fig:HST-Sp-cf}
\end{figure*}

Figure \ref{fig:jwst} shows the best-fit spectrum and temperature
profile when we use temperature Parametrization I with five
free parameters (see Appendix \ref{sec:Line} for possible {\math{T-P} shapes
using this parametrization). The spectrum and temperature
structure is very similar to the case when all synthetic data
({\em Spitzer, HST, and JWST}) are included (see Appendix \ref{sec:all}).
The posterior distribution (Figure \ref{fig:jwst-hist}) of the temperature
parameters looks well constrained, except for the parameter \math{\kappa}.
This parameter hits the wall around the same value as in the
case when all the data are included, Appendix \ref{sec:all}.

As shown in Appendix \ref{sec:Line}, larger \math{\kappa} values cause the bottom
of the temperature profile to go to very high temperatures,
above 3000 K, which are the current boundaries for the
HITRAN partition functions. Although the MCMC has a
tendency to explore higher values of parameter κ, these steps
are excluded from the MCMC exploration.

In the left panel of Figure \ref{fig:jwst-cf}, we plot the suite of our initial
3D profiles, with both averages from Section \ref{sec:aver}, overplotted
with the best-fit retrieved temperature profile. We also overplot
the curves with the theoretical contribution functions μ-weighted
peak-weighted average and the contribution function envelope
for the {\em JWST}. In the right panel of Figure \ref{fig:jwst-cf}, we can see the
contribution functions calculated from the retrieved best-fit
model. In both cases, we plot only the averages from Sections \ref{sec:theo-cf}
and \ref{sec:ret-cf} that we believe represent the best the overall contribution
function trend.

As shown, both theoretical contribution functions and
retrieved contribution functions sample similar pressure intervals,
emphasizing that the bottom part of the temperature profile is
not too reliable. The black curve does not match either the red
curve (the arithmetic average) or the blue curve (\math{\mu}-weighted
average), but is more similar to the red curve. Returning to
Figure \ref{fig:MC3-PT-fits}, we can see that this parametrization does not have
capabilities to reproduce the complex shapes of any of these two
profiles, or any of the inversion profiles seen in orange that
come from the longitudes and latitudes close to the substellar
point (0.6 < \math{\mu} < 1.0).

To explore a wider range of thermal profile shapes, we run the
analysis using the temperature Parametrization II (Figures \ref{fig:jwst-madhu} and \ref{fig:jwst-hist-madhu}. We see a very nice match between the
best-fit model and the 3D arithmetic average
(Figure \ref{fig:jwst-cf-madhu}). Although the posterior histograms in
Figure \ref{fig:jwst-hist-madhu} look somewhat worse than in the case
when we had all the data included, they are still very nicely
constrained. Similarly to Appendix \ref{sec:all}, the {\em JWST}
contribution function of the retrieved model samples lower pressures
than the theoretical contribution functions, suggesting that the most
believable part of the temperature profile is actually the inversion
peak.

In the literature \citep{KnutsonEtal2007natHD189733b, SwainEtal2008NatureHD189733bspec, LineEtal2014-Retrieval-II,
BennekeEtal2015ApjCOratios} the dayside emission of HD 189733b is
always presented without the presence of a thermal inversion. However,
these conclusions are based on the data coming from the {\em HST} and or
{\em Spitzer} observations. In the next section, we explore the
best-fit model using the synthetic data for these two instruments, simultaneously.

\subsection{HST and Spitzer}
\label{sec:HST-Sp}

To explore how well the currently available space telescopes for
exoplanetary observations can reproduce the intrinsically 3D thermal
structure, we take only the {\em HST} and {\em Spitzer} data as the input
for retrieval. We have two data points for the {\em Spitzer} observations at
3.6 and 4.5 {\micron}, and 31 data points for the {\em HST} observations
between the wavelength region of 1.1 and 1.7 {\micron}. In this case,
we use only temperature Parametrization I (Appendix \ref{sec:Line}) to
explore the shapes of the temperature profiles, as the data lead MCMC
exploration to the pure non-inversion temperature profile shapes, well
represented by this temperature parametrization. We have also tried
Parametrization II (Appendix \ref{sec:Madhu}), but the results are
quite similar.

The best-fit temperature profile is fully non-inverted
(Figure \ref{fig:HST-Sp}), often seen in the
literature \citep{KnutsonEtal2007natHD189733b, SwainEtal2008NatureHD189733bspec,Dobbs-DixonAgol2013-RHD, 
LineEtal2014-Retrieval-II,
BennekeEtal2015ApjCOratios}. The posterior
histograms in Figure \ref{fig:HST-Sp-hist} are somewhat nicely constrained, but the 1\math{\sigma} and 2\math{\sigma}
regions of the temperature profile are rather wide.

The retrieved contribution functions for the best-fit model, Figure \ref{fig:HST-Sp-cf}, left panel, reveal
that only the pressure region around 1 bar is well probed by these
observations. This is at much higher pressures than the contribution
functions from the theoretical model, (Figure \ref{fig:HST-Sp-cf}, left panel) suggesting that the best-fit
model is loosely constrained by the data. The number of data points
and/or the wavelength coverage is not enough to lead MCMC to models
similar to any of the averages of the 3D thermal structure. The
retrieved temperature profile falls in the cold 3D temperature region,
close to the terminator. In
Appendices \ref{sec:HST} and \ref{sec:Spitzer} we discuss the analyses done using
{\em HST} and {\em Spitzer} synthetic data separately.

\newpage
\section{Conclusions}
\label{sec:conc}

In this paper, we investigate how well the hemispherical-average
temperature model retrieved by a reverse statistically robust modeling
approach compares to the non-uniform 3D dayside temperature structure
coming from a hydrodynamic simulation. We take the output from the 3D
radiative-hydrodynamic simulation of HD 189733b and calculate the
emergent spectra on the dayside atmosphere of the planet. This spectra
serves as a synthetic high-resolution model that we pass through the
emission spectrum simulator to generate the data points with
uncertainties for the retrieval, assuming we have used {\em Spitzer},
{\em HST}, and {\em JWST} observations. Starting with a 1D temperature--pressure model, the retrieval framework retrieves the best-fit spectra
and temperature structure. In the main body of our manuscript, we
present the retrieval results when we include the synthetic data
points for the {\em JWST} only and for the {\em HST} and {\em Spitzer} together,
focusing our analysis on the combination of data most often used in
the literature. In Appendix \ref{sec:Appendix-B}, we discuss our
findings using all the synthetic data points and uncertainties for the 
{\em JWST}, {\em HST}, and {\em Spitzer} together, as well as the results for
the {\em HST} and {\em Spitzer} separately.  We explore several methods
of averaging our 3D \math{T-P} profiles to facilitate comparison to
our retrieval results. To asses which particular 3D temperature
profile matches the retrieved one, we average the initial 3D structure
by performing the \math{\mu}-weighted average and the arithmetic
average. We test two temperature parametrizations that are commonly used in
retrieval to thoroughly explore the possible shapes of the temperature
profiles and find the best match with the 3D averages in the case
when we use the {\em JWST} synthetic data alone.  To assess which part of the
atmosphere would be mostly probed with the 3D structure and which part
is revealed with the retrieved best-fit models, we perform a thorough
exploration of theoretical contribution functions (coming from the 3D
structure) and the retrieved contribution functions. We average the
complex 3D contribution functions of {\em Spitzer}, {\em HST}, and {\em
JWST} in a way similar to the original 3D thermal profiles, and in
addition, we calculate the \math{\mu}-weighted and peak-weighted average
of the {\em JWST} and {\em HST} contribution functions coming from
different bins. We then compare the averaged theoretical and retrieved
contribution functions to determine which part of the retrieved
temperature profile we should trust the most.

Our results are strongly affected by the spectral resolution and the
wavelength region covered by a particular instrument. When combining
the data points from different instruments, our results are strongly
influenced by the number of data points included in the analysis. In
addition, the results are affected to some extent by the particular
temperature parametrization used in retrieval. The possible shapes of
the temperature profiles produced with each parametrization are
slightly different, particularly in the inversion case.

As shown, when {\em JWST} simulated data are included in the analysis,
based on both theoretical and retrieved contribution functions, most
of the emergent spectra come from the regions between 10\sp{1} and
10\sp{-2} bars, sampling the middle region of the atmosphere. In
addition, the extent of the best-fit 1\math{\sigma} and 2\math{\sigma}
temperature profile confidence regions infer that pressure regions
below 1 bar and above 10\sp{-3} bars, are less well constrained.  The
best-fit thermal profiles infer weak thermal inversion present in the
dayside atmosphere of HD 189733b, almost matching the averaged 3D
temperature profile, with a better match when temperature
Parametrization II (Appendix \ref{sec:Madhu}) is used.
Particularly, when we use temperature Parametrization II (Appendix \ref{sec:Madhu}) and we include only {\em JWST} synthetic
data points, we see that the 3D arithmetic average closely matches the
best-fit temperature profile within the pressure layers where the
retrieved {\em JWST} contribution functions probe the atmosphere
(Figure \ref{fig:jwst-cf-madhu}). This is not seen when all the
data are included (see Appendix \ref{sec:all}). We speculate
that either the additional 31 {\em HST} data points drag the MCMC to a
different phase space, affecting (and even better constraining) the
H\sb{2}O abundance at the same time (the {\em HST} wavelength
region of 1.1 to 1.7 {\micron} is mostly affected by the H\sb{2}O
spectral features), or that the {\em HST} and {\em Spitzer} data are not as
sensitive to temperature inversions as the {\em JWST} data, perhaps
due to their limited spectral range and resolving
power. The \math{\mu}-weighted 3D average could not be closely matched
(retrieved) with any analysis, although in our opinion this would be
the most realistic representation of the initial 3D
structure. However, both the \math{\mu}-weighted average and the
arithmetic average fall within the 1\math{\sigma} and 2\math{\sigma} regions of
the best-fit profiles, suggesting that both solutions are plausible.

When the {\em JWST} data are excluded from the analysis, the
best-fit thermal profile confidence regions are much wider. They become even less constrained when fewer data points are included. The
retrieved contribution functions do not probe the same part of the
planet atmosphere as the theoretical functions, and the best-fit thermal
profile is non-inverted, which does not match the shape of the averaged 3D
thermal profiles. Particularly when we include both the {\em
HST} and {\em Spitzer} data points, the \math{\mu}-weighted and
arithmetic averages do not even fall within the 1\math{\sigma} and 2\math{\sigma}
region of the best-fit profile.  However, the retrieved contribution
functions suggest that the pressure layers between 1 and 10 bars are
probed with this set of observations, and both \math{\mu}-weighted and
arithmetic averages closely match the best-fit profile in this
part of the atmosphere.

As discussed in \citet{Dobbs-DixonAgol2013-RHD}, the two-stream
approximation used in the RHD simulator has some limitations.  The
resolution required to compute an accurate temperature profile at
depth using this approach far exceeds the computation capability. As
the density increases with depth, the integrated optical depth becomes
very large and the profile becomes independent of the numerical
resolution, resulting in the isothermal temperature--pressure
profile \citep[e.g.,][]{rauscher2012}.  As this effect is purely
numerical and not motivated by the physically correct isothermal plateau \citep[e.g][]{hubeny2003}, the RHD code applies a diffusive
scheme at optical depth greater than 2.5, allowing a non-isothermal
temperature profile below \math{\sim}1 bar.

On the other hand, the two temperature parametrizations that we used
in this analysis expose different limitations in different parts of
the atmosphere.  Temperature Parametrization I (Appendix \ref{sec:Line}) is unable to allow the full exploration of
the parameter \math{\kappa}, because of the temperature boundaries of the
HITRAN database above 3000 K. It is also incapable of generating
curvatures in the middle part of the atmosphere, which would be due to the
presence of temperature inversion deeper in the planetary atmosphere.
In contrast, temperature Parametrization II does not allow a
slope in the bottom part of the atmosphere, allowing the full
exploration of the parameter space and therefore a better match with
the observations in the mid part of the atmosphere. However, the slope
seen in the bottom part of the 3D temperature profiles coming from the
RHD simulation could not be reproduced with this
parametrization. These results suggest that temperature
Parametrization II (Appendix \ref{sec:Madhu}) works better,
but also has certain restrictions. It is capable of reproducing more
curvature in the middle part of the atmosphere often seen in the outputs
of the hydrodynamic simulations (often probed with observations), but
it is incapable of creating a curvature in the bottom part of the
atmosphere (rarely probed with the observations). Nevertheless, it is
worth noting that retrieved profiles from both temperature parametrizations are close to each other in the middle part of the atmosphere,
where they show only marginal difference, and that all (except in the
case when we use both {\em HST} and {\em Spitzer} synthetic data)
averages fall within the 1\math{\sigma} and 2\math{\sigma} regions of the
best-fit profiles, allowing that both averages and other similar
shapes are equally plausible.

The restrictions of both approaches could be overcome by introducing
an additional free parameter in the middle part of the atmosphere for
temperature Parametrization I (Appendix \ref{sec:Line}) and below the
1 bar level for temperature Parametrization II (Appendix \ref{sec:Madhu}). However, including a new free parameter
introduces a significant computational penalty when exploring the
additional phase space, with even a possibility that the new parameter
could not be constrained if the data are not good enough. When we use the real or synthetic {\em JWST} observations, the additional
free parameters could be easily justified, as the quality of data will
support good constraints of the confidence regions.

In general, the results (the retrieved temperature and pressure
profile) of our comprehensive retrieval analysis using different
simulated data and their combinations match in the range of
pressures where the retrieved contribution functions are sampling the
atmosphere. The recovered temperature and pressure profile most
closely matches the arithmetic average of the initial 3D thermal
structure.  Although the recovered \math{T-P} profiles differ
significantly in the different parts of the atmosphere depending on
the combination of the data used in retrieval, they agree fairly well 
within the parts of the atmosphere that are probed by the
observations. Thus, we can say that retrieval works well to retrieve the same \math{T-P} profile regardless of which data are used. The {\em JWST} synthetic data provide the best match to the
averaged temperature profile using the temperature Parametrization II because of its higher flexibility to cover a wider range of possible
thermal profile shapes.

Although in this analysis we start from the hydrodynamic 3D model that
we believe represents a realistic dayside model of the HD 189733b
atmosphere, the results of this approach are valid regardless of
whether we believe in the initial model. This method can be
applied to any set of temperature--pressure profiles, in an attempt to
test how well the 1D retrieval can match the initial 3D structure.

We have released the software written for this analysis along with all
inputs and outputs under the reproducibility-research license,
allowing anybody to use it and modify it. At the same time, we require
that future users/developers attach the same license to their
additions to this code. We wish to ensure the reproducibility of our
results, and to support the efficient progress of science. The RRC for this paper, including all the packages and documentation, is available at \href{
https://github.com/dzesmin/RRC-BlecicEtal-2017-ApJ-3Dretriev}{github.com/dzesmin/RRC-BlecicEtal-2017-ApJ-3Dretriev}.

\begin{figure*}[t!]
\centering
\includegraphics[width=.85\textwidth, clip=True]{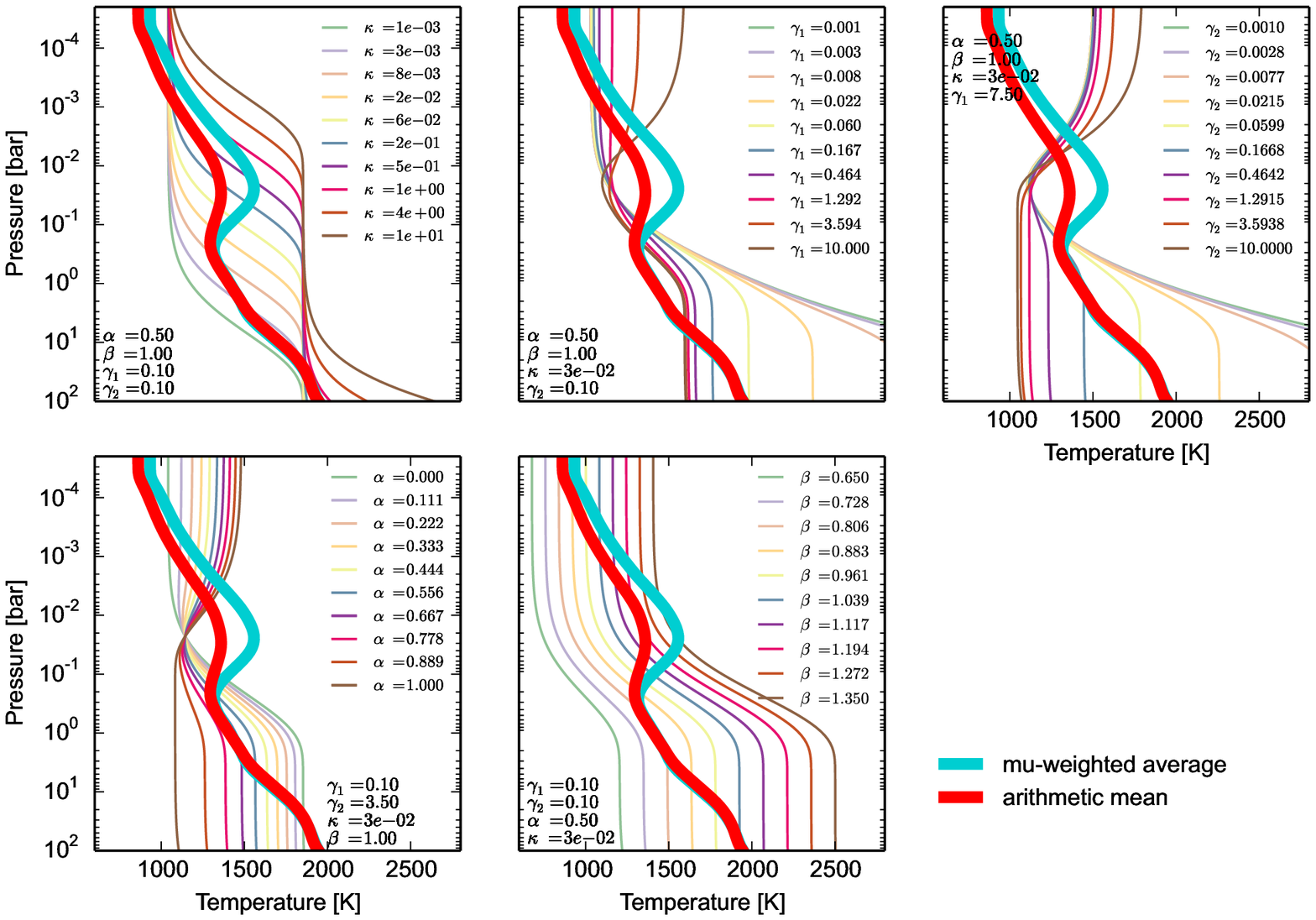}
\vspace{-20pt}
\caption{Possible temperature profile shapes generated using the temperature Parametrization I given by Equation (\ref{LineTP}), Appendix \ref{sec:Line}. This parametrization has five free parameters, \math{\kappa\sb{\rm{IR}}}, \math{\kappa\sb{\rm{\upsilon{1}}}}, \math{\kappa\sb{\rm{\upsilon{2}}}}, \math{\alpha},
and \math{\beta}. In the
figures \math{\gamma\sb{\rm{1}}}=\math{\kappa\sb{\rm{\upsilon{1}}}}/\math{\kappa\sb{\rm{IR}}}
and \math{\gamma\sb{\rm{2}}}=\math{\kappa\sb{\rm{\upsilon{2}}}}/\math{\kappa\sb{\rm{IR}}}. In
each panel, we vary different parameter and fix the remaining parameters. The profiles
are overplotted with the \math{\mu}-weighted average and the
arithmetic average from RHD for comparison.}
\label{fig:LineProfiles}
\end{figure*}

\acknowledgments We would like to thank the contributors to SciPy, NumPy,
Matplotlib, and the Python Programming Language; the
open-source development website GitHub.com; and other
contributors to the free and open-source community. J.B and
I.D.D are supported by NASA trough the NASA ROSES-
2016/Exoplanets Research Program, grant NNX17AC03G.
The research was carried out, in part, on the High Performance
Computing resources at New York University, Abu Dhabi. \\

\begin{appendices}

\section{Appendix}
\label{sec:Appendix-A}

In this section we describe the two temperature parametrizations that are 
commonly used in retrieval. We explore their possible shapes with a
goal to reveal their advantages and limitations.

\subsection{Parametrization Scheme I}
\label{sec:Line}

This parametrization scheme was originally formulated
by \citet{Guillot2010A-LinePTprofile} and subsequently modified
by \citet{ParmentierGuillot2014-LinesPTprofile,
LineEtal2013-Retrieval-III}, and \citet{HengEtal2012LinePTprofile} to
include more freedom for the case when a temperature inversion is
present in a planetary atmosphere. The approach is usually denoted as
the three-channel approximation, where the planet temperature is
given as

\begin{equation}
\label{LineTP}
T^4(\tau) = \frac{3T_{\rm{int}}^4}{4}\big(\frac{2}{3} + \tau\big)
+ \frac{3T_{\rm{irr}}^4}{4}(1 -\alpha)\,\xi_{\gamma_{\rm 1}}(\tau)
+ \frac{3T_{\rm{irr}}^4}{4}\,\alpha\,\xi_{\gamma_{\rm 2}}(\tau)\, ,
\end{equation}

\noindent with \math{\xi\sb{\gamma_{\rm i}}} defined as

\begin{equation}
\xi_{\gamma_{\rm i}} = \frac{2}{3} + \frac{2}{3\gamma_{\rm i}}\big[1 + \big(\frac{\gamma_{\rm i}\tau}{2} - 1\big)\,e^{-\gamma_{\rm i}\tau}\big] + \frac{2\gamma_{\rm i}}{3} \big(1 - \frac{\tau^2}{2}\big)\,E_2(\gamma_{\rm i}\tau)\, .
\end{equation}

\noindent \math{\gamma\sb{\rm 1}} and \math{\gamma\sb{\rm 2}} are the ratios of the mean opacities in the visible to the ratio in the infrared, given as \math{\gamma\sb{\rm 1} = \kappa\sb{\upsilon\sb{1}}/\kappa\sb{IR}} and \math{\gamma\sb{\rm 2} = \kappa\sb{\upsilon\sb{2}}/\kappa\sb{IR}}. The parameter \math{\alpha} ranges between 0 and 1 and describes the relative weight of the two visible streams, \math{\kappa\sb{\rm{\upsilon{1}}}} and \math{\kappa\sb{\rm{\upsilon{2}}}}. \math{E\sb{2}(\gamma\tau)} is the second-order exponential integral function. The irradiation that the planet receives is given as

\begin{equation}
T\sb{\rm{irr}} = \beta\,\big(\frac{R_{*}}{2a})^{1/2}\, T_{*} \,
\end{equation}

\noindent where \math{R\sb{*}} and \math{T\sb{*}} are the stellar radius and temperature, and \math{a} is the semimajor axis. The internal planetary flux is denoted as \math{T\sb{\rm{int}}}. Its value is usually estimated to \math{\sim}100 K and fixed, as it has little impact on the spectra. The parameter \math{\beta} has a value of around 1 and accounts for albedo, emissivity, and day-night redistribution.  The parameter \math{\tau} is the infrared optical depth calculated using the mean infrared opacity, \math{\kappa\sb{\rm{IR}}}, pressure \math{P}, and the planet surface gravity \math{g} at one the 1 bar level:

\begin{equation}
\tau = \frac{\kappa_{\rm{IR}}\,P}{g} \, ,
\end{equation}

Each \math{T-P} profile has five free
parameters: \math{\kappa\sb{\rm{IR}}}, \math{\kappa\sb{\rm{\upsilon{1}}}}, \math{\kappa\sb{\rm{\upsilon{2}}}}, \math{\alpha},
and \math{\beta}. The energy balance at the top of the atmosphere is
accounted for with the parameter \math{\beta}. The existence of a
temperature inversion is allowed through the
parameters \math{\kappa\sb{\rm{\upsilon{1}}}}
and \math{\kappa\sb{\rm{\upsilon{2}}}}. To explore the parameter
phase space, we follow \citet{LineEtal2013-Retrieval-I}, Section 3.2,
when imposing boundaries.

In Figure \ref{fig:LineProfiles}, we plotted possible profile shapes
using this parametrization, allowing one parameter to vary and fixing
the remaining parameters. Our initial parameters are chosen to reproduce the best-fit
model of the dayside atmopshere of HD 189733b
from \citet{SwainEtal2008NatureHD189733bspec}. We overplotted each
case with the \math{\mu}-weighted average and arithmetic averages from
Figure \ref{fig:averages}.

\subsection{Parametrization Scheme II}
\label{sec:Madhu}

The second temperature parametrization was originally developed
by \citet{MadhusudhanSeager2009ApJ-AbundanceMethod}. In this scheme,
the profiles are generated for inverted and non-inverted atmospheres
separately, see Equations (\ref{invert}) and (\ref{NONinvert}). We
make minor changes to this approach as described below.

The atmosphere is divided into three layers based on the physical
constraints expected in hot Jupiters, as shown in
Figure \ref{fig:MadhuPT}. Layer 3, the deep isothermal layer, exists
due to the strong irradiation from the parent star, which shifts the
radiative-convective boundary deep in the planetary atmosphere
(usually thought to be around several 100 bars). Most of the radiation
is absorbed higher in the atmosphere and cannot reach the deep
atmospheric layers. Layer 2, the stratospheric-radiative layer, is the
zone where radiation is the dominant transport mechanism. Depending on
the level of irradiation from the host star, a thermal inversion can
occur. Most of the spectral features come from this layer. Layer 1,
the mesospheric layer, is the layer below 10\sp{-5} bars, important
for atmospheric escape and photochemistry. This layer is transparent
to the incoming and outgoing radiation in the infrared and optical, and does not affect the emergent spectra. It is heated from the lower layers and
cools with increasing altitude.

\begin{figure}[h!]
\centering
\hspace{-15pt}\includegraphics[clip,width=0.55\textwidth]{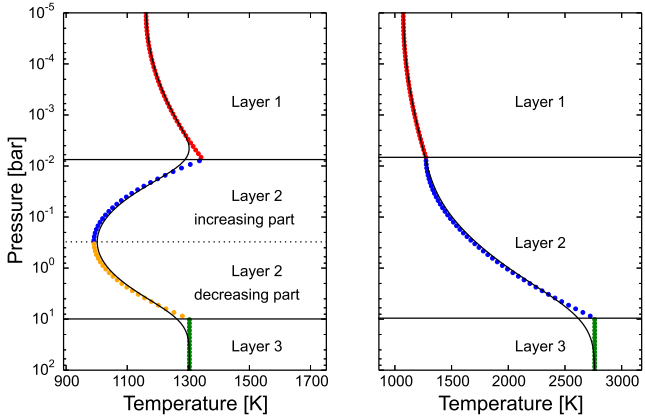}
\caption{Parametric \math{T-P} profiles for the inverted (left panel) and non-inverted (right panel) atmosphere based on \citet{MadhusudhanSeager2009ApJ-AbundanceMethod}. The profiles include three layers: the isothermal layer (Layer 3), the 'stratospheric' layer (Layer 2), the 'mesospheric' layer (Layer 1). Layer 2 can consist of two parts: one where temperature decreases with height, and the other where temperature increases with height, i.e., thermal inversion occurs. Different colors depict partial profiles for each layer generated using Equations (\ref{invert}) and (\ref{NONinvert}). The dots display the number of levels in the atmosphere, which is equally spaced in log-pressure space. The thin black lines show the smoothed profile using a 1D Gaussian filter.}
\label{fig:MadhuPT}
\end{figure}

The following set of equations, as given
by \citet{MadhusudhanSeager2009ApJ-AbundanceMethod}, describes the
behavior in each atmospheric layer:

\begin{align}
\label{MadhuPT}
  P_0 &< P < P_1& &P = P_0e^{\alpha_1(T - T_0)^{\beta_1}} &\hspace{0.2cm} \rm{layer\,\,1} \nonumber \\
  P_1 &< P < P_3& &P = P_2e^{\alpha_2(T - T_2)^{\beta_2}} &\hspace{0.2cm} \rm{layer\,\,2}  \\
  P~&> P_3& &T = T_3 &\hspace{0.2cm} \rm{layer\,\,3}  \nonumber
\end{align}

\begin{figure*}[t!]
\centering
\includegraphics[width=.85\textwidth, clip=True]{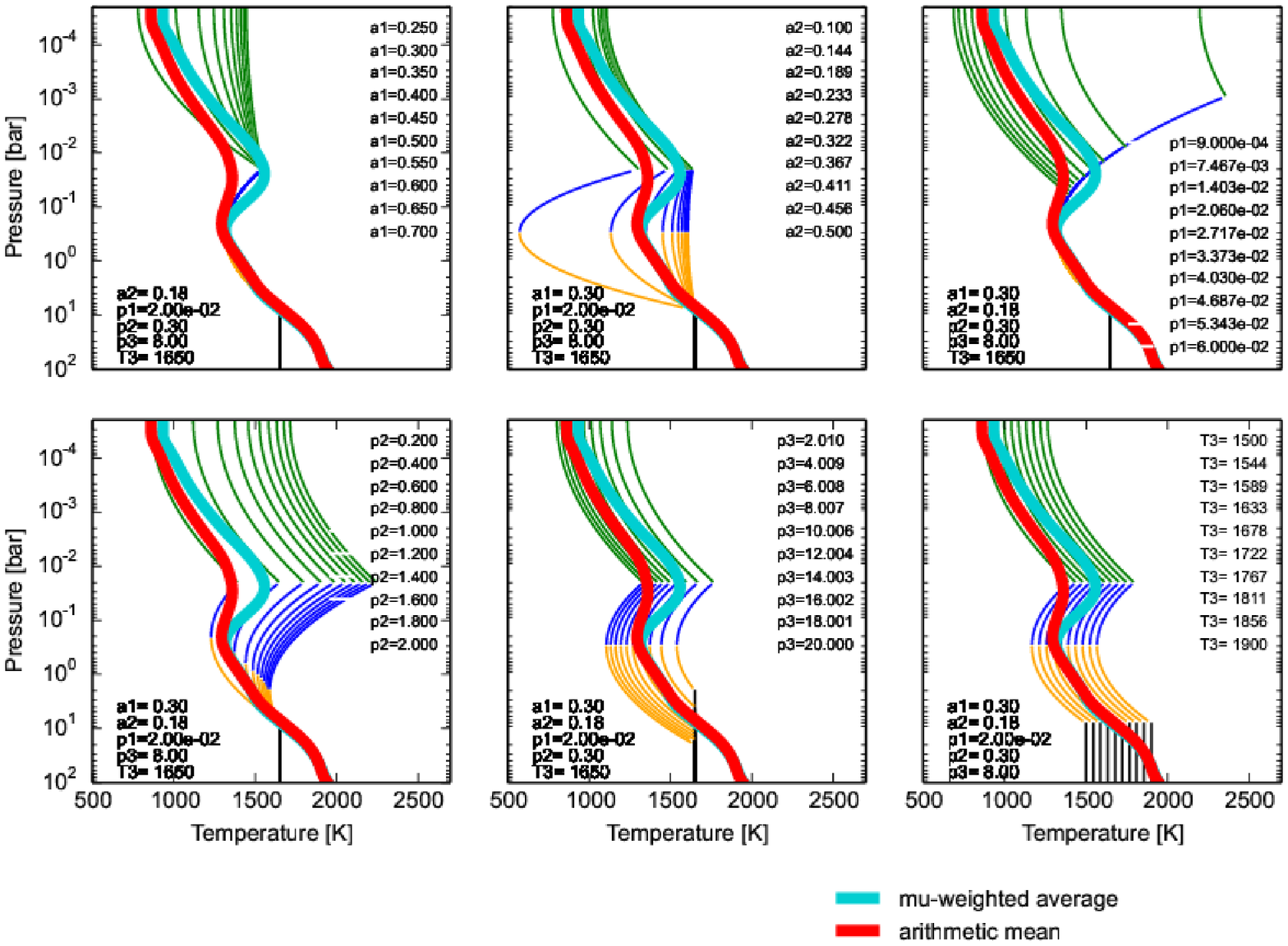}
\includegraphics[width=.85\textwidth, clip=True]{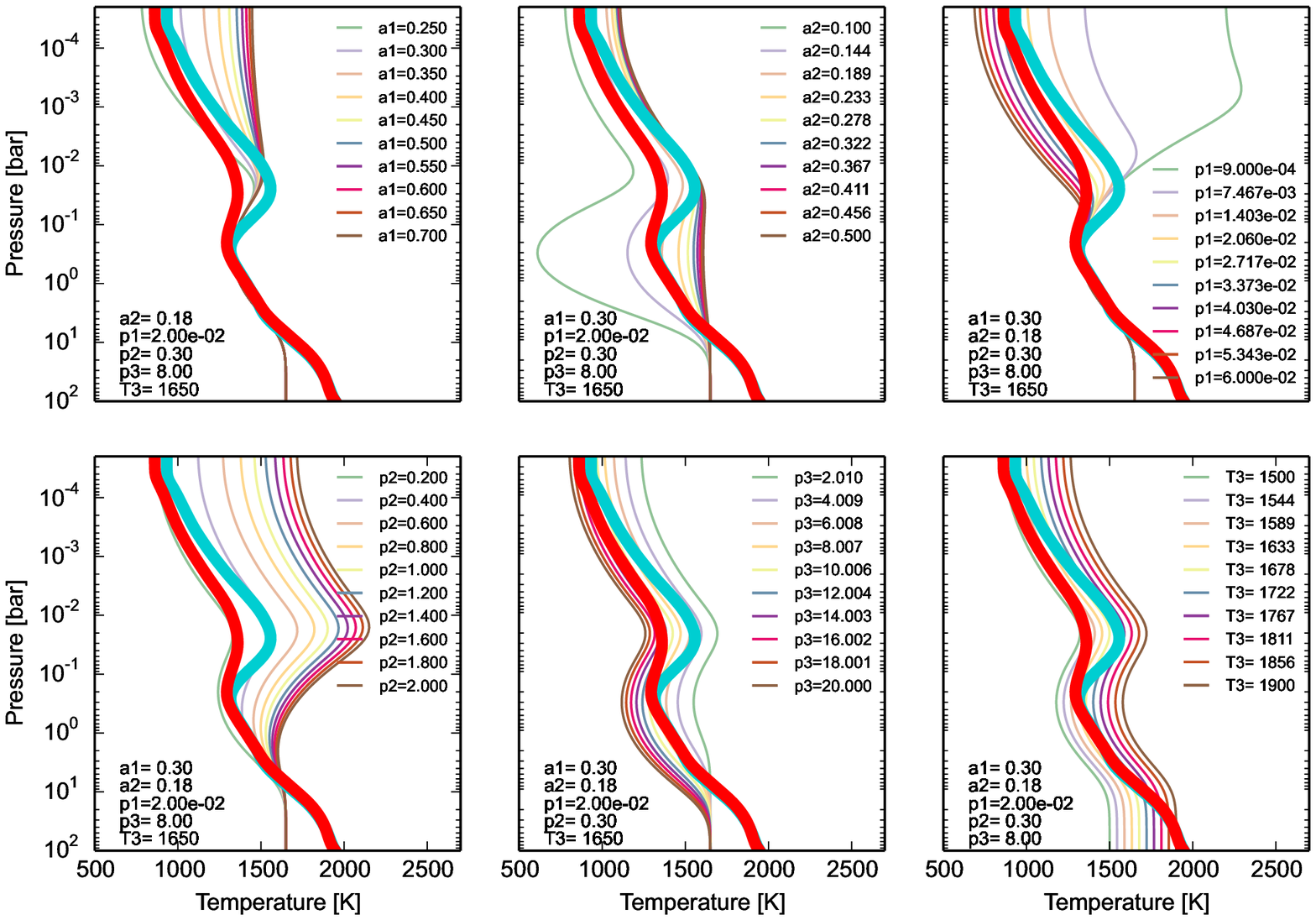}
\caption{Possible temperature profile shapes generated using the temperature parametrization II given by Equation (\ref{invert}), Appendix \ref{sec:Madhu}. This parametrization has six free parameters for the inverted atmosphere, \math{P}\sb{1}, \math{P}\sb{2}, \math{P}\sb{3}, \math{T}\sb{3}, \math{\alpha\sb{1}},
and \math{\alpha\sb{2}}. In each panel, we vary a different parameter
and fix the remaining parameters. The top six panels display each layer separately,
while the bottom panels show smoothed profiles. The profiles are
overplotted with the \math{\mu}-weighted average and the arithmetic
average from RHD for comparison.}
\label{fig:MadhuProfiles}
\end{figure*}

This set of equations reveals 12 unknown
parameters: \math{P}\sb{0}, \math{P}\sb{1}, \math{P}\sb{2}, \math{P}\sb{3}, 
\math{T}\sb{0}, \math{T}\sb{1}, {T}\sb{2},
{T}\sb{3}, \math{\alpha\sb{1}}, \math{\alpha\sb{2}}, \math{\beta\sb{1}},
and \math{\beta\sb{2}}. To decrease the number of free parameters, we made
minor changes to this method. We first set \math{P}\sb{0} to the
pressure at the top of the atmosphere (2x10\sp{-4}). The
parameters \math{\beta\sb{1}} and \math{\beta\sb{2}} are empirically
determined to be \math{\beta\sb{1}} = \math{\beta}\sb{2} =
0.5 \citep[see ][]{MadhusudhanSeager2009ApJ-AbundanceMethod}. Two of
the parameters can be eliminated based on the requirement of
continuity between two layer boundaries, i.e., layers 1--2 and layers
2--3. The initial guess of temperature \math{T\sb{3}} is estimated
based on the equilibrium temperature of the planet. Based on the
energy balance equation, the planetary equilibrium temperature is
given as:.

\begin{equation}
\label{Teq}
T_{\rm eq}^4 = f\,\, T_{\rm eff}^{*\,\,\,4}\, \Big(\frac{R}{a}\Big)^2\,(1-A)\, ,
\end{equation}

\noindent where the factor \math{f} describes the energy redistribution from the day-
to the night side. \math{f} = 1/4 defines the uniform redistribution
of energy between the day- and the nightside of the planet. Since we
are observing the planet dayside during secondary eclipse, we are
interested in the case when none of the energy is transferred to the
night side. In that case, the factor \math{f} is 1/2. For a zero albedo,
Equation (\ref{Teq}) becomes

\begin{equation}
\label{PlanetTq}
T_{\rm eq}^4 = \frac{1}{2}\,T_{\rm
eff}^{*\,\,\,4}\, \Big(\frac{R}{a}\Big)^2\, .
\end{equation}

The parametric profile for an {\bf inverted atmosphere} has six free
parameters, and these
are \math{P}\sb{1}, \math{P}\sb{2}, \math{P}\sb{3}, \math{T}\sb{3}, \math{\alpha\sb{1}},
and \math{\alpha\sb{2}}. We calculate the \math{T\sb{0}}, \math{T\sb{1}},
and \math{T\sb{2}} temperatures as

\begin{align}
\label{invert}
  T_2 = T_3 - \big(\frac{log(P_3/P_2)}{\alpha_2}\big)^2   \nonumber \\
  T_0 = T_2 - \big(\frac{log(P_1/P_0) }{\alpha_1}\big)^2 + \big(\frac{log(P_1/P_2)}{-\alpha_2}\big)^2  &\hspace{0.4cm} \\
  T_1 = T_0 + \big(\frac{log(P_1/P_0)}{\alpha_1}\big)^2 &\hspace{0.2cm} \nonumber
\end{align}

For a {\bf non-inverted atmosphere}, we assume that the Layer 2
follows an adiabatic temperature profile and exclude \math{P\sb{2}} as
a free parameter. Thus, the parametric profile for the non-inverted
atmosphere has five free
parameters: \math{P}\sb{1}, \math{P}\sb{3}, \math{T}\sb{3}, \math{\alpha\sb{1}},
and \math{\alpha\sb{2}}.  We calculate \math{T\sb{0}}
and \math{T\sb{1}} as

\begin{align}
\label{NONinvert}
  T_1 = T_3 - \big(\frac{log(P_3/P_1)}{\alpha_2}\big)^2   \\
  T_0 = T_1 - \big(\frac{log(P_1/P_0)}{\alpha_1}\big)^2  &\hspace{0.1cm} \nonumber
\end{align}

An example of an inverted and a non-inverted profile is shown in
Figure \ref{fig:MadhuPT}. To smooth the profiles so they do not have
sharp kinks on the layer boundaries, we used a 1D Gaussian filter
function, where 10\% of total number of data points are used to smooth
the data.

Figure \ref{fig:MadhuProfiles} shows different profiles generated
using Equation (\ref{invert}) by varying one parameter and fixing the
rest. The top panel displays the exact solution, while the bottom
panel shows smoothed profiles. We again overplot each case with
the \math{\mu}-weighted average and arithmetic averages from
Figure \ref{fig:averages}.

\section{Appendix}
\label{sec:Appendix-B}

In this section we discuss the retrieval results when we include
all three instruments together ({\em JWST}, {\em HST}, and {\em
Spitzer}), and {\em HST} and {\em Spitzer}, separately. We test
both temperature parametrizations when {\em JWST} data are included
to investigate the more complex thermal shapes that occur in the middle
region of the planetary atmosphere. The following sections elaborate
on each of these cases.

\subsection{JWST, HST, and Spitzer}
\label{sec:all}

In this section we present results when all data are included. First
we discuss the results when we use temperature Parametrization I
(Appendix \ref{sec:Line}) and then we present the results using the
temperature Parametrization II
(Appendix \ref{sec:Madhu}). Figure \ref{fig:all} shows the best-fit
spectrum and \math{T-P} profile when we use temperature
Parametrization I with five free parameters (see
Figure \ref{fig:LineProfiles} for possible \math{T-P} shapes using
this parametrization). Figure \ref{fig:all-hist} shows the posterior
histograms of the temperature profile's free parameters. As seen, all
parameters except \math{\kappa} are well
constrained. Parameter \math{\kappa} hits the wall
around \math{\log\sb{10}(1.3)}, although the boundaries for this
parameter are set between \math{\log\sb{10}(-1)}
and \math{\log\sb{10}(4)}. As shown in Figure \ref{fig:LineProfiles},
higher \math{\kappa} values cause the bottom of the profile to extend to
very high temperatures. However, the current boundaries of the HITRAN
database forbid these steps. The HITRAN partition functions are not
defined above temperatures of 3000 K, and these steps are excluded
from the MCMC exploration. However, it is obvious that the MCMC has a
tendency to explore higher values of the \math{\kappa} parameter,
pushing the bottom shape of the temperature profile toward the higher
temperatures, and shifting the top kink-part to lower pressures.

In the left panel of Figure \ref{fig:all-cf}, we plot the suite of our
initial 3D profiles, with both averages from Section \ref{sec:aver},
overplotted with the best-fit retrieved temperature profile. We also
overplot the curves with the theoretical the contribution
functions \math{\mu}-weighted peak-weighted average and the contribution
function envelope for each instrument. In the right panel of Figure \ref{fig:all-cf}, we show the contribution functions calculated from the
retrieved best-fit model, again for each instrument separately. In
both cases, we plot only the averages from Sections \ref{sec:theo-cf}
and \ref{sec:ret-cf} that we believe represent the best the overall
contribution function trend. By comparing these two panels, our goal
is to show the pressure layers where each of our instruments (1) had the 
theoretical potential to probe the atmosphere (left panel) and (2) is
actually probing the atmosphere based on the best-fit model. We also
wish to see which part of the temperature--pressure profile is best
represented by the data, i.e., which part of the retrieved profile we
can believe.

As shown, both theoretical contribution functions and retrieved
contribution functions sample similar pressure intervals, emphasizing
that the bottom part of the temperature profile is not
reliable. However, the black curve does not match either the red curve
(the arithmetic average) or the blue curve (\math{\mu}-weighted
average). Returning to Figure \ref{fig:LineProfiles}, we can see that
this parametrization does not have capabilities to really reproduce
the complex shapes of any of these two profiles, or any of the
inversion profiles seen in orange that come from the the longitudes
and latitudes close to the substellar point (0.6 < \math{\mu} < 1.0).

The limitations of this parametrization approach led us to explore
the temperatures shapes of the parametrization II,
Appendix \ref{sec:Madhu}. This parametrization has six free parameters
for the inversion case. We set the boundaries of these parameters to
account for all possible plausible scenarious and let the MCMC to explore
the possible parameter space. In Figure \ref{fig:all-madhu} we show
the best-fit spectrum and temperature
profile. Figure \ref{fig:all-hist-madhu} shows the posterior
distribution of all six parameters. As shown, all parameters are nicely
constrained. When we compare the 3D averages with our best-fit
profile, we see a similar trend as in the case when we used
Parametrization I (Figure \ref{fig:all-cf}). However, this temperature
profile has more curvatures and matches the arithmetic
average (red curve) somewhat better.  Surprisingly, the contribution functions of the
retrieved best-fit model are shifted to lower pressures compared to
the theoretical contribution functions, suggesting that these
observations mostly probe the pressure around 10\sp{-2} bar (Figure \ref{fig:all-cf-madhu}).

\subsection{HST}
\label{sec:HST}

When we only have {\em HST} data available, the retrieved
best-fit spectrum is influenced by the
synthetic data points coming from a small region between 1.1 and 1.7
{\micron} (Figure \ref{fig:HST}. Again, we present the results using the Parametrization I
(Appendix \ref{sec:Line}), because the results generated using
Parametrization II (Appendix \ref{sec:Madhu}) lead to the same
conclusion.

The posterior histograms are not well constrained (Figure \ref{fig:HST-hist}, revealing that the
number of data points are not nearly enough to constrain the
temperature parameters fully. The 1\math{\sigma} and 2\math{\sigma} temperature regions are rather wide, which allows many possible temperature
scenarios. Still, the best-fit \math{T-P} profile is suggestive of a
non-inverted atmosphere, again falling in the cold terminator region
of our 3D temperature structure. The contribution functions for the
retrieved model are well below the averaged theoretical 3D
contribution functions, which prevents any conclusions about
the pressure range where inversion can occur (Figure \ref{fig:HST-cf}

\begin{figure*}[t!]
\centering
\hspace{-30pt}\includegraphics[width=.75\textwidth, height=6.5cm]{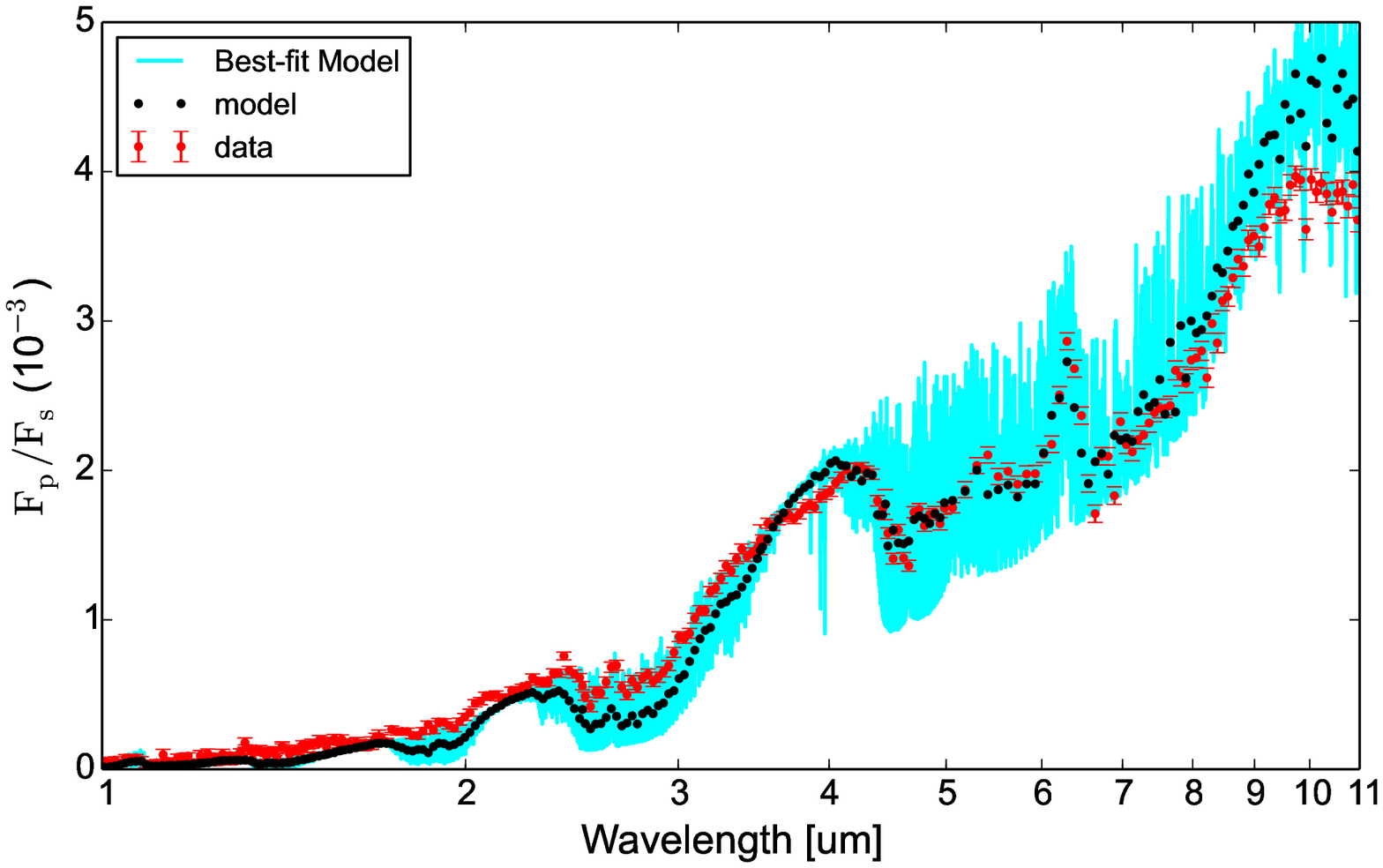}\hspace{-30pt}
\includegraphics[width=.28\textwidth, height=6.5cm]{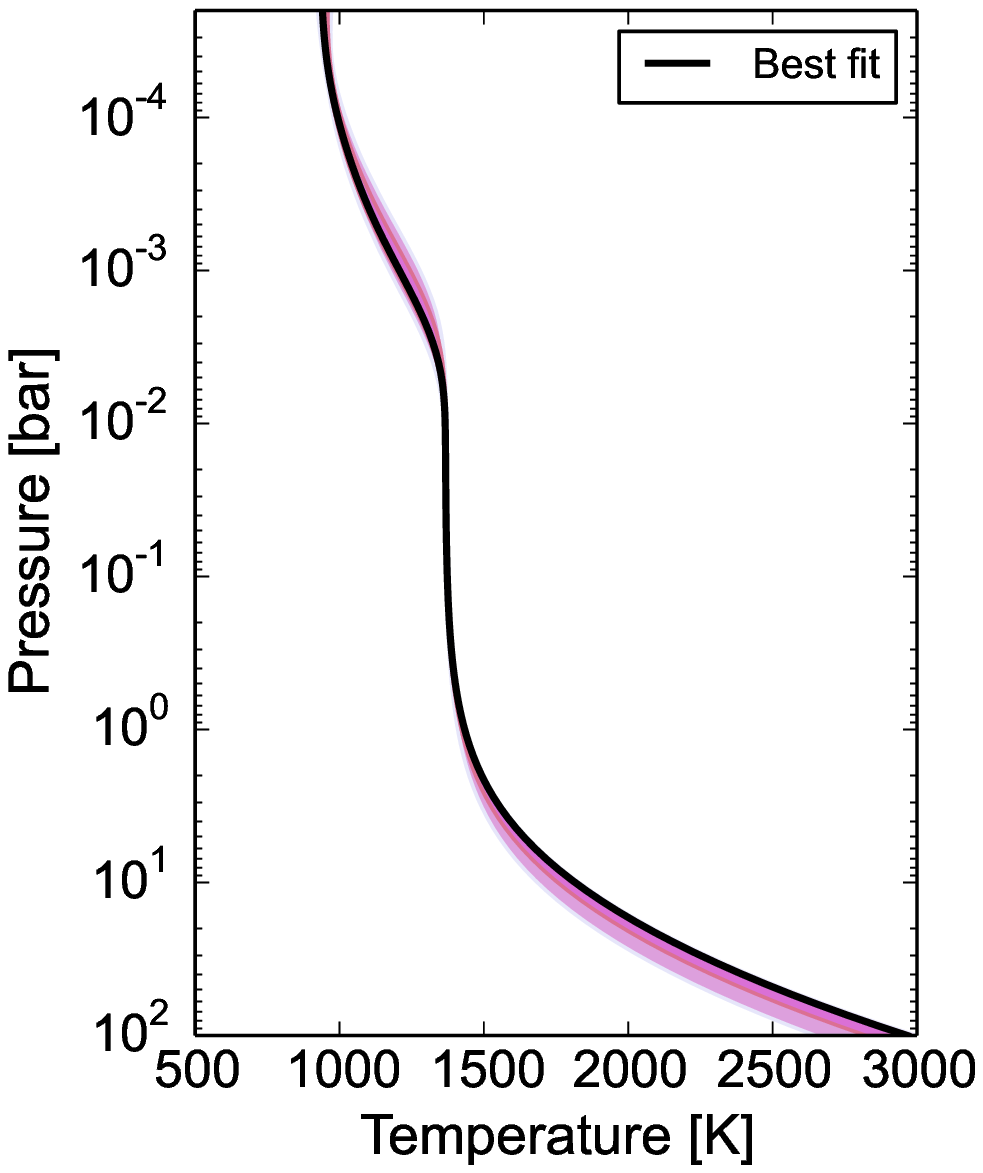}
\caption{Left: the retrieved best-fit spectra (blue) for the case when {\em all data are included, the {\em JWST}, {\em HST} and {\em Spitzer}} and the temperature profile is generated using the temperature {\em Parametrization I}, Appendix \ref{sec:Line}. In red are plottedcthe data points (eclipse depths) with error bars. In black we show the model points integrated over the bandpasses of our synthetic model. Right: the best-fit \math{T-P} profile with 1\math{\sigma} and 2\math{\sigma} confidence regions.}
\label{fig:all}
\end{figure*}

\begin{figure*}[t!]
\centering
\includegraphics[height=3.2cm, clip=True]{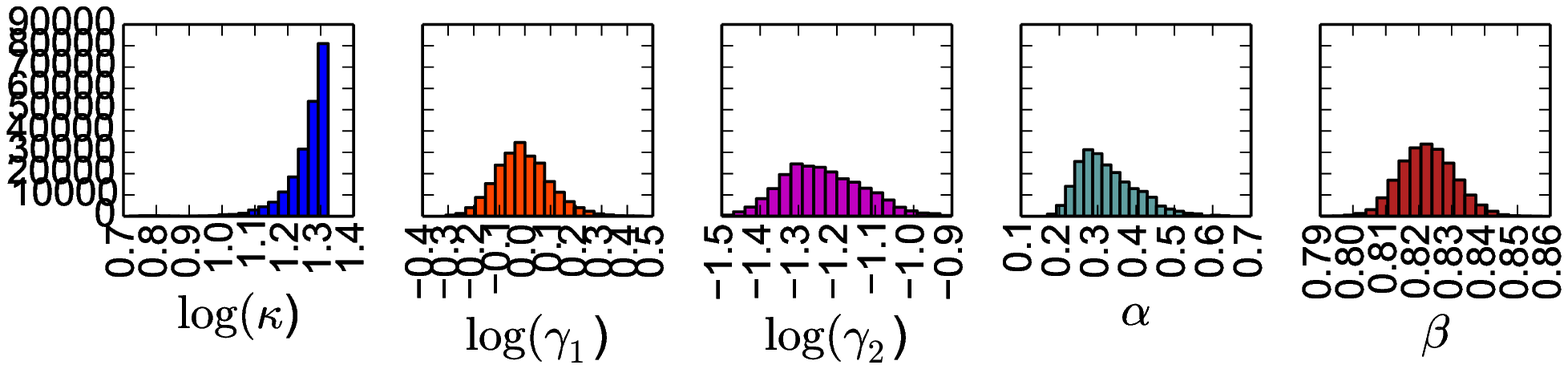}
\caption{Histograms of the temperature profile parameters for the case when {\em the {\em JWST}, {\em HST}, and {\em Spitzer}} synthetic data points are included and the temperature profile is generated using the temperature {\em Parametrization I}, Appendix \ref{sec:Line}. The panels show the \math{T-P} profile parameters, where some of them are expressed as \math{log\sb{10}(X)}, with \math{X} being the free parameter of the model.}
\label{fig:all-hist}
\end{figure*}


\begin{figure*}[t!]
\vspace{7pt}
\centering
\hspace{-15pt}\includegraphics[height=7.4cm, clip=True]{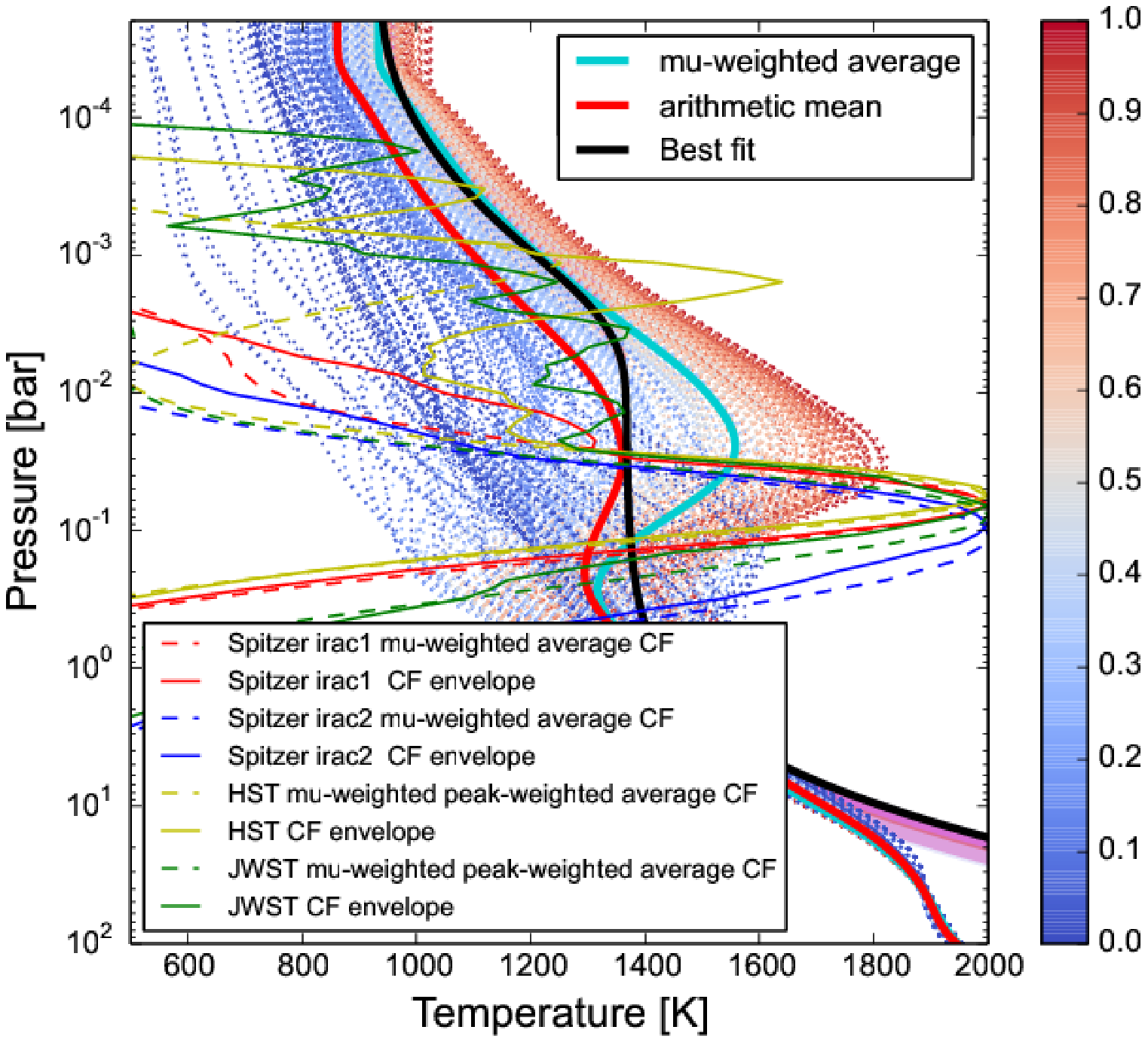}\hspace{-35pt}
\includegraphics[height=7.4cm, clip=True]{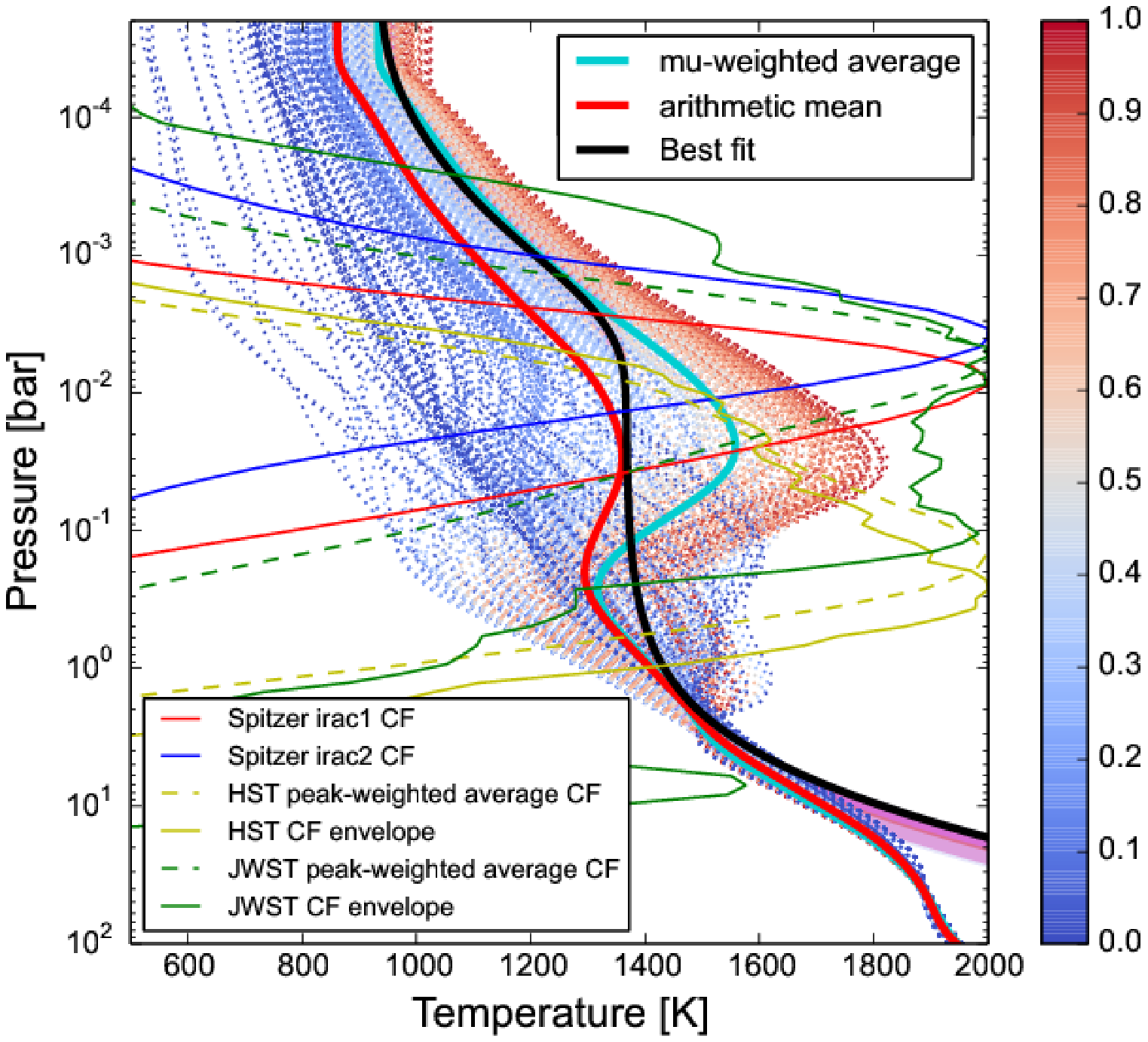}
\caption{Left: the 3D \math{T-P} profile dayside structure of HD 189733b, with the retrieved best-fit temperature profile (black curve) from Figure \ref{fig:all}, right panel, and the 3D thermal structure averages (red and turquoise curves), overplotted with {\em the {\em JWST}, {\em HST}, and {\em Spitzer} theoretical} contribution functions, normalized to 2000, and generated using {\em Parametrization I}, Appendix \ref{sec:Line}. Red and turquoise curves are the {\em Spitzer} theoretical contribution functions, while the yellow dotted curve is the {\em HST} theoretical \math{\mu}-weighted peak-weighted average, and the solid yellow curve is the {\em HST} contribution function envelope. The dotted green curve is the {\em JWST} \math{\mu}-weighted peak-weighted average, while the solid green curve is the {\em JWST} contribution function envelope (see Figure \ref{fig:Theor-CF-aver}).  Right: the 3D \math{T-P} profile dayside structure of HD 189733b, with the retrieved best-fit temperature profile (black curve) and 3D thermal structure averages (red and turquoise curves), overplotted with {\em the {\em JWST}, {\em HST}, and {\em Spitzer} retrieved} contribution functions, normalized to 2000, and generated using {\em Parametrization I}, Appendix \ref{sec:Line}. Red and turquoise curves are {\em Spitzer} contribution functions, while the yellow dotted curve is the {\em HST} peak-weighted average, and the solid yellow curve is the {\em HST} contribution function envelope. The dotted green curve is the {\em JWST} peak-weighted average, while the solid green curve is the {\em JWST} contribution function envelope.}
\vspace{12pt}
\label{fig:all-cf}
\end{figure*}

\begin{figure*}[t!]
\centering
\hspace{-30pt}\includegraphics[width=.75\textwidth, height=6.5cm]{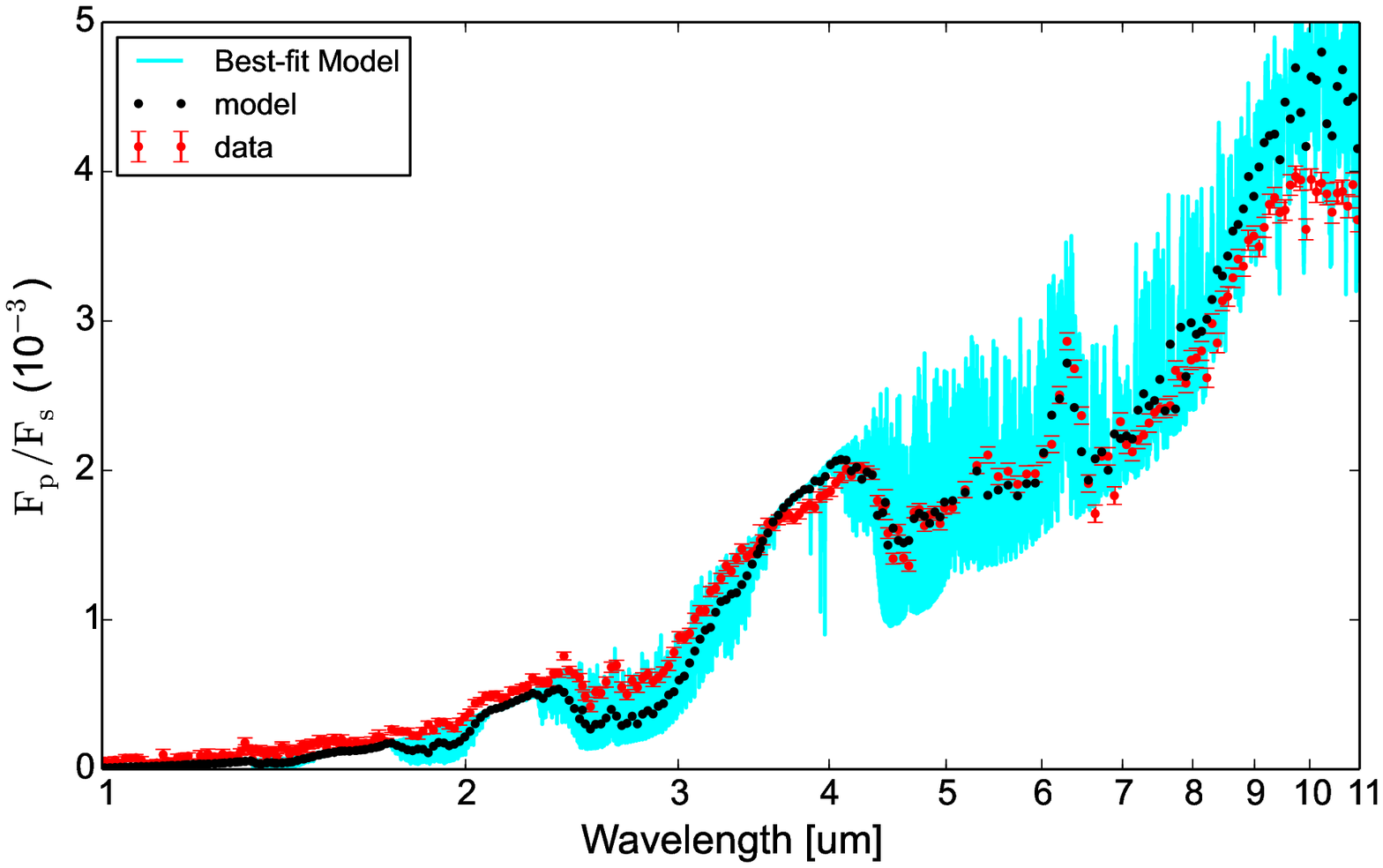}\hspace{-30pt}
\includegraphics[width=.28\textwidth, height=6.5cm]{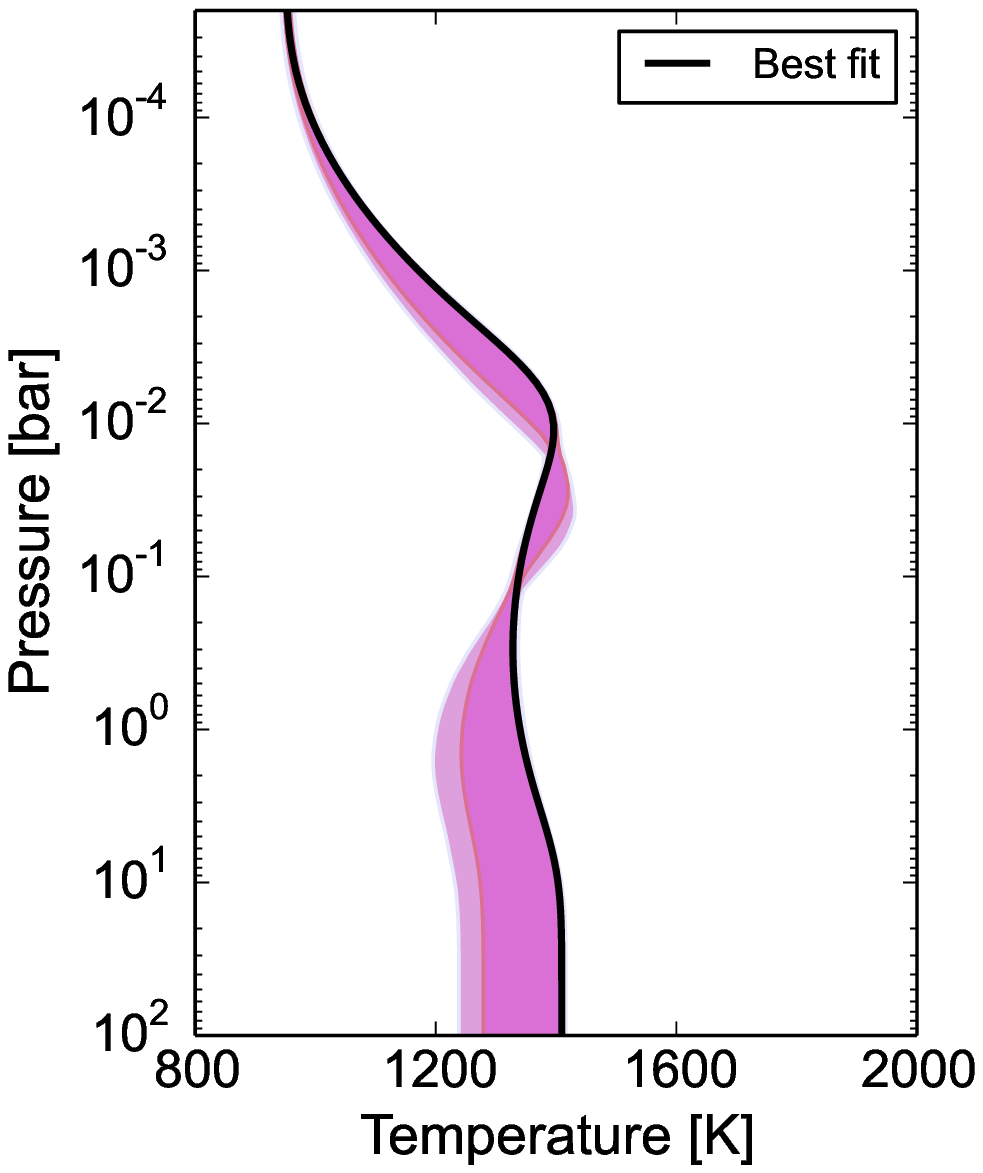}
\caption{Left: the retrieved best-fit spectra (blue) for the case when {\em all data are included, the {\em JWST}, {\em HST} and {\em Spitzer}} and the temperature profile is generated using the temperature {\em Parametrization II}, Appendix \ref{sec:Madhu}. In red are plotted the data points (eclipse depths) with error bars. In black we show the model points integrated over the bandpasses of our synthetic model. Right: the best-fit \math{T-P} profile with 1\math{\sigma} and 2\math{\sigma} confidence regions.}
\label{fig:all-madhu}
\end{figure*}

\begin{figure*}[t!]
\vspace{-5pt}
\centering
\includegraphics[height=3cm, clip=True]{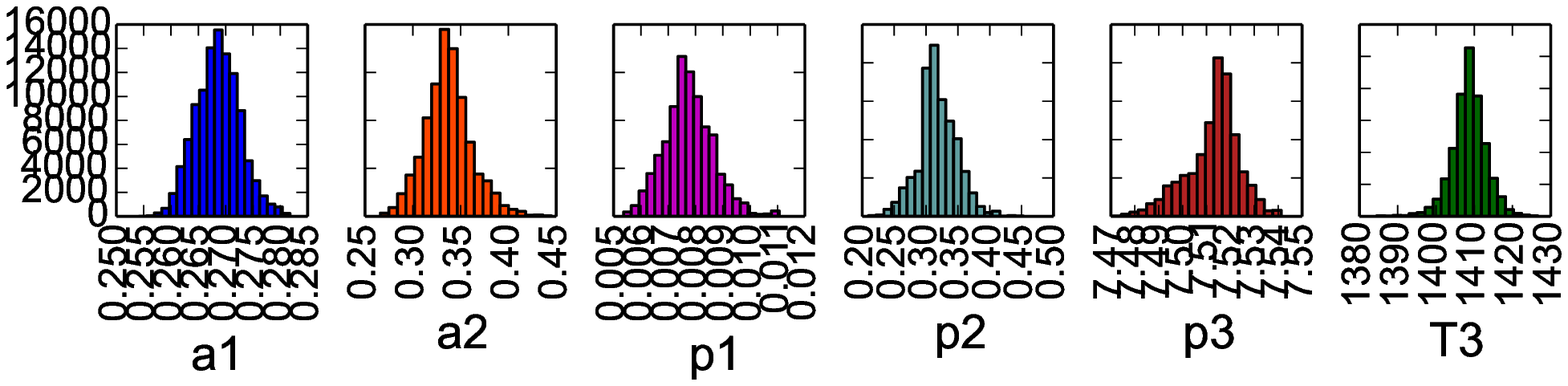}
\vspace{-5pt}
\caption{Histograms of the temperature profile parameters for the case when {\em the {\em JWST}, {\em HST}, and {\em Spitzer}} synthetic data points are included and the temperature profile is generated using the temperature {\em Parametrization II}, Appendix \ref{sec:Madhu}. The panels show the \math{T-P} profile parameters.}
\vspace{-10pt}
\label{fig:all-hist-madhu}
\end{figure*}


\begin{figure*}[t!]
\centering
\hspace{-15pt}\includegraphics[height=7.0cm, clip=True]{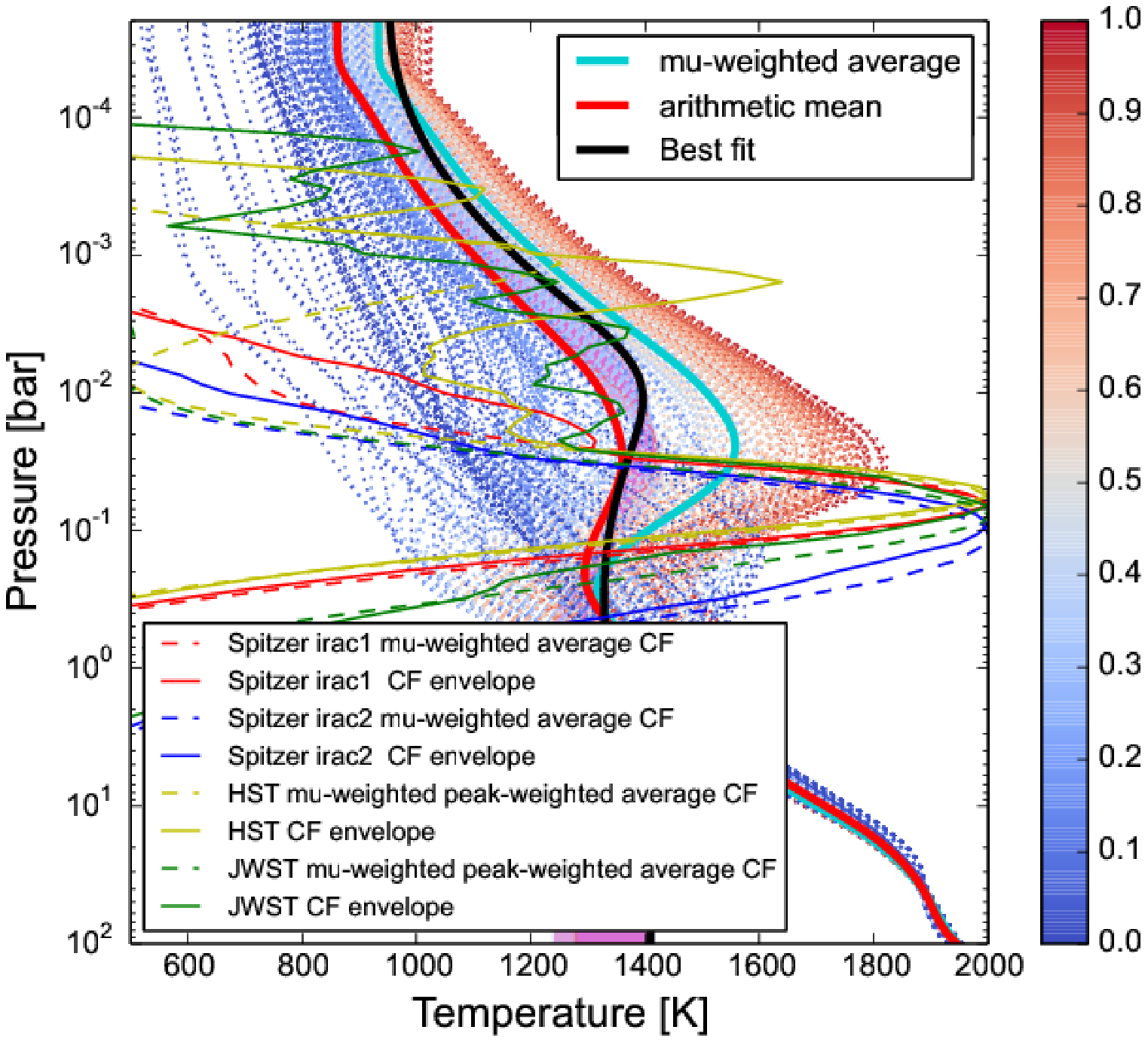}\hspace{-35pt}
\includegraphics[height=7.0cm, clip=True]{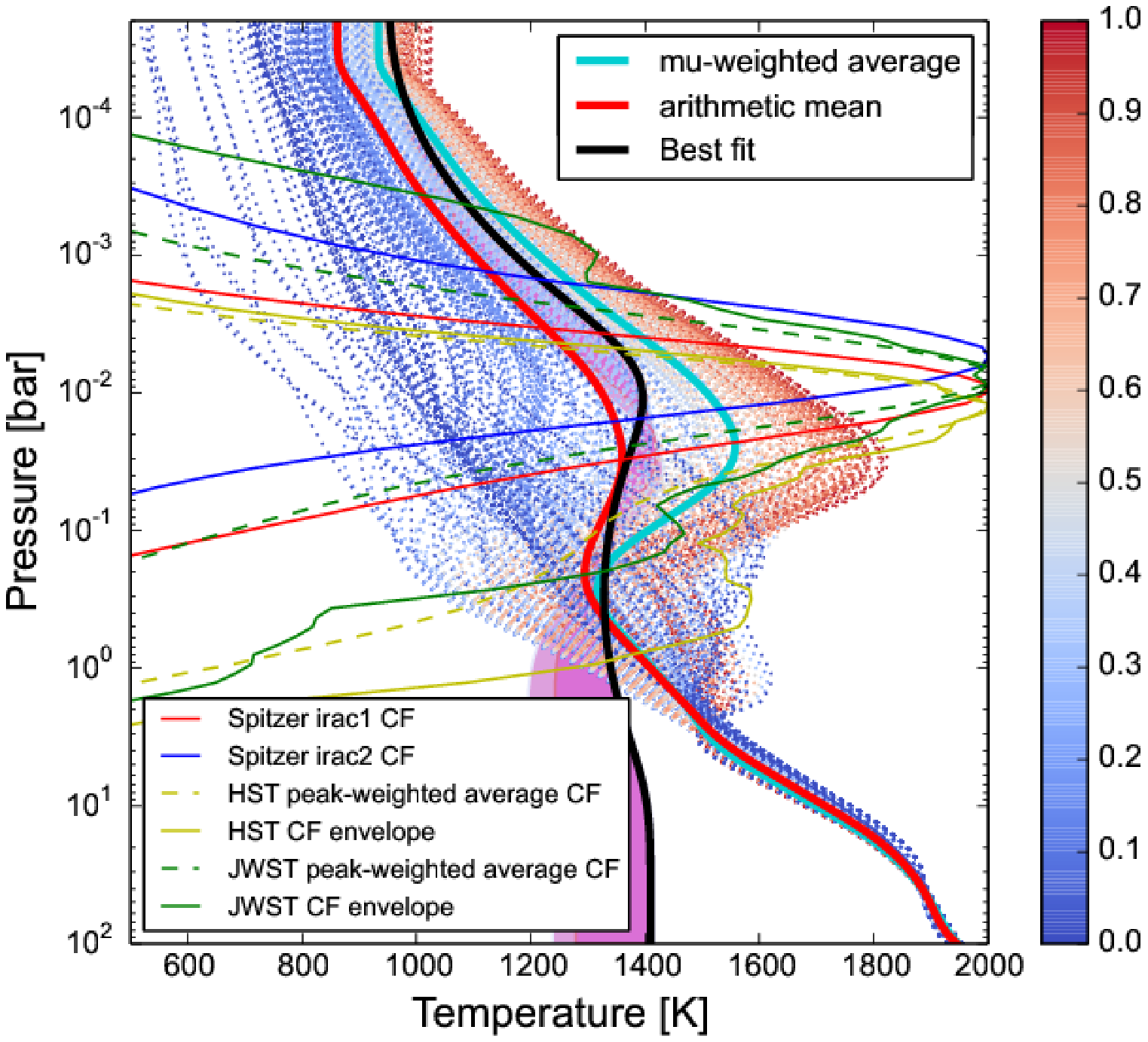}
\caption{Left: the 3D \math{T-P} profile dayside structure of HD 189733b, with the retrieved best-fit temperature profile (black curve) from Figure \ref{fig:all-madhu}, right panel, and the 3D thermal structure averages (red and turquoise curves), overplotted with {\em the {\em JWST}, {\em HST}, and {\em Spitzer} theoretical} contribution functions, normalized to 2000, and generated using {\em Parametrization II}, Appendix \ref{sec:Madhu}. Red and turquoise curves are the {\em Spitzer} theoretical contribution functions, while the yellow dotted curve is the {\em HST} theoretical \math{\mu}-weighted peak-weighted average, and the solid yellow curve is the {\em HST} contribution function envelope. The dotted green curve is the {\em JWST} \math{\mu}-weighted peak-weighted average, while the solid green curve is the {\em JWST} contribution function envelope (see Figure \ref{fig:Theor-CF-aver}). Right: the 3D \math{T-P} profile dayside structure of HD 189733b, with the retrieved best-fit temperature profile (black curve) and 3D thermal structure averages (red and turquoise curves), overplotted with {\em the {\em JWST}, {\em HST}, and {\em Spitzer} retrieved} contribution functions, normalized to 2000, and generated using {\em Parametrization II}, Appendix \ref{sec:Madhu}. Red and turquoise curves are {\em Spitzer} contribution functions, while the yellow dotted curve is the {\em HST} peak-weighted average, and the solid yellow curve is the {\em HST} contribution function envelope. The dotted green curve is the {\em JWST} peak-weighted average, while the solid green curve is the {\em JWST} contribution function envelope.}
\label{fig:all-cf-madhu}
\end{figure*}

\begin{figure*}[ht!]
\centering
\hspace{-30pt}\includegraphics[width=.75\textwidth, height=6.5cm]{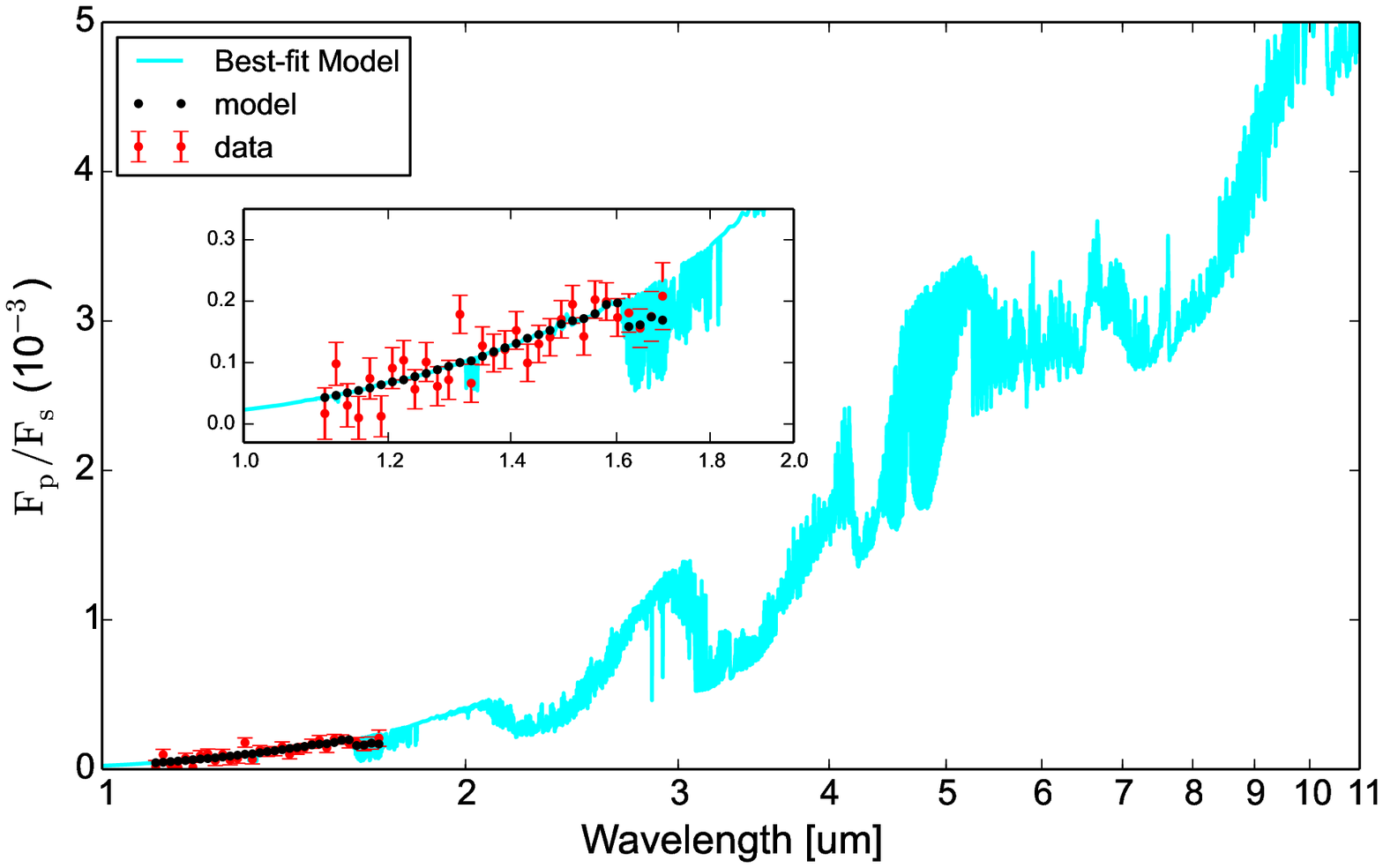}\hspace{-30pt}
\includegraphics[width=.28\textwidth, height=6.5cm]{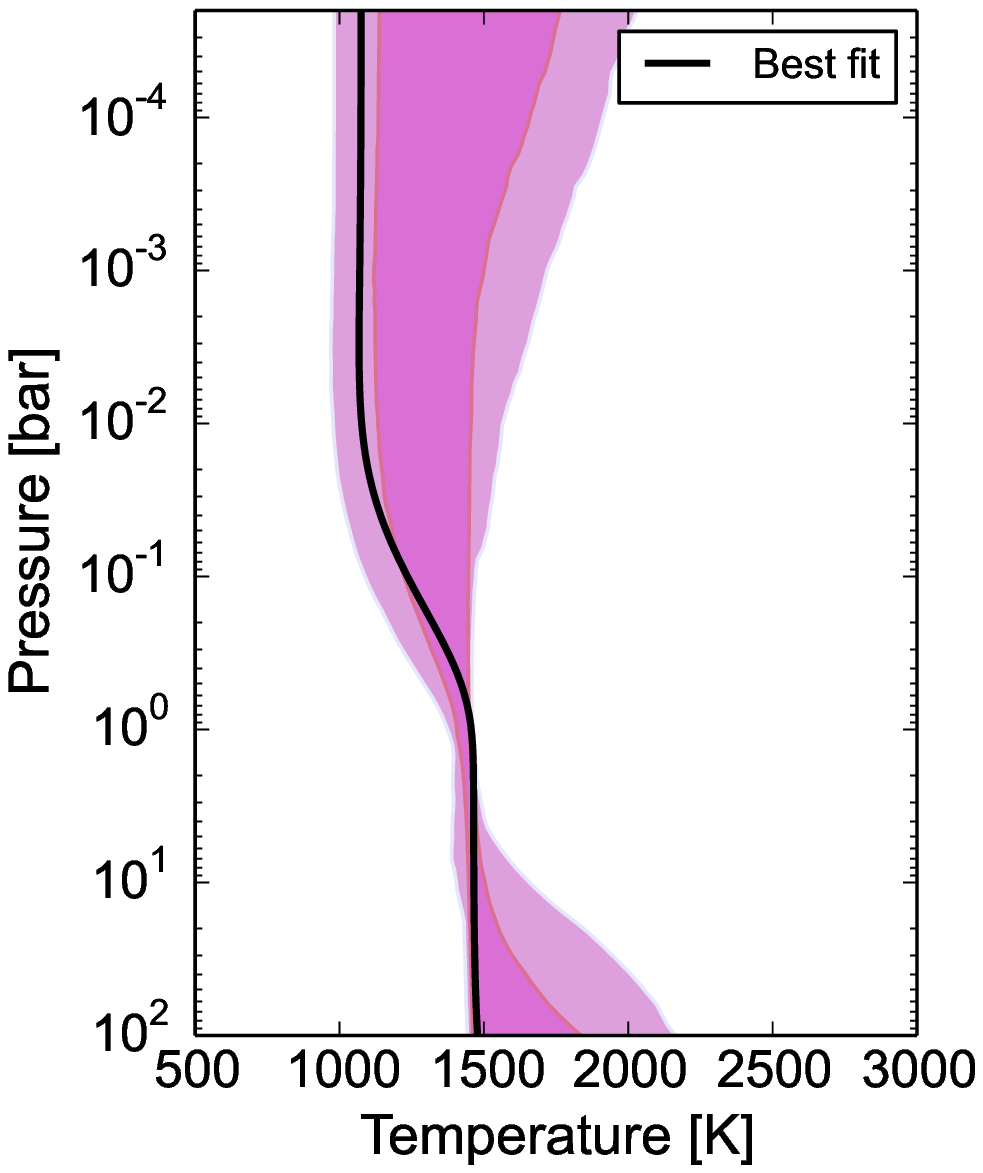}
\caption{Left: the retrieved best-fit spectra (blue) for the case when {\em only the {\em HST}} synthetic data are included and the temperature profile is generated using the temperature {\em Parametrization I}, Appendix \ref{sec:Line}. In red are plotted the data points (eclipse depths) with error bars. In black we show the model points integrated over the bandpasses of our synthetic model. Right: the best-fit \math{T-P} profile with 1\math{\sigma} and 2\math{\sigma} confidence regions.}
\label{fig:HST}
\end{figure*}

\begin{figure*}[t!]
\vspace{-5pt}
\centering
\includegraphics[height=3cm, clip=True]{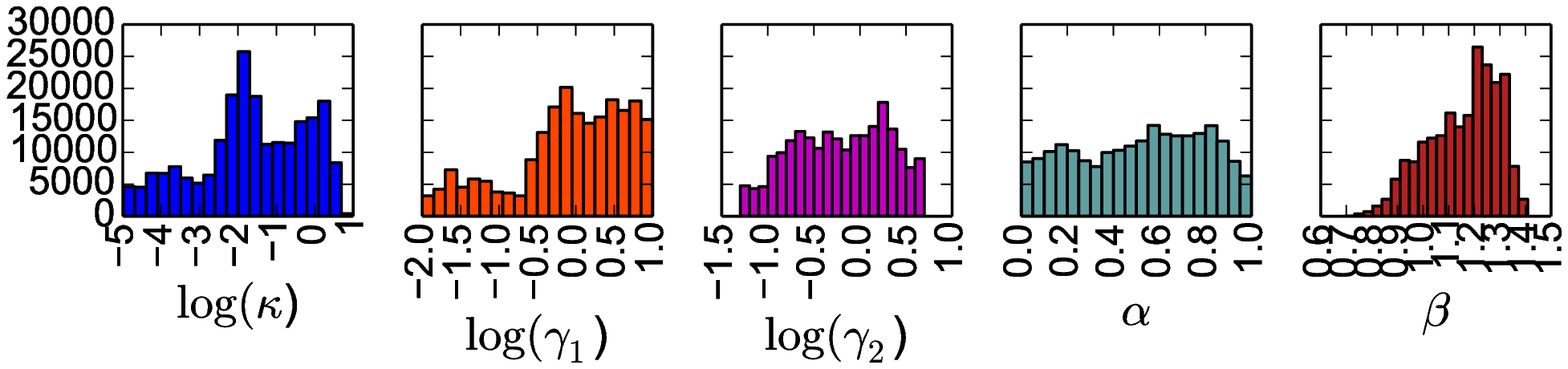}
\vspace{-5pt}
\caption{Histograms of the temperature profile parameters for the case when {\em only the {\em HST}} synthetic data points are included and the temperature profile is generated using the temperature {\em Parametrization I}, Appendix \ref{sec:Line}. Figures show the \math{T-P} profile parameters, where some of them are expressed as \math{log\sb{10}(X)}, with \math{X} being the free parameter of the model.}
\vspace{-10pt}
\label{fig:HST-hist}
\end{figure*}


\begin{figure*}[t!]
\centering
\hspace{-15pt}\includegraphics[height=7.0cm, clip=True]{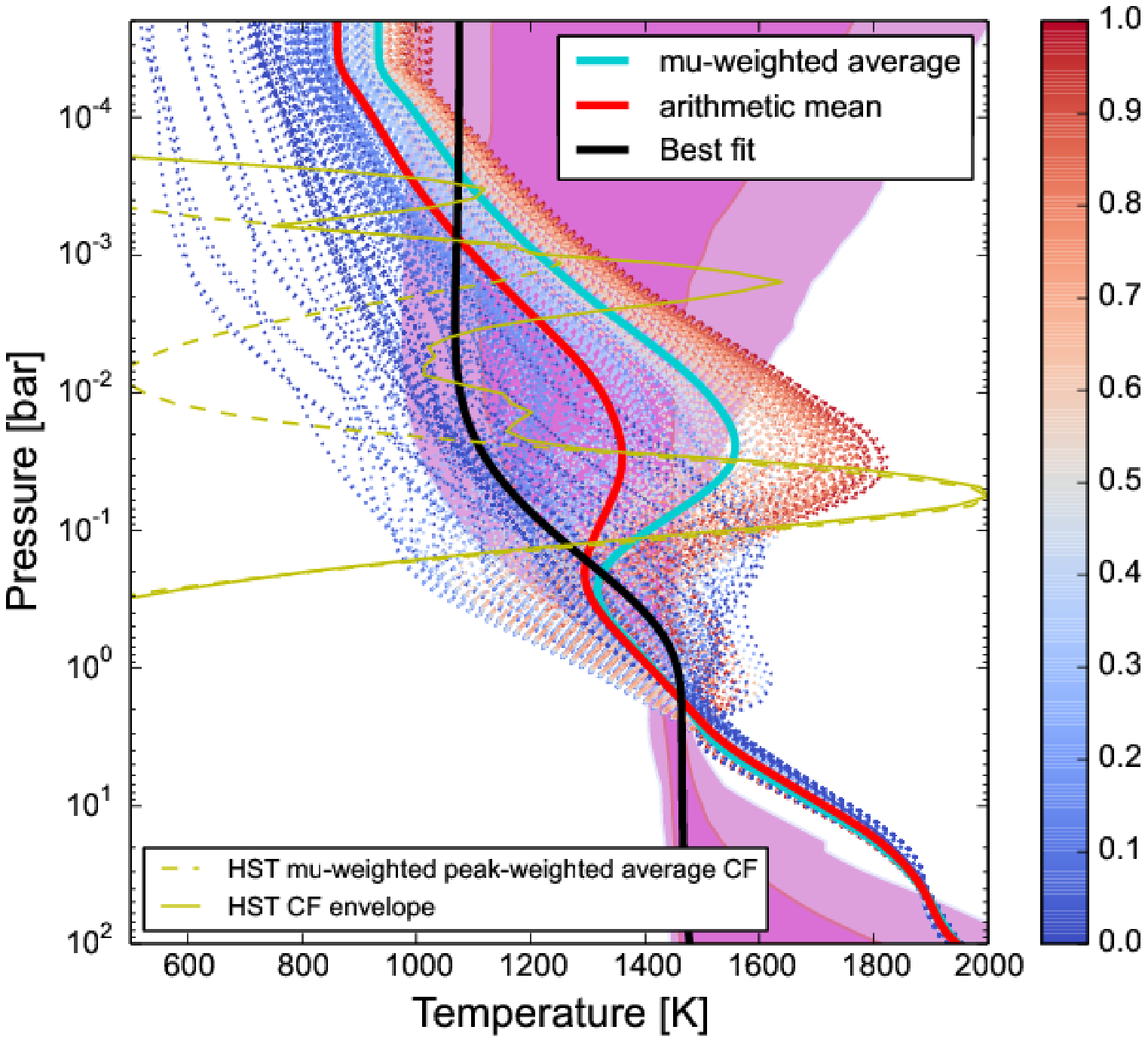}\hspace{-35pt}
\includegraphics[height=7.0cm, clip=True]{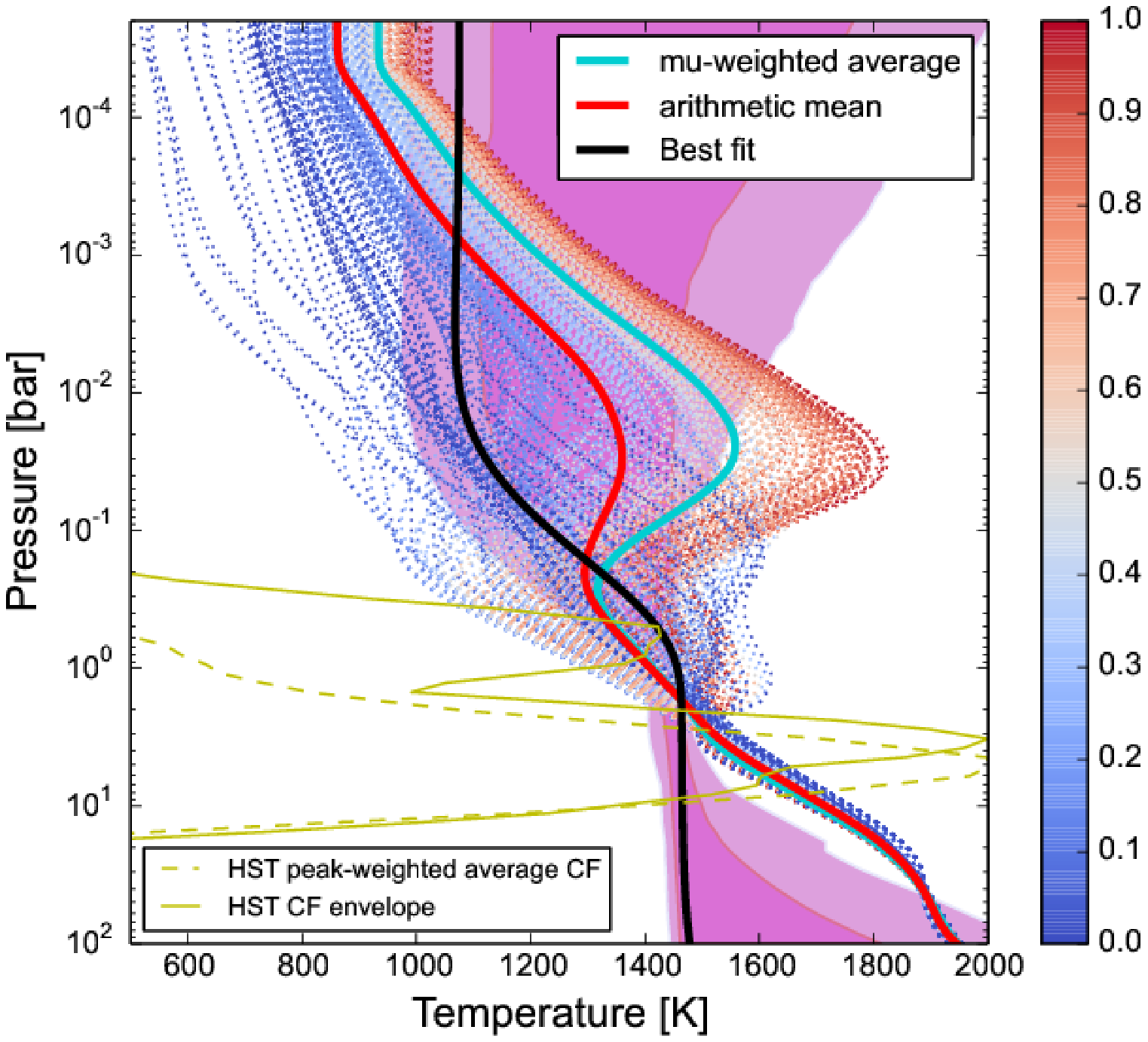}
\caption{Left: the 3D \math{T-P} profile dayside structure of HD 189733b, with the retrieved best-fit temperature profile (black curve) from Figure \ref{fig:HST}, right panel, and the 3D thermal structure averages (red and turquoise curves), overplotted with {\em only the {\em HST} theoretical} contribution functions, normalized to 2000, and generated using {\em Parametrization I}, Appendix \ref{sec:Line}. The yellow dotted curve is the {\em HST} theoretical \math{\mu}-weighted peak-weighted average, and the solid yellow curve is the {\em HST} contribution function envelope (see Figure \ref{fig:Theor-CF-aver}). Right: the 3D \math{T-P} profile dayside structure of HD 189733b, with the retrieved best-fit temperature profile (black curve) and 3D thermal structure averages (red and turquoise curves), overplotted with {\em only the {\em HST} retrieved} contribution functions, normalized to 2000, and generated using {\em Parametrization I}, Appendix \ref{sec:Line}. The yellow dotted curve is the {\em HST} peak-weighted average, and the solid yellow curve is the {\em HST} contribution function envelope.}
\label{fig:HST-cf}
\end{figure*}

\subsection{Spitzer}
\label{sec:Spitzer}

\begin{figure*}[ht!]
\centering
\hspace{-30pt}\includegraphics[width=.75\textwidth, height=6.5cm]{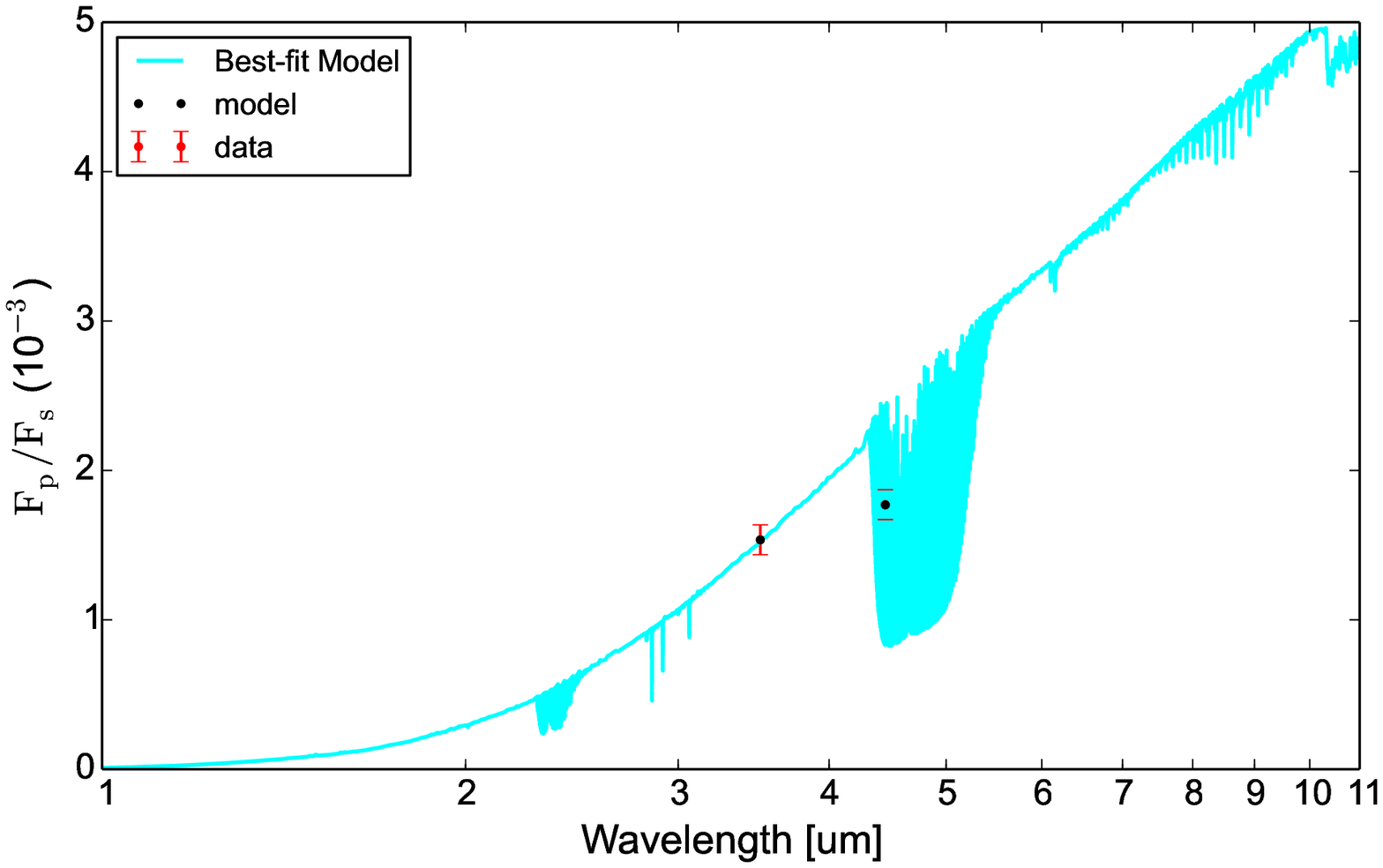}\hspace{-30pt}
\includegraphics[width=.28\textwidth, height=6.5cm]{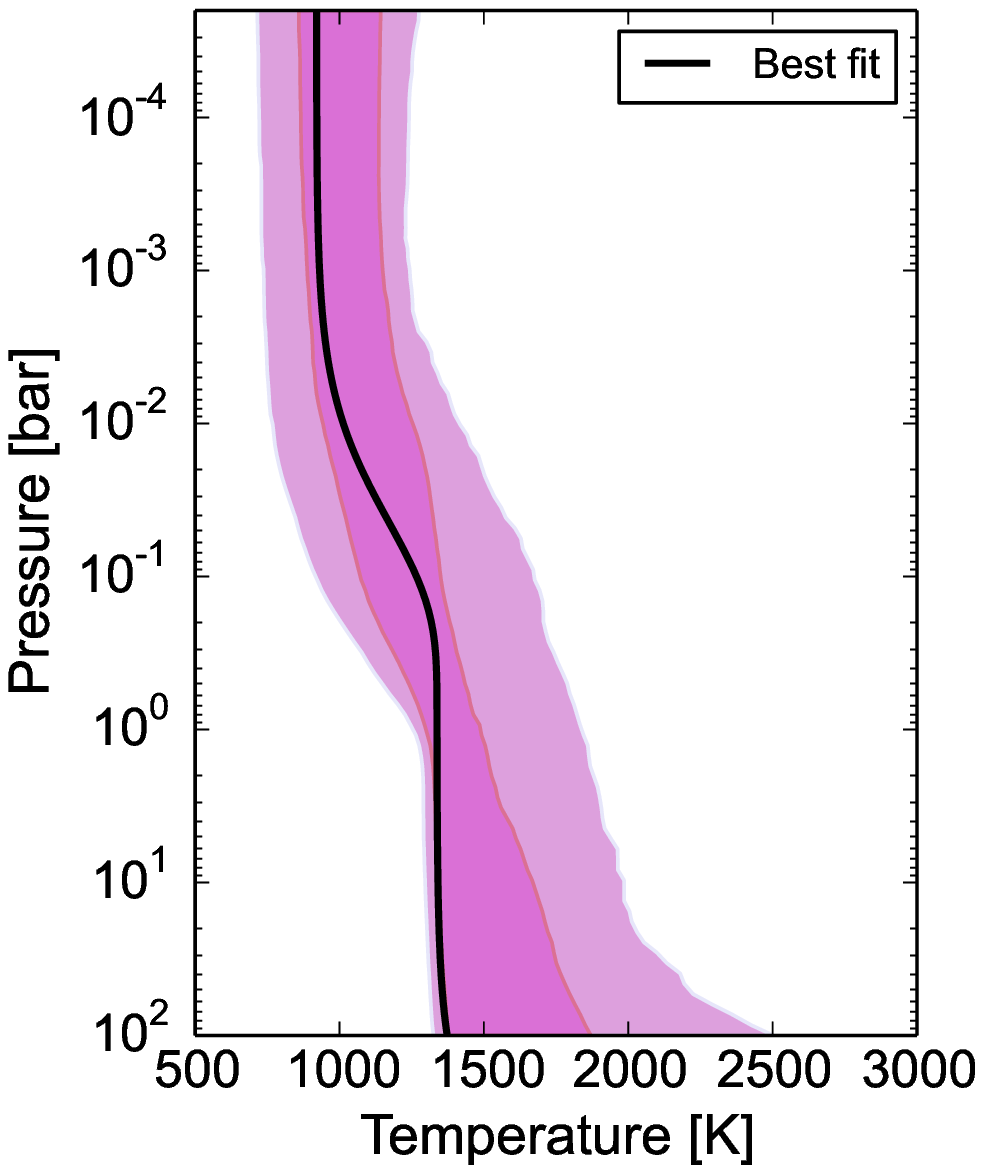}
\caption{Left: the retrieved best-fit spectra (blue) for the case when {\em only the {\em Spitzer}} synthetic data are included and the temperature profile is generated using the temperature {\em Parametrization I}, Appendix \ref{sec:Line}. In red are plotted the data points (eclipse depths) with error bars. In black we show the model points integrated over the bandpasses of our synthetic model. Right: the best-fit \math{T-P} profile with 1\math{\sigma} and 2\math{\sigma} confidence regions.}
\label{fig:spit}
\end{figure*}

\begin{figure*}[t!]
\vspace{-5pt}
\centering
\includegraphics[height=3cm, clip=True]{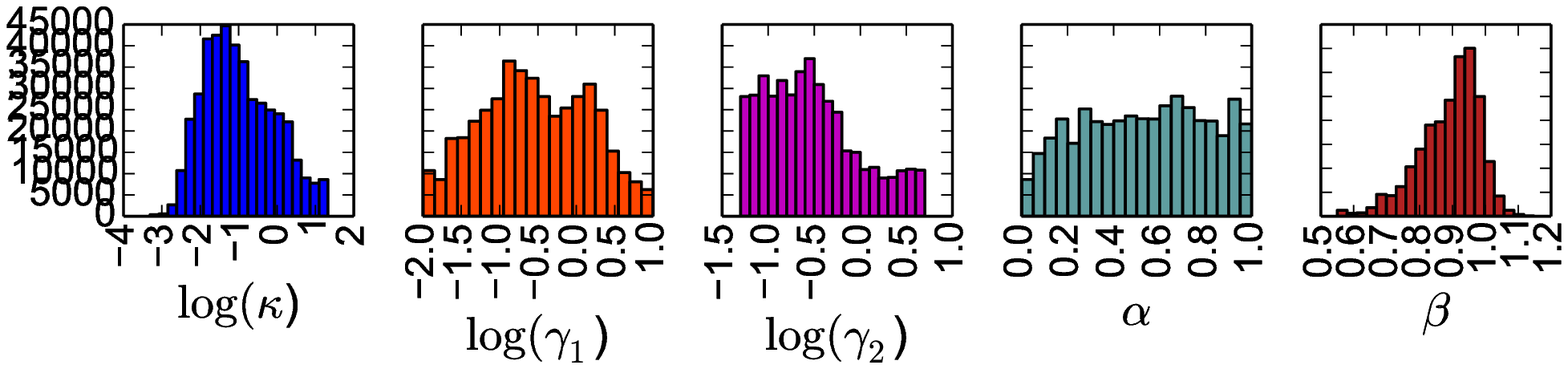}
\vspace{-5pt}
\caption{Histograms of the temperature profile parameters for the case when {\em only the {\em Spitzer}} synthetic data points are included and the temperature profile is generated using the temperature {\em Parametrization I}, Appendix \ref{sec:Line}. The panels show the \math{T-P} profile parameters, where some of them are expressed as \math{log\sb{10}(X)}, with \math{X} being the free parameter of the model.}
\vspace{-10pt}
\label{fig:spit-hist}
\end{figure*}

\begin{figure*}[t!]
\centering
\hspace{-15pt}\includegraphics[height=7.0cm, clip=True]{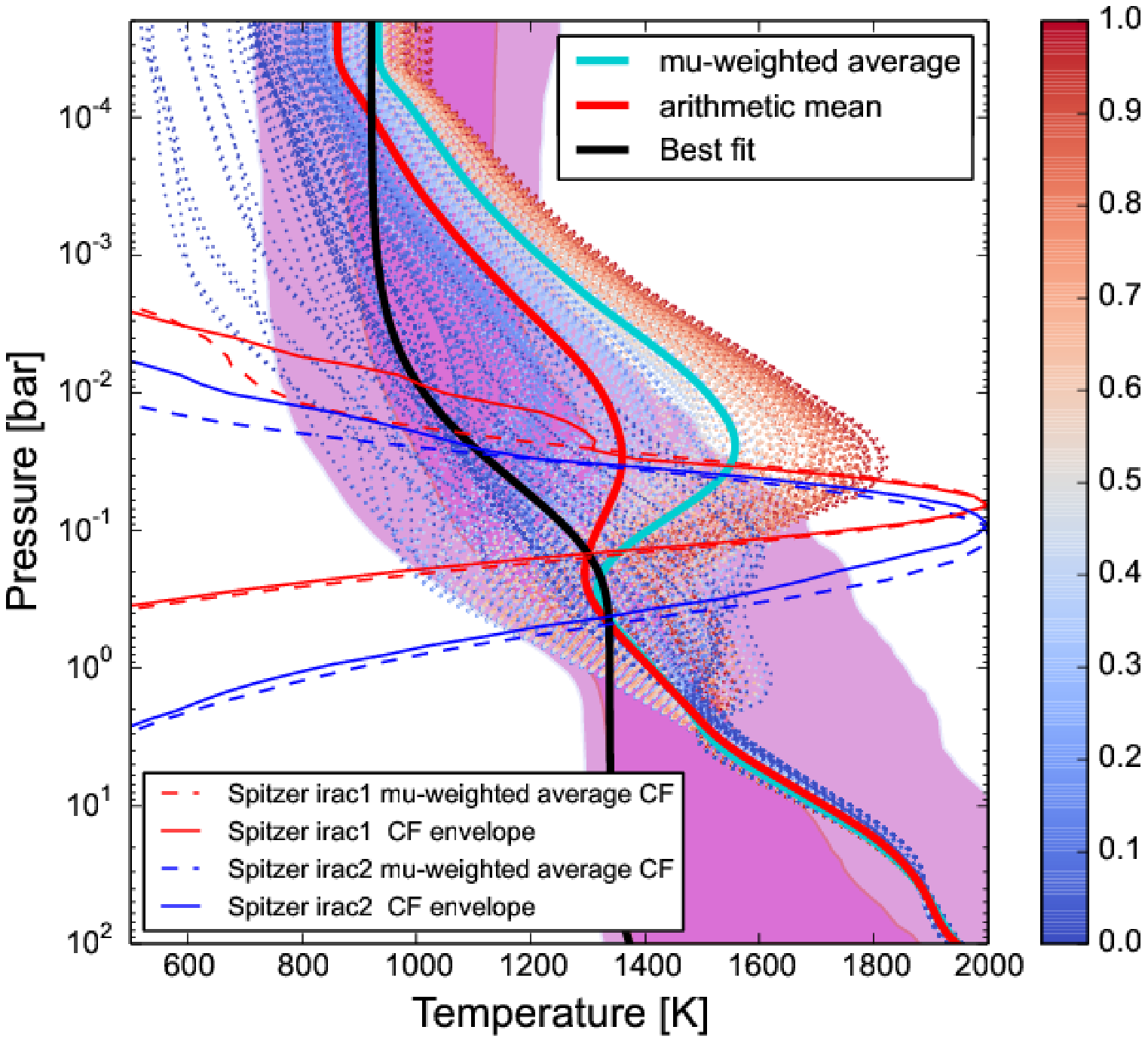}\hspace{-35pt}
\includegraphics[height=7.0cm, clip=True]{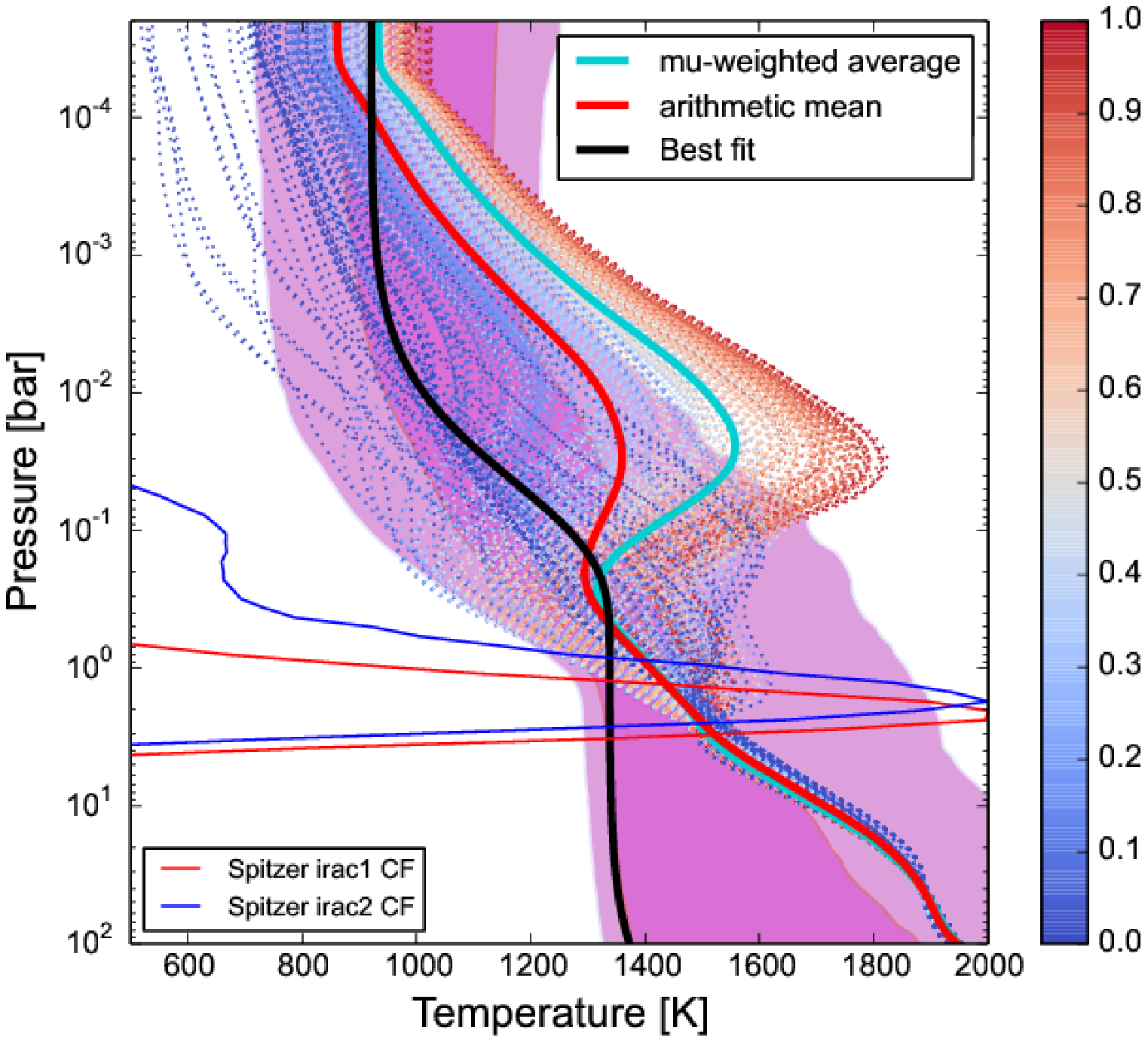}
\caption{Left: the 3D \math{T-P} profile dayside structure of HD 189733b, with the retrieved best-fit temperature profile (black curve) from Figure \ref{fig:spit}, right panel, and the 3D thermal structure averages (red and turquoise curves), overplotted with {\em only the {\em Spitzer} theoretical} contribution functions, normalized to 2000, and generated using {\em Parametrization I}, Appendix \ref{sec:Line}. Red and turquoise curves are the {\em Spitzer} theoretical contribution functions (see Figure \ref{fig:Theor-CF-aver}). Right: the 3D \math{T-P} profile dayside structure of HD 189733b, with the retrieved best-fit temperature profile (black curve) and 3D thermal structure averages (red and turquoise curves), overplotted with {\em only the {\em Spitzer} retrieved} contribution functions, normalized to 2000, and generated using {\em Parametrization I}, Appendix \ref{sec:Line}. Red and turquoise curves are the {\em Spitzer} contribution functions.}
\label{fig:spit-cf}
\end{figure*}

As a final test, we use only the {\em Spitzer} data. Attempts have been
made in the literature to make conclusions about the presence of the
temperature inversion and energy redistribution (energy budget) just
based on the two {\em Spitzer} points \citep[e.g.,][]{KnutsonEtal2007natHD189733b, BlecicEtal2013apjWASP14b}. When there
is a great difference in flux values between the channels 1 and 2, and
the channel 2 flux has higher brightness temperature, a thermal
inversion in the planetary atmosphere is inferred.

We present here the results using temperature Parametrization
I, but again we tried Parametrization II, which confirmed the
same result. The spectrum looks rather flat  (Figure \ref{fig:spit}), with the lines coming mostly from the region where most of the CO\sb{2}/CO and
H\sb{2}O lines can be seen. The 1\math{\sigma} and 2\math{\sigma}  regions cover a wide
range of temperatures and shapes, again falling around the cold
terminator region of our 3D model. The posterior
histograms are not well constrained (Figure \ref{fig:spit-hist}). The retrieved contribution
functions are placed low in the planetary atmosphere, which does
not provide valuable information about the inverted/non-inverted
part of the atmosphere (Figure \ref{fig:spit-cf}). However, the best-fit profile
infers a non-inverted atmosphere.

\end{appendices}

\clearpage
\bibliography{3Dretriev}

\begin{thebibliography}{74}
\expandafter\ifx\csname natexlab\endcsname\relax\def\natexlab#1{#1}\fi

\bibitem[{{Amundsen} {et~al.}(2014){Amundsen}, {Baraffe}, {Tremblin},
  {Manners}, {Hayek}, {Mayne}, \& {Acreman}}]{Amundsen2014}
{Amundsen}, D.~S., {Baraffe}, I., {Tremblin}, P., {Manners}, J., {Hayek}, W.,
  {Mayne}, N.~J., \& {Acreman}, D.~M. 2014, \aap, 564, A59,
  \adsurl{http://adsabs.harvard.edu/abs/2014A\%26A...564A..59A},
  \eprint{1402.0814}

\bibitem[{{Amundsen} {et~al.}(2016){Amundsen}, {Mayne}, {Baraffe}, {Manners},
  {Tremblin}, {Drummond}, {Smith}, {Acreman}, \& {Homeier}}]{Amundsen2016}
{Amundsen}, D.~S. {et~al.} 2016, \aap, 595, A36,
  \adsurl{http://adsabs.harvard.edu/abs/2016A\%26A...595A..36A},
  \eprint{1608.08593}

\bibitem[{{Asplund} {et~al.}(2009){Asplund}, {Grevesse}, {Sauval}, \&
  {Scott}}]{AsplundEtal2009-SunAbundances}
{Asplund}, M., {Grevesse}, N., {Sauval}, A.~J., \& {Scott}, P. 2009, \araa, 47,
  481, \adsurl{http://adsabs.harvard.edu/abs/2009ARA\%26A..47..481A},
  \eprint{0909.0948}

\bibitem[{{Barstow} {et~al.}(2015){Barstow}, {Aigrain}, {Irwin}, {Kendrew}, \&
  {Fletcher}}]{Barstow2015}
{Barstow}, J.~K., {Aigrain}, S., {Irwin}, P.~G.~J., {Kendrew}, S., \&
  {Fletcher}, L.~N. 2015, \mnras, 451, 1306,
  \adsurl{http://adsabs.harvard.edu/abs/2015MNRAS.451.1306B}

\bibitem[{{Batalha} {et~al.}(2015){Batalha}, {Kalirai}, {Lunine}, {Clampin}, \&
  {Lindler}}]{Batalha2015}
{Batalha}, N., {Kalirai}, J., {Lunine}, J., {Clampin}, M., \& {Lindler}, D.
  2015, ArXiv e-prints,
  \adsurl{http://adsabs.harvard.edu/abs/2015arXiv150702655B},
  \eprint{1507.02655}

\bibitem[{{Batalha} \& {Line}(2017)}]{BatalhaLine2017AJ-JWST}
{Batalha}, N.~E., \& {Line}, M.~R. 2017, \aj, 153, 151,
  \adsurl{http://adsabs.harvard.edu/abs/2017AJ....153..151B},
  \eprint{1612.02085}

\bibitem[{{Beichman} {et~al.}(2014){Beichman}, {Benneke}, {Knutson}, {Smith},
  {Lagage}, {Dressing}, {Latham}, {Lunine}, {Birkmann}, {Ferruit}, {Giardino},
  {Kempton}, {Carey}, {Krick}, {Deroo}, {Mandell}, {Ressler}, {Shporer},
  {Swain}, {Vasisht}, {Ricker}, {Bouwman}, {Crossfield}, {Greene}, {Howell},
  {Christiansen}, {Ciardi}, {Clampin}, {Greenhouse}, {Sozzetti}, {Goudfrooij},
  {Hines}, {Keyes}, {Lee}, {McCullough}, {Robberto}, {Stansberry}, {Valenti},
  {Rieke}, {Rieke}, {Fortney}, {Bean}, {Kreidberg}, {Ehrenreich}, {Deming},
  {Albert}, {Doyon}, \& {Sing}}]{Beichman2014}
{Beichman}, C. {et~al.} 2014, \pasp, 126, 1134,
  \adsurl{http://adsabs.harvard.edu/abs/2014PASP..126.1134B}

\bibitem[{{Benneke}(2015)}]{BennekeEtal2015ApjCOratios}
{Benneke}, B. 2015, ArXiv e-prints,
  \adsurl{http://adsabs.harvard.edu/abs/2015arXiv150407655B},
  \eprint{1504.07655}

\bibitem[{{Benneke} \& {Seager}(2012)}]{BennekeSeager2012-Retrieval}
{Benneke}, B., \& {Seager}, S. 2012, \apj, 753, 100,
  \adsurl{http://adsabs.harvard.edu/abs/2012ApJ...753..100B},
  \eprint{1203.4018}

\bibitem[{{Blecic}(2016)}]{Blecic2016arXiv-dissertation}
{Blecic}, J. 2016, ArXiv e-prints,
  \adsurl{http://adsabs.harvard.edu/abs/2016arXiv160402692B},
  \eprint{1604.02692}

\bibitem[{{Blecic} {et~al.}(2016){Blecic}, {Harrington}, \&
  {Bowman}}]{BlecicEtal2016-TEA}
{Blecic}, J., {Harrington}, J., \& {Bowman}, M.~O. 2016, \apjs, 225, 4,
  \adsurl{http://adsabs.harvard.edu/abs/2016ApJS..225....4B},
  \eprint{1505.06392}

\bibitem[{{Blecic} {et~al.}(2017){Blecic}, {Harrington}, {Cubillos},
  {Maddison}, \& {Foster}}]{BlecicEtal2017-BART}
{Blecic}, J., {Harrington}, J., {Cubillos}, P., {Maddison}, S., \& {Foster}, A.
  2017, in prep

\bibitem[{{Blecic} {et~al.}(2014){Blecic}, {Harrington}, {Madhusudhan},
  {Stevenson}, {Hardy}, {Cubillos}, {Hardin}, {Bowman}, {Nymeyer}, {Anderson},
  {Hellier}, {Smith}, \& {Collier Cameron}}]{BlecicEtal2014apjWASP43b}
{Blecic}, J. {et~al.} 2014, \apj, 781, 116,
  \adsurl{http://adsabs.harvard.edu/abs/2014ApJ...781..116B},
  \eprint{1302.7003}

\bibitem[{{Blecic} {et~al.}(2013){Blecic}, {Harrington}, {Madhusudhan},
  {Stevenson}, {Hardy}, {Cubillos}, {Hardin}, {Campo}, {Bowman}, {Nymeyer},
  {Loredo}, {Anderson}, \& {Maxted}}]{BlecicEtal2013apjWASP14b}
---. 2013, \apj, 779, 5,
  \adsurl{http://adsabs.harvard.edu/abs/2013ApJ...779....5B},
  \eprint{1111.2363}

\bibitem[{{Borysow}(2002)}]{Borysow2002-H2H2lowT}
{Borysow}, A. 2002, \aap, 390, 779,
  \adsurl{http://adsabs.harvard.edu/abs/2002A\%26A...390..779B}

\bibitem[{{Borysow} {et~al.}(2001){Borysow}, {Jorgensen}, \&
  {Fu}}]{BorysowEtal2001-H2H2highT}
{Borysow}, A., {Jorgensen}, U.~G., \& {Fu}, Y. 2001, \jqsrt, 68, 235,
  \adsurl{http://adsabs.harvard.edu/abs/2001JQSRT..68..235B}

\bibitem[{{Bouchy} {et~al.}(2005){Bouchy}, {Udry}, {Mayor}, {Moutou}, {Pont},
  {Iribarne}, {da Silva}, {Ilovaisky}, {Queloz}, {Santos}, {S{\'e}gransan}, \&
  {Zucker}}]{Bouchy2005-HD189733b}
{Bouchy}, F. {et~al.} 2005, \aap, 444, L15,
  \adsurl{http://adsabs.harvard.edu/abs/2005A\%26A...444L..15B},
  \eprint{astro-ph/0510119}

\bibitem[{Braak(2006)}]{terBraak2006markov}
Braak, C. J.~T. 2006, Statistics and Computing, 16, 239

\bibitem[{{Burrows} \& {Sharp}(1999)}]{BurrowsSharp1999apjchemeq}
{Burrows}, A., \& {Sharp}, C.~M. 1999, \apj, 512, 843

\bibitem[{{Burrows} {et~al.}(2006){Burrows}, {Sudarsky}, \&
  {Hubeny}}]{BurrowsSudarskyHubeny2006apjTheorySpectra}
{Burrows}, A., {Sudarsky}, D., \& {Hubeny}, I. 2006, \apj, 650, 1140,
  \adsurl{http://adsabs.harvard.edu/abs/2006ApJ...650.1140B},
  \eprint{astro-ph/0607014}

\bibitem[{{Burrows}(2014)}]{Burrows2014-review}
{Burrows}, A.~S. 2014, Proceedings of the National Academy of Science, 111,
  12601, \adsurl{http://adsabs.harvard.edu/abs/2014PNAS..11112601B},
  \eprint{1312.2009}

\bibitem[{Castelli \& Kurucz(2004)}]{CastelliKurucz-2004new}
Castelli, F., \& Kurucz, R. 2004, arXiv preprint astro-ph/0405087

\bibitem[{{Cubillos} {et~al.}(2017{\natexlab{a}}){Cubillos}, {Harrington},
  {Blecic}, {Rojo}, {Loredo}, {Bowman}, {Foster}, {Stemm}, \&
  {Lust}}]{CubillosEtal2017-BART}
{Cubillos}, P. {et~al.} 2017{\natexlab{a}}, in prep

\bibitem[{{Cubillos} {et~al.}(2017{\natexlab{b}}){Cubillos}, {Harrington},
  {Loredo}, {Lust}, {Blecic}, \& {Stemm}}]{CubillosEtal2017apjRednoise}
{Cubillos}, P., {Harrington}, J., {Loredo}, T.~J., {Lust}, N.~B., {Blecic}, J.,
  \& {Stemm}, M. 2017{\natexlab{b}}, \aj, 153, 3,
  \adsurl{http://adsabs.harvard.edu/abs/2017AJ....153....3C},
  \eprint{1610.01336}

\bibitem[{{Cubillos}(2016)}]{Cubillos2016arXiv-dissertation}
{Cubillos}, P.~E. 2016, ArXiv e-prints,
  \adsurl{http://adsabs.harvard.edu/abs/2016arXiv160401320C},
  \eprint{1604.01320}

\bibitem[{{Deming} {et~al.}(2009){Deming}, {Seager}, {Winn}, {Miller-Ricci},
  {Clampin}, {Lindler}, {Greene}, {Charbonneau}, {Laughlin}, {Ricker},
  {Latham}, \& {Ennico}}]{Deming2009}
{Deming}, D. {et~al.} 2009, \pasp, 121, 952,
  \adsurl{http://adsabs.harvard.edu/abs/2009PASP..121..952D},
  \eprint{0903.4880}

\bibitem[{{Dobbs-Dixon} \& {Agol}(2013)}]{Dobbs-DixonAgol2013-RHD}
{Dobbs-Dixon}, I., \& {Agol}, E. 2013, \mnras, 435, 3159,
  \adsurl{http://adsabs.harvard.edu/abs/2013MNRAS.435.3159D},
  \eprint{1211.1709}

\bibitem[{{Doyon} {et~al.}(2012){Doyon}, {Hutchings}, {Beaulieu}, {Albert},
  {Lafreni{\`e}re}, {Willott}, {Touahri}, {Rowlands}, {Maszkiewicz},
  {Fullerton}, {Volk}, {Martel}, {Chayer}, {Sivaramakrishnan}, {Abraham},
  {Ferrarese}, {Jayawardhana}, {Johnstone}, {Meyer}, {Pipher}, \&
  {Sawicki}}]{Doyon2012}
{Doyon}, R. {et~al.} 2012, in \procspie, Vol. 8442, Space Telescopes and
  Instrumentation 2012: Optical, Infrared, and Millimeter Wave, 84422R,
  \adsurl{http://adsabs.harvard.edu/abs/2012SPIE.8442E..2RD}

\bibitem[{{Eriksson}(1971)}]{Eriksson1971}
{Eriksson}, G. 1971, Acta Chem.Scand.

\bibitem[{{Feng} {et~al.}(2016){Feng}, {Line}, {Fortney}, {Stevenson}, {Bean},
  {Kreidberg}, \& {Parmentier}}]{FengLineEtal2016ApJ-2TPs}
{Feng}, Y.~K., {Line}, M.~R., {Fortney}, J.~J., {Stevenson}, K.~B., {Bean}, J.,
  {Kreidberg}, L., \& {Parmentier}, V. 2016, \apj, 829, 52,
  \adsurl{http://adsabs.harvard.edu/abs/2016ApJ...829...52F},
  \eprint{1607.03230}

\bibitem[{{Fortney} {et~al.}(2006){Fortney}, {Cooper}, {Showman}, {Marley}, \&
  {Freedman}}]{FortneyEtal2006apjAtmDynamics}
{Fortney}, J.~J., {Cooper}, C.~S., {Showman}, A.~P., {Marley}, M.~S., \&
  {Freedman}, R.~S. 2006, \apj, 652, 746,
  \adsurl{http://adsabs.harvard.edu/abs/2006ApJ...652..746F},
  \eprint{astro-ph/0608235}

\bibitem[{{Fortney} {et~al.}(2005){Fortney}, {Marley}, {Lodders}, {Saumon}, \&
  {Freedman}}]{FortneyEtal2005apjlhjmodels}
{Fortney}, J.~J., {Marley}, M.~S., {Lodders}, K., {Saumon}, D., \& {Freedman},
  R. 2005, \apjl, 627, L69,
  \adsurl{http://adsabs.harvard.edu/cgi-bin/nph-bib_query?bibcode=2005ApJ...627L..69F&db_key=AST}

\bibitem[{{Fraine} {et~al.}(2014){Fraine}, {Deming}, {Benneke}, {Knutson},
  {Jord{\'a}n}, {Espinoza}, {Madhusudhan}, {Wilkins}, \&
  {Todorov}}]{FraineEtal2014natHATP11bH2O}
{Fraine}, J. {et~al.} 2014, \nat, 513, 526,
  \adsurl{http://adsabs.harvard.edu/abs/2014Natur.513..526F},
  \eprint{1409.8349}

\bibitem[{Gelman \& Rubin(1992)}]{GelmanRubin1992}
Gelman, A., \& Rubin, D. 1992, Statistical Science, 7, 457

\bibitem[{Gordon \& McBride(1994)}]{GordonMcBride:1994}
Gordon, S., \& McBride, B.~J. 1994, Computer Program for Calculation of Complex
  Chemical Equilibrium Compositions and Applications. {I}. Analysis, Reference
  Publication RP-1311, NASA, describes theory and numerical algorithms behind
  CEA computer program

\bibitem[{{Greene} {et~al.}(2016){Greene}, {Line}, {Montero}, {Fortney},
  {Lustig-Yaeger}, \& {Luther}}]{Greene2016}
{Greene}, T.~P., {Line}, M.~R., {Montero}, C., {Fortney}, J.~J.,
  {Lustig-Yaeger}, J., \& {Luther}, K. 2016, \apj, 817, 17,
  \adsurl{http://adsabs.harvard.edu/abs/2016ApJ...817...17G},
  \eprint{1511.05528}

\bibitem[{{Guillot}(2010)}]{Guillot2010A-LinePTprofile}
{Guillot}, T. 2010, \aap, 520, A27,
  \adsurl{http://adsabs.harvard.edu/abs/2010A\%26A...520A..27G},
  \eprint{1006.4702}

\bibitem[{Hansen {et~al.}(2014)Hansen, Schwartz, \& Cowan}]{Hansen2014features}
Hansen, C.~J., Schwartz, J.~C., \& Cowan, N.~B. 2014, \mnras, 444, 3632

\bibitem[{{Harrington} {et~al.}(2017){Harrington}, {Cubillos}, {Blecic},
  {Rojo}, {Lust}, {Challener}, {Bowman}, D., E., {Foster}, J., J., J., \&
  {Stemm}}]{HarringtonEtal2017-BART}
{Harrington}, J. {et~al.} 2017, in prep

\bibitem[{{Heng} {et~al.}(2012){Heng}, {Hayek}, {Pont}, \&
  {Sing}}]{HengEtal2012LinePTprofile}
{Heng}, K., {Hayek}, W., {Pont}, F., \& {Sing}, D.~K. 2012, \mnras, 420, 20,
  \adsurl{http://adsabs.harvard.edu/abs/2012MNRAS.420...20H},
  \eprint{1107.1390}

\bibitem[{{Heng} \& {Tsai}(2016)}]{HengTsai2016ApJ-analuticalModels}
{Heng}, K., \& {Tsai}, S.-M. 2016, \apj, 829, 104,
  \adsurl{http://adsabs.harvard.edu/abs/2016ApJ...829..104H},
  \eprint{1603.05418}

\bibitem[{{Howe} {et~al.}(2016){Howe}, {Burrows}, \& {Deming}}]{HoweEtal2016}
{Howe}, A.~R., {Burrows}, A., \& {Deming}, D. 2016, ArXiv e-prints,
  \adsurl{http://adsabs.harvard.edu/abs/2016arXiv161201245H},
  \eprint{1612.01245}

\bibitem[{{Hubeny} {et~al.}(2003){Hubeny}, {Burrows}, \&
  {Sudarsky}}]{hubeny2003}
{Hubeny}, I., {Burrows}, A., \& {Sudarsky}, D. 2003, \apj, 594, 1011,
  \adsurl{http://adsabs.harvard.edu/abs/2003ApJ...594.1011H},
  \eprint{astro-ph/0305349}

\bibitem[{{Kataria} {et~al.}(2015){Kataria}, {Showman}, {Fortney}, {Stevenson},
  {Line}, {Kreidberg}, {Bean}, \& {D{\'e}sert}}]{kataria2015atmospheric}
{Kataria}, T., {Showman}, A.~P., {Fortney}, J.~J., {Stevenson}, K.~B., {Line},
  M.~R., {Kreidberg}, L., {Bean}, J.~L., \& {D{\'e}sert}, J.-M. 2015, \apj,
  801, 86, \adsurl{http://adsabs.harvard.edu/abs/2015ApJ...801...86K},
  \eprint{1410.2382}

\bibitem[{{Kendrew} {et~al.}(2015){Kendrew}, {Scheithauer}, {Bouchet},
  {Amiaux}, {Azzollini}, {Bouwman}, {Chen}, {Dubreuil}, {Fischer}, {Glasse},
  {Greene}, {Lagage}, {Lahuis}, {Ronayette}, {Wright}, \&
  {Wright}}]{Kendrew2015P}
{Kendrew}, S. {et~al.} 2015, \pasp, 127, 623,
  \adsurl{http://adsabs.harvard.edu/abs/2015PASP..127..623K},
  \eprint{1512.03000}

\bibitem[{{Knutson} {et~al.}(2007){Knutson}, {Charbonneau}, {Allen}, {Fortney},
  {Agol}, {Cowan}, {Showman}, {Cooper}, \&
  {Megeath}}]{KnutsonEtal2007natHD189733b}
{Knutson}, H.~A. {et~al.} 2007, \nat, 447, 183,
  \adsurl{http://adsabs.harvard.edu/abs/2007Natur.447..183K},
  \eprint{0705.0993}

\bibitem[{{Knutson} {et~al.}(2009){Knutson}, {Charbonneau}, {Cowan}, {Fortney},
  {Showman}, {Agol}, {Henry}, {Everett}, \&
  {Allen}}]{KnutsonEtal2009ApJ-redistribution}
---. 2009, \apj, 690, 822,
  \adsurl{http://adsabs.harvard.edu/abs/2009ApJ...690..822K},
  \eprint{0802.1705}

\bibitem[{{Kreidberg} {et~al.}(2014){Kreidberg}, {Bean}, {D{\'e}sert}, {Line},
  {Fortney}, {Madhusudhan}, {Stevenson}, {Showman}, {Charbonneau},
  {McCullough}, {Seager}, {Burrows}, {Henry}, {Williamson}, {Kataria}, \&
  {Homeier}}]{Kreidberg2014}
{Kreidberg}, L. {et~al.} 2014, \apjl, 793, L27,
  \adsurl{http://adsabs.harvard.edu/abs/2014ApJ...793L..27K},
  \eprint{1410.2255}

\bibitem[{{Laraia} {et~al.}(2011){Laraia}, {Gamache}, {Lamouroux}, {Gordon}, \&
  {Rothman}}]{LaraiaEtal2011TIPS}
{Laraia}, A.~L., {Gamache}, R.~R., {Lamouroux}, J., {Gordon}, I.~E., \&
  {Rothman}, L.~S. 2011, Icarus, 215, 391,
  \adsurl{http://adsabs.harvard.edu/abs/2011Icar..215..391L}

\bibitem[{{Laughlin} \& {Lissauer}(2015)}]{LaughlinLissauer2015}
{Laughlin}, G., \& {Lissauer}, J.~J. 2015, ArXiv e-prints,
  \adsurl{http://adsabs.harvard.edu/abs/2015arXiv150105685L},
  \eprint{1501.05685}

\bibitem[{Lee {et~al.}(2012)Lee, Fletcher, \& Irwin}]{LeeEtal2012-CF}
Lee, J.-M., Fletcher, L.~N., \& Irwin, P.~G. 2012, \mnras, 420, 170

\bibitem[{{Line} {et~al.}(2014){Line}, {Knutson}, {Wolf}, \&
  {Yung}}]{LineEtal2014-Retrieval-II}
{Line}, M.~R., {Knutson}, H., {Wolf}, A.~S., \& {Yung}, Y.~L. 2014, \apj, 783,
  70, \adsurl{http://adsabs.harvard.edu/abs/2014ApJ...783...70L},
  \eprint{1309.6663}

\bibitem[{{Line} {et~al.}(2013){Line}, {Wolf}, {Zhang}, {Knutson}, {Kammer},
  {Ellison}, {Deroo}, {Crisp}, \& {Yung}}]{LineEtal2013-Retrieval-I}
{Line}, M.~R. {et~al.} 2013, \apj, 775, 137,
  \adsurl{http://adsabs.harvard.edu/abs/2013ApJ...775..137L},
  \eprint{1304.5561}

\bibitem[{{Line} \& {Yung}(2013)}]{LineEtal2013-Retrieval-III}
{Line}, M.~R., \& {Yung}, Y.~L. 2013, \apj, 779, 3,
  \adsurl{http://adsabs.harvard.edu/abs/2013ApJ...779....3L},
  \eprint{1309.6679}

\bibitem[{{Madhusudhan} \&
  {Seager}(2009)}]{MadhusudhanSeager2009ApJ-AbundanceMethod}
{Madhusudhan}, N., \& {Seager}, S. 2009, \apj, 707, 24,
  \adsurl{http://adsabs.harvard.edu/abs/2009ApJ...707...24M},
  \eprint{0910.1347}

\bibitem[{{Madhusudhan} \& {Seager}(2010)}]{MadhusudhanSeager2010}
---. 2010, \apj, 725, 261,
  \adsurl{http://adsabs.harvard.edu/abs/2010ApJ...725..261M},
  \eprint{1010.4585}

\bibitem[{{Mihalas} {et~al.}(1978){Mihalas}, {Auer}, \&
  {Mihalas}}]{Mihalas1978}
{Mihalas}, D., {Auer}, L.~H., \& {Mihalas}, B.~R. 1978, \apj, 220, 1001,
  \adsurl{http://adsabs.harvard.edu/abs/1978ApJ...220.1001M}

\bibitem[{{Molli{\`e}re} {et~al.}(2017){Molli{\`e}re}, {van Boekel}, {Bouwman},
  {Henning}, {Lagage}, \& {Min}}]{Molliere2017-JWST}
{Molli{\`e}re}, P., {van Boekel}, R., {Bouwman}, J., {Henning}, T., {Lagage},
  P.-O., \& {Min}, M. 2017, \aap, 600, A10,
  \adsurl{http://adsabs.harvard.edu/abs/2017A\%26A...600A..10M}

\bibitem[{{Parmentier} \&
  {Guillot}(2014)}]{ParmentierGuillot2014-LinesPTprofile}
{Parmentier}, V., \& {Guillot}, T. 2014, \aap, 562, A133,
  \adsurl{http://adsabs.harvard.edu/abs/2014A\%26A...562A.133P},
  \eprint{1311.6597}

\bibitem[{{Rauscher} \& {Menou}(2012)}]{rauscher2012}
{Rauscher}, E., \& {Menou}, K. 2012, \apj, 750, 96,
  \adsurl{http://adsabs.harvard.edu/abs/2012ApJ...750...96R},
  \eprint{1112.1658}

\bibitem[{{Richard} {et~al.}(2012){Richard}, {Gordon}, {Rothman}, {Abel},
  {Frommhold}, {Gustafsson}, {Hartmann}, {Hermans}, {Lafferty}, {Orton},
  {Smith}, \& {Tran}}]{RichardEtal2012-HITRAN-CIA}
{Richard}, C. {et~al.} 2012, \jqsrt, 113, 1276,
  \adsurl{http://adsabs.harvard.edu/abs/2012JQSRT.113.1276R}

\bibitem[{{Rojo}(2006)}]{Rojo2006PhDtransit}
{Rojo}, P. 2006, {PhD} dissertation, Cornell University

\bibitem[{{Rojo} {et~al.}(2009){Rojo}, {Harrington}, {Deming}, \&
  {Fortney}}]{RojoEtal-HD209}
{Rojo}, P., {Harrington}, J., {Deming}, D., \& {Fortney}, J. 2009, in
  Astronomical Society of the Pacific Conference Series, Vol. 420, Bioastronomy
  2007: Molecules, Microbes and Extraterrestrial Life, ed. K.~J. {Meech}, J.~V.
  {Keane}, M.~J. {Mumma}, J.~L. {Siefert}, \& D.~J. {Werthimer}, 321,
  \adsurl{http://adsabs.harvard.edu/abs/2009ASPC..420..321R}

\bibitem[{Rothman {et~al.}(2013)Rothman, Gordon, Babikov, Barbe, Benner,
  Bernath, Birk, Bizzocchi, Boudon, Brown, {et~al.}}]{rothman2013-hitran2012}
Rothman, L. {et~al.} 2013, \jqsrt, 130, 4

\bibitem[{Rothman {et~al.}(2010)Rothman, Gordon, Barber, Dothe, Gamache,
  Goldman, Perevalov, Tashkun, \& Tennyson}]{rothman2010-hitemp}
---. 2010, \jqsrt, 111, 2139

\bibitem[{{Sharp} \& {Burrows}(2007)}]{sharp2007}
{Sharp}, C.~M., \& {Burrows}, A. 2007, \apjs, 168, 140,
  \adsurl{http://ads.ari.uni-heidelberg.de/abs/2007ApJS..168..140S},
  \eprint{astro-ph/0607211}

\bibitem[{{Showman} {et~al.}(2009){Showman}, {Fortney}, {Lian}, {Marley},
  {Freedman}, {Knutson}, \& {Charbonneau}}]{showman2009}
{Showman}, A.~P., {Fortney}, J.~J., {Lian}, Y., {Marley}, M.~S., {Freedman},
  R.~S., {Knutson}, H.~A., \& {Charbonneau}, D. 2009, \apj, 699, 564

\bibitem[{{Southworth}(2010)}]{Southworth2010}
{Southworth}, J. 2010, \mnras, 408, 1689,
  \adsurl{http://adsabs.harvard.edu/abs/2010MNRAS.408.1689S},
  \eprint{1006.4443}

\bibitem[{Swain {et~al.}(2013)Swain, Deroo, Tinetti, Hollis, Tessenyi, Line,
  Kawahara, Fujii, Showman, \& Yurchenko}]{Swain2013-WASP12b}
Swain, M. {et~al.} 2013, Icarus, 225, 432

\bibitem[{{Swain} {et~al.}(2008){Swain}, {Vasisht}, \&
  {Tinetti}}]{SwainEtal2008NatureHD189733bspec}
{Swain}, M.~R., {Vasisht}, G., \& {Tinetti}, G. 2008, \nat, 452, 329,
  \adsurl{http://adsabs.harvard.edu/abs/2008Natur.452..329S}

\bibitem[{ter Braak \& Vrugt(2008)}]{terBraak2008differential}
ter Braak, C.~J., \& Vrugt, J.~A. 2008, Statistics and Computing, 18, 435

\bibitem[{{Triaud} {et~al.}(2009){Triaud}, {Queloz}, {Bouchy}, {Moutou},
  {Collier Cameron}, {Claret}, {Barge}, {Benz}, {Deleuil}, {Guillot},
  {H{\'e}brard}, {Lecavelier Des {\'E}tangs}, {Lovis}, {Mayor}, {Pepe}, \&
  {Udry}}]{Triaud2009-HD189733b}
{Triaud}, A.~H.~M.~J. {et~al.} 2009, \aap, 506, 377,
  \adsurl{http://adsabs.harvard.edu/abs/2009A\%26A...506..377T},
  \eprint{0907.2956}

\bibitem[{{Waldmann} {et~al.}(2015){Waldmann}, {Tinetti}, {Rocchetto},
  {Barton}, {Yurchenko}, \& {Tennyson}}]{Waldmann2015-TAU}
{Waldmann}, I.~P., {Tinetti}, G., {Rocchetto}, M., {Barton}, E.~J.,
  {Yurchenko}, S.~N., \& {Tennyson}, J. 2015, \apj, 802, 107,
  \adsurl{http://adsabs.harvard.edu/abs/2015ApJ...802..107W},
  \eprint{1409.2312}

\bibitem[{{White} {et~al.}(1958){White}, {Johnson}, \&
  {Dantzig}}]{WhiteJohnsonDantzig1958JGibbs}
{White}, W.~B., {Johnson}, S.~M., \& {Dantzig}, G.~B. 1958, \jcp, 28, 751,
  \adsurl{http://adsabs.harvard.edu/abs/1958JChPh..28..751W}

\end{thebibliography}

\end{document}